%% file: Paper.tex
\documentclass{article}
\usepackage{graphicx} % Required for inserting images
\usepackage{amsmath,amssymb}
\usepackage{amsfonts}
\usepackage{amsthm}
\usepackage{natbib}
\usepackage{tikz}
\usepackage{appendix}
\usetikzlibrary{arrows.meta}

\usepackage[margin=1in]{geometry}
\input{preamble.tex}  % added by Hiro
\title{Sticky Manifold Piecewise Deterministic Monte Carlo for Bayesian Model Averaging}
\author[1,2]{Hirofumi Shiba}
\author[3]{Luke Hardcastle}
\author[1,2]{Kengo Kamatani}
\affil[1]{Institute of Statistical Mathematics, Tokyo, Japan}
\affil[2]{Graduate University for Advanced Studies (SOKENDAI), Tokyo, Japan}
\affil[3]{MRC Biostatistics Unit, University of Cambridge, Cambridge, UK}
\begin{document}
\maketitle
\begin{abstract}
    Assigning positive prior probability to lower-dimensional submodels enriches the range of uncertainty that Bayesian models can represent. The resulting posterior distributions combine continuous parameter uncertainty with discrete model uncertainty. Efficient movement between these submodels, however, can be difficult to design. We propose the sticky manifold construction as a principled, modular framework for combining piecewise deterministic Markov processes (PDMPs) across different submodels. The integration is rejection-free, requires no additional likelihood evaluations, and retains the non-reversible and ballistic nature of the component PDMP dynamics. We prove that the resulting process leaves the intended mixed-dimensional target invariant.
    We also develop a manifold PDMP algorithm based on coordinate charts to provide the within-submodel dynamics and a blow-up procedure that resolves singular sets into codimension-one sticky boundaries. This geometric construction makes the remaining design choices explicit; component dynamics and unsticking kernels can be chosen freely, subject to easily verifiable velocity compatibility conditions. Finally, we demonstrate the sticky manifold algorithms in Bayesian model averaging examples, propagating uncertainty over pure versus mixed states in quantum state tomography, rank in Bayesian principal component analysis, and parametric families in survival extrapolation.
\end{abstract}

\section{Introduction}\label{sec;1Introducion}

Statistical models are commonly defined such that the number of model parameters is unknown. In the Bayesian paradigm, with a prior on all unknown quantities, Markov Chain Monte Carlo (MCMC) samplers are then required to sample across several submodels. This is most commonly achieved using the celebrated reversible jump MCMC algorithm \citep{Green1995,Green2005}. Example settings include finite mixture models \citep{Richardson1997}, variable selection \citep{Dellaportas+2002}, autoregressive models \citep{Vermaak+2004}, Bayesian neural networks \citep{Andrieu+2001}, and matrix factorisation \citep{lopes2004bayesian, fruhwirth2025sparse}, with application areas in finance, biology and epidemiology. For a recent comprehensive review see \citealt[Chapter~7]{craiu2026handbook}. 

Efficiency of reversible jump methods primarily depends on the design of efficient across-model jumps, a non-trivial task in general settings \citep{ehlers2008adaptive}. In the case of nested submodels (e.g., finite mixture models and variable selection) a useful perspective is to recast the problem as sampling from nested subspaces. This motivates the widely used split-merge proposals of \cite{Richardson1997}, the design of local across-model moves around centring points where likelihoods cancel \citep{Brooks2003}, and the same, gain, and lose proposals of \cite{Holmes-Cerfon2020} that can be constructed without explicit manifold charts. Recent works have combined this perspective with Piecewise Deterministic Markov Processes (PDMPs) in the context of variable selection, designing samplers that traverse a sequence of axis-aligned hyperplanes \citep{Chevallier2023, Bierkens2023a} using tractable continuous-time dynamics. In this paper, we extend this approach to design PDMP algorithms that traverse a general sequence of nested submanifolds and models embedded in a common encompassing model, generalising this approach to a wider class of problems beyond the case of variable selection.

Piecewise Deterministic Markov Processes are a class of Markov processes defined by deterministic dynamics interrupted by jumps at random event times \citep{Davis1984}. When interest is in sampling from a target measure $\pi$ on the state space $\mathbb{R}^d$, recent works introduce PDMPs defined on the augmented state space $(x_t,v_t) \in \mathbb{R}^{2d}$, such that the marginal stationary distribution of $x_t$ is $\pi$. Example processes include the bouncy particle sampler \citep{Bouchard2018}, the zig-zag sampler \citep{Bierkens2019}, forward event-chain Monte Carlo \citep{Michel2020}, the boomerang sampler \citep{Bierkens+2020}, and randomised Hamiltonian Monte Carlo \citep{Bou2017,Kleppe2022}. These samplers have attracted considerable attention in the MCMC literature in recent years, as velocity persistence allows for non-reversible, ballistic exploration of the state space, improving sampling efficiency \citep{Bierkens-Roberts2017,Andrieu2021}. Further, they exhibit attractive properties such as robustness to length-scale differences \citep{Bierkens-Kamatani-Roberts2024,Power2025,Lu-Luo2026} and scalability to large datasets \citep{Bierkens2019,Bierkens+2020,Agrawal2024}, in comparison to other gradient-informed MCMC schemes.

Recent works have shown that these processes can be designed to sample from target measures consisting of mixtures of atomic and continuous components \citep{Chevallier2023, Bierkens2023a,Bierkens2023b}. These targets are common in Bayesian variable selection with spike and slab priors where the prior is constructed coordinate-wise as a mixture of a Dirac mass at zero and a continuous density \citep{Mitchell-Beauchamp1988}. PDMPs with tractable deterministic dynamics can target these measures by deterministically sticking to the spike when the process intersects it, and then unsticking after a random time \citep{Chevallier2023, Bierkens2023a}. We refer to these dynamics throughout as sticky PDMPs following the terminology of \cite{Bierkens2023a}. Under this construction trans-dimensional moves involve no computation of the likelihood and require no additional tuning parameters, unlike most reversible jump MCMC algorithms. Similar considerations have motivated the design of PDMP methods for piecewise continuous densities and state spaces with boundaries \citep{Bierkens2023b, Chevallier+2024, Bierkens2018}, with several applications of PDMP samplers in the statistical literature focusing on target distributions defined on non-trivial spaces \citep{Koskela2022,Hardcastle2024,Hardcastle2025,VanDenBergh-Marsman2026,Arnstad+2026}. 

%Existing trans-dimensional samplers broadly take one of two approaches: transforming the mixed-dimensional target into a continuous one over an augmented space or adapting the sampling dynamics to the given geometry. In the former approach, \cite{Carlin-Chib1995} collect all the parameters into a common product space with the aid of `pseudo-priors', while \cite{Petris-Tardella2003}, \cite{Nishimura+2020}, and \cite{Chin-Nishimura2026} replace lower-dimensional probability mass with auxiliary continuous distributions. In the latter, the main challenge lies in constructing valid and efficient trans-dimensional transitions. Reversible-jump MCMC \citep{Green1995} is a general framework to achieve this through dimension-matched proposals, which is later adapted to stratified spaces by \cite{Holmes-Cerfon2020}. \cite{Gottardo-Raftery2008} generalised the standard Metropolis--Hastings framework to mixed-dimensional targets using a common dominating measure. We propose a new framework that belongs to the latter class.

\subsection{Our contributions}\label{subsec;Introduction-contribution}

In its simplest form, we consider target measures of the form
\begin{equation}
    \label{eq;6manifoldtarget}
    \pi(dx) \propto \exp(-U(x))\bigg(\omega\pi_{\mathbb{R}^d}(x)\,dx + (1-\omega)\pi_{\mathcal{M}}(x)\mathcal{H}^{d-1}_{\mathcal{M}}(dx)\bigg),
\end{equation}
where $\pi_{\mathbb{R}^d}$ is a density with respect to the $d$-dimensional Lebesgue measure (denoted $dx$) and $\pi_{\mathcal{M}}$ is a density on the embedded Riemannian manifold $\mathcal{M}$ with respect to the normalised Hausdorff measure %induced by the ambient Euclidean metric 
(denoted $\mathcal{H}^{d-1}_{\mathcal{M}}$). 

\begin{figure}[!ht]
    \centering
    \includegraphics[width=\linewidth]{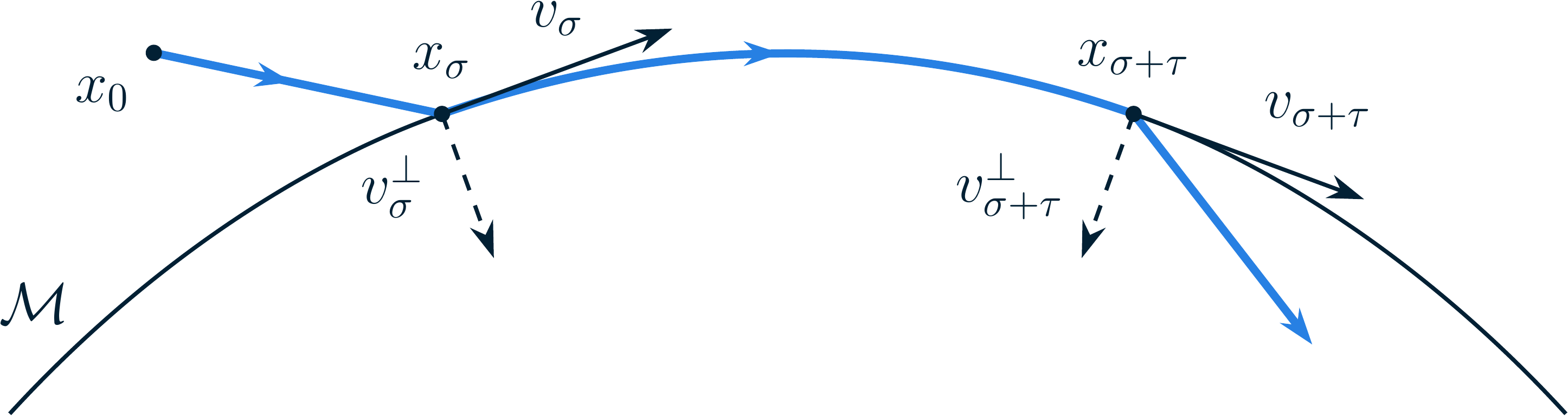}
    \caption{An illustration of the sticky manifold piecewise deterministic Markov process, whose trajectory is in blue. When it intersects $\mathcal{M}$, the dynamics are confined within $\mathcal{M}$, with the normal velocity component frozen and parallel-transported along $\mathcal{M}$, until the first arrival time $\tau$ from a Poisson process with rate $\Lambda$ \eqref{eq;6Manifoldthawingrate} elapses. See Section~\ref{sec;3StickyManifold} for details.}
    \label{fig:Sticking}
\end{figure}

Our primary contribution is to show that the following process, that we term the sticky manifold piecewise deterministic Markov process (SM-PDMP) and illustrate in Figure \ref{fig:Sticking}, is $\pi$-invariant:
\begin{enumerate}
    \item In the ambient space, the process is given by a standard Euclidean PDMP with $\exp(-U(x))\pi_{\mathbb{R}^d}(x)\,dx$ as the marginal invariant measure for $x$.
    \item When the ambient process intersects $\mathcal{M}$, it deterministically sticks to the manifold, $\mathcal{M}$.
    \item On $\mathcal{M}$, the process is given by a transformation of a standard PDMP sampler on the embedded manifold, with $\exp(-U(x))\pi_{\mathcal{M}}(x)\,\mathcal{H}^{d-1}_{\mathcal{M}}$ as the marginal invariant measure for $x$; see Algorithm~\ref{alg;ManifoldBPS} in Section~\ref{sec;2ManifoldPDMPs}.
    %Although the natural choice of deterministic dynamics here may be given by the geodesic flow, we also show that these dynamics can be described by a standard PDMP sampler defined in a coordinate system for $\mathcal{M}$ in Section~\ref{sec;2ManifoldPDMPs}.
    \item The process leaves the manifold with an unsticking rate and kernel that preserves the velocity of the process relative to $\mathcal{M}$ before the sticking event, allowing the process to move from $\mathcal{M}$ to the ambient space. This rate is given by
        \begin{equation}
        \label{eq;6Manifoldthawingrate}
            \Lambda(x) = \|v^\perp_\sigma\|\frac{\omega}{1-\omega}\frac{\pi_{\mathbb{R}^d}(x)}{\pi_{\mathcal{M}}(x)},
        \end{equation}
    where $v^\perp_\sigma$ is the component of the velocity that is orthogonal to the manifold at the hitting time $\sigma$.
\end{enumerate}

The above construction admits the original sticky PDMP algorithm \citep{Bierkens2023a} as a special case and, in particular, it retains the attractive property that the unsticking rate \eqref{eq;6Manifoldthawingrate} does not depend on the negative log-likelihood $U$ and can be computed only through priors. Our full algorithm (Algorithm~\ref{alg;TheAlgorithm}) further allows stochastic transformation upon sticking and unsticking and applies beyond the target \eqref{eq;6manifoldtarget} to the cases with a nested sequence of manifolds, in which the sampler can iteratively stick to the adjacent lower dimensional submanifolds, and cases with multiple manifolds. We also propose the blow-up construction as a key to this generalisation, while also using it to turn a singular subset into sticky boundaries. %and apply it to nested rank strata in matrix spaces (Section~\ref{subsec;BayesianPCA}).
This geometric treatment also enables the clear separation of the geometric construction of state spaces (Sections~\ref{subsec;preliminaries} and~\ref{subsec-blowup}) and invariance proofs (Sections~\ref{subsec;pullback-PDMP} and~\ref{subsec;invarianceProof}) from the remaining algorithmic choices of specific dynamics and unsticking kernels (Sections~\ref{subsec;TheCoordinateBasedMPDMP} and~\ref{subsec;theAlgorithm}). Our invariance results centre on Theorem~\ref{thm;NewTheorem}, whose proof rests on a flow-wise Gauss--Green formula that provides a common basis for further generalisations. Extending the extended-generator approach of \cite{Chevallier2023,Chevallier+2024}, the formula applies to non-linear flows, test functions that are only absolutely continuous along the flow, and general curved boundaries.

%To support such a generic procedure, two technical ingredients underlie these constructions. First, as a core engine to prove the main invariance Theorem~\ref{thm;NewTheorem}, we establish a flow-wise Gauss--Green lemma that provides a single concise invariance argument applicable to a broad class of PDMP samplers used in the ambient space or within the submanifold. Following the extended generator approach of \cite{Chevallier2023,Chevallier+2024}, the argument applies directly to test functions in the extended generator domain, thereby avoiding the functional-analytic detour of establishing the Feller property or identifying the strong generator core. Second, we reframe the state space construction in \cite{Bierkens2023a} in terms of the geometric notion of radial blow-up. This generalised construction accommodates a wide range of submanifolds, including such singular sets as treated in Section~\ref{subsec;illustration-blowup-cone}, to be integrated within our methodology. Such a purely geometric treatment also has the advantage of making the remaining design choices explicit.

Before introducing our sticky manifold framework in Section~\ref{sec;3StickyManifold}, we separately treat its subroutine that does manifold sampling within $\mathcal{M}$ in Section \ref{sec;2ManifoldPDMPs}. This method extends existing PDMPs to target measures defined on a Riemannian manifold $\mathcal{M}$ based on their coordinate charts. When a suitable chart is available, our method provides a largely plug-and-play route to manifold sampling, allowing existing PDMP methodology to be reused rather than redesigned for each manifold.

\subsection{A motivating example}\label{subsec-GMM-illustration}

Target measures in the form of \eqref{eq;6manifoldtarget} commonly arise in Bayesian model averaging. To motivate this, we consider the simple example of fitting a Gaussian mixture model with a prior on the number of components, 
\begin{equation*}
    f(y)=\sum_{k=1}^K\frac{1}{K}h(y\,|\,\mu_k,\sigma^2),
\end{equation*}
where $h(y\,|\, \mu,\sigma^2)$ denotes a Gaussian density function with mean $\mu$ and variance $\sigma^2$. For simplicity of exposition, we assume that components are equally weighted with shared variance $\sigma^2$, and that the prior on $K$ is restricted to $K \in \{1,2\}$. This then translates to the implied prior on $\mu$
\begin{equation*}
    \pi(d\mu)=\omega\pi_{\mathbb{R}^2}(\mu)\,d\mu+(1-\omega)\pi_{\mathcal{M}}(\mu)\,\mathcal{H}^1_{\mathcal{M}}(d\mu), \quad \mathcal{M} = \{\mu \in \mathbb{R}^2\mid\mu_1 = \mu_2\}.
\end{equation*}
When the density $\pi_{\mathcal{M}}$ is the normalised restriction of $\pi_{\mathbb{R}^2}$ on $\mathcal{M}$,
the unsticking rate \eqref{eq;6Manifoldthawingrate} becomes constant
\begin{equation*}
    \Lambda= \frac{\omega}{1-\omega}\|v^\perp_\sigma\|Z,\qquad Z:=\int_{\mathcal{M}}\pi_{\mathbb{R}^2}(x)\,\mathcal{H}^1(dx).
\end{equation*}

\begin{figure}[H]
    \centering
    \includegraphics[width=\linewidth]{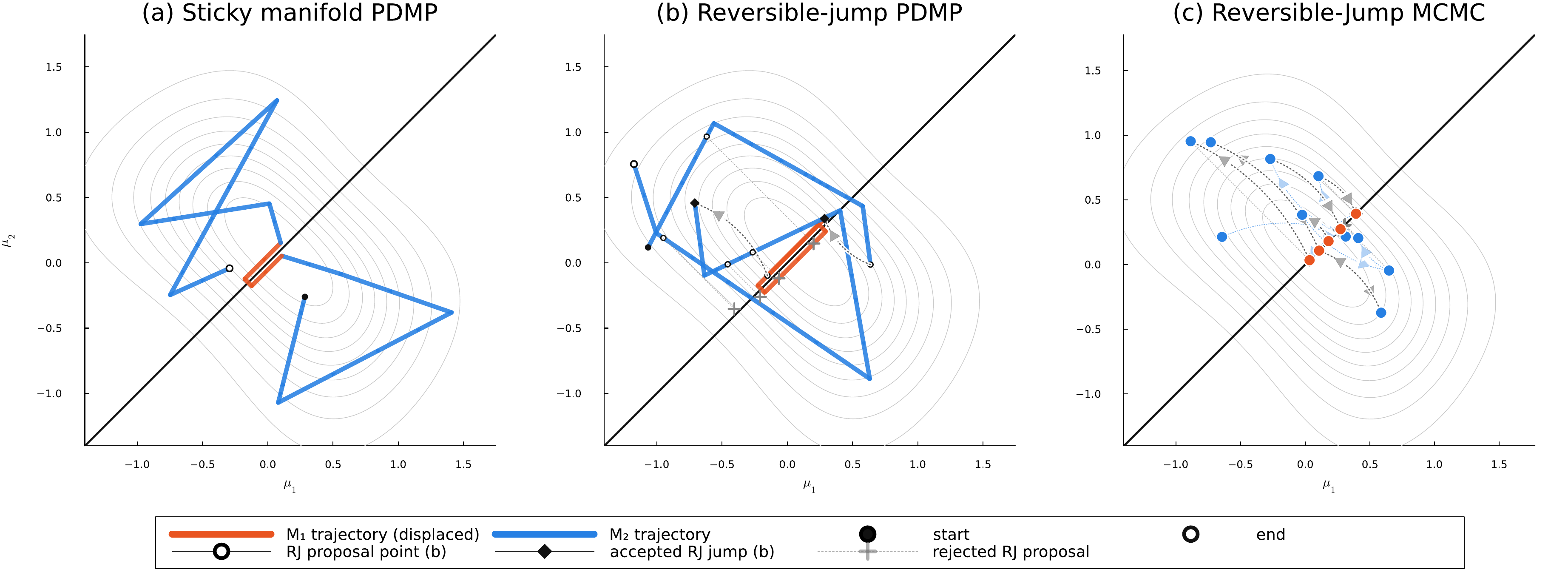}
    \caption{Representative trajectories on the common posterior geometry: (a) sticky manifold forward event-chain Monte Carlo, (b) reversible-jump forward event-chain Monte Carlo in the sense of \cite{Hardcastle2024}, and (c) reversible-jump random walk Metropolis--Hastings.
    In panel (b), the trajectory crosses the manifold without sticking, whereas in panel (a), it sticks every time it intersects the manifold.
    The dotted black arrows in panel (b) and (c) indicate the reversible-jump moves, while the dotted blue arrows in panel (c) indicate the random-walk moves. In (b) and (c), there are some rejected reversible-jump proposals depicted by the plus mark (+), while the sticking-and-unsticking mechanism is rejection-free. Manifold trajectories are displaced slightly in the normal direction for visibility.% The contours show the ambient posterior component.
    }
    \label{fig:6PDMPGMM}
\end{figure}

Employing the forward event-chain Monte Carlo dynamics \citep{Michel2020} in both the ambient space and the submanifold, our method produces a single continuous non-reversible trajectory traversing the two components, as shown in Figure~\ref{fig:6PDMPGMM} (a). For comparison, panel (c) shows the standard reversible-jump MCMC with a random walk proposal, and (b) a simple PDMP analogue where the same trans-dimensional jumps are triggered at a constant Poisson rate $\lambda>0$ within the PDMP dynamics; see also \cite{Hardcastle2024}.
As we demonstrate with a variety of examples in Section~\ref{sec;4NumericalExamples}, the main strength of our construction is that the same principle extends beyond this linear example, regardless of the geometry of the submanifold or the particular choice of within-model PDMP dynamics.

\section{The coordinate-based manifold PDMP}
\label{sec;2ManifoldPDMPs}

In this Section we introduce a mechanism that adapts standard PDMP samplers with piecewise linear dynamics to sample from target measures defined on Riemannian manifolds. Existing constructions either retain Hamiltonian dynamics, as in the randomised Riemannian manifold HMC \citep{Whalley+2023}, or are targeted at hyperspheres, such as the spherical bouncy particle sampler \citep{Yang2024} and the event-chain Monte Carlo for continuous-spin models \citep{Michel+2015,Nishikawa+2015}. In contrast, our approach based on coordinate charts can be applied to any PDMP sampler with tractable deterministic dynamics when the manifold admits a global chart.  %A general mechanism for transferring standard piecewise linear PDMP samplers, including BPS and zig-zag samplers, to general manifolds appears to be largely absent. In this section, we address this gap by considering a simple method based on coordinate charts.

\subsection{Euclidean PDMP algorithms}

We briefly recall the Euclidean PDMP dynamics used as a building block. A piecewise linear sampler uses the Euclidean geodesic flow $\phi_t(x,v)=(x+tv,v)$ between events. The bouncy particle sampler \citep{Bouchard2018}, the zig-zag sampler \citep{Bierkens2019}, and forward event-chain Monte Carlo \citep{Michel2020} belong to this class, while the boomerang sampler \citep{Bierkens+2020} and randomised Hamiltonian Monte Carlo \citep{Bou2017} do not. Stochastic events occur with state-dependent rate $\lambda(x,v)$, and the state is updated according to a transition kernel $Q$, in most instances only updating the velocity components. Different choices of $\lambda$ and $Q$ lead to different dynamics and algorithms.

A key computational challenge is simulating the event times $\{\tau\}_{i=1}^n$. Theoretically, given $\tau_{i-1}$, the next event time is given by
\begin{equation}\label{eq;inversion}
    \tau_{i}:=\tau_{i-1}+\operatorname*{arg\,min}_{s\ge0}\left\{\int^{s}_{0}\lambda\circ\phi_t(x_{\tau_{i-1}},v_{\tau_{i-1}})\,dt=E\right\},
\end{equation}
for $E\sim \operatorname{Exp}(1)$. When a closed-form solution is unavailable, however, this equation must be solved numerically \citep{Pagani+2024,Chevallier+2026}. Alternatively, when there is an upper bound of $\lambda$ from  which event times can be simulated exactly, one obtains exact samples of $\tau_{i}$ by accepting or rejecting the candidate event time. This latter approach is called Poisson thinning \citep{Lewis-Shedler1979,Lemaire+2018}, and some methods aim to automatically generate the upper bound \citep{Corbella2022,Sutton2023,Andral-Kamatani2024}. See \citealt[Chapter~5]{Fearnhead+2025} for an accessible overview.

\subsection{The manifold PDMP algorithm}\label{subsec;TheCoordinateBasedMPDMP}

Given a target probability measure $\pi^g(x)\mathcal{H}^p_g(dx)$ on the Riemannian manifold $(\mathcal{M},g)$ that admits a global chart $\varphi:\mathcal{M}\to\mathbb{R}^p$, we consider the following two-step procedure
\begin{enumerate}
    \item Sample from the pushforward density
    \begin{equation}\label{eq;weighted-target}
        \pi^\varphi(x):=\pi^g(\xi)J^g_\varphi(\xi)^{-1},\qquad \xi=\varphi^{-1}(x),
    \end{equation}
    where $J^g_\varphi$ is the Jacobian of $\varphi$ with respect to the Riemannian metric $g$ of $\mathcal{M}$.
    \item Transform the output via $\varphi^{-1}$ to recover samples on $\mathcal{M}$.
\end{enumerate}
Since the weighted target $\pi^\varphi$ is now defined on a Euclidean region $\varphi(\mathcal{M})$, one can use standard PDMP algorithms, such as those of \cite{Bierkens2023b,Bierkens2018}, to obtain samples $X_t\sim\pi^\varphi$. Proposition~\ref{prop;pullback} below then ensures that the transformed samples $\varphi^{-1}(X_t)$ obey the original position law $\pi^g$.

\begin{algorithm}[H]
\caption{Manifold PDMP with a global chart $\varphi$.}
\label{alg;ManifoldBPS}
\begin{algorithmic}[1]
\State Initialise $(\xi_0,\eta_0) \in T(\mathcal{M})$
\State Initialise $X_0\gets\varphi(\xi_0)$ and $V_0\gets D\varphi(\xi_0)\eta_0$
\State Run a standard PDMP on $\varphi(\mathcal{M})\subset\mathbb{R}^p$ targeting $\pi^\varphi$ \eqref{eq;weighted-target}
\State Recover the sample of $\pi^g$ by $\xi_t\gets\varphi^{-1}(X_t)$
\end{algorithmic}
\end{algorithm}

When we use this as a subroutine within the sticky manifold PDMP algorithm (Algorithm~\ref{alg;TheAlgorithm}), the metric $g$ is set to the induced metric from the embedding $\mathcal{M}\subset\mathbb{R}^d$ so that $\pi^g$ corresponds to $\pi_{\mathcal{M}}$ in \eqref{eq;6manifoldtarget}. The Jacobian $J^g_\varphi(\xi)^{-1}=J^g_{\varphi^{-1}}(\xi)$ accounts for the artificial acceleration caused by the transformation $\varphi^{-1}$ at $\xi\in\mathcal{M}$. In our numerical examples of survival analysis (Section~\ref{subsec;survival-extrapolation}), we use Algorithm~\ref{alg;ManifoldBPS} to sample from four parametric subfamilies, while to sample from the unit sphere $\mathbb{S}^2$ (Sections~\ref{subsec;ConcentricSpheres} and \ref{subsec;QuantumStateTomography}) or from the Stiefel manifold $O(20)$ (Section~\ref{subsec;BayesianPCA}), we simulate the geodesic flow explicitly.

\begin{remark}[Extension to general manifolds]
    \label{remark;GeneralManifolds}
    In this work, we mainly consider manifolds with a global chart or explicit geodesic flows. We however suggest that a global chart may not be necessary by working with an atlas, $\{\varphi_i\}_{i=1}^n$, of $\mathcal{M}$ and dividing $\mathcal{M}$ into a disjoint sum of $U_i$'s. The above process can then transition between local charts using the transition map $\varphi_j\circ\varphi_i^{-1}$ and its derivative to transform the position and velocity, respectively.
    %The global-chart assumption can, in principle, be relaxed by working with an atlas; however we only give a brief remark on this point.
    %Given an atlas $\{\varphi_i\}_{i=1}^n$ of $\mathcal{M}$, we can divide $\mathcal{M}$ into a disjoint sum of $U_i$'s. Within each $U_i$, the coordinate-based PDMP described above can be run on $\varphi_i(U_i)\subset\mathbb R^d$.
    %Additional care is required when a trajectory reaches an interface between these patches. A natural approach is to use the transition map $\varphi_j\circ\varphi_i^{-1}$ and its derivative to transform the position and velocity, respectively.
    More generally, one may include the chart index as a discrete component of the state and introduce jumps between coordinate patches, with their rates and transition kernels chosen to preserve the global target distribution. This leads to a setting structurally similar to the reversible-jump PDMP framework of \citet{Chevallier2023}.%, which connects PDMP dynamics on different Euclidean regions of possibly different dimensions through appropriately designed inter-component moves, primarily in the context of Bayesian variable selection. Developing such a multi-chart construction is beyond the scope of the present work.
\end{remark}

\subsection{Geometric foundations: pullback construction of the state space}\label{subsec;preliminaries}

% Before establishing the invariance of the manifold PDMP (Algorithm~\ref{alg;ManifoldBPS}) in Proposition~\ref{prop;pullback}, we review the pullback construction for tangent bundles, which also underlies the mathematical structure of the sticky manifold PDMP discussed in Section~\ref{sec;3StickyManifold}.
% %In short, the natural state space for PDMPs is the tangent bundle $T(\mathbb{R}^d)\simeq\mathbb{R}^{2d}$, while the state spaces of manifold PDMPs and sticky manifold PDMPs are the pullback bundles of the Euclidean bundles. To define the dynamics of sticky manifold PDMPs, we also need to lift the manifold PDMP dynamics along the blow-down map derived from the blow-up construction; see Section~\ref{subsec-blowup}.
% This construction for state spaces can be done in purely geometric terms. Thus, part of the key properties are automatic once correctly formulated. This leads to the modular nature of our framework, since we separate geometric constructions from the specificity of PDMP dynamics, thereby providing easily verifiable sufficient conditions for compatibility of dynamics that can be employed (Section~\ref{subsec;invarianceProof}). Here, we focus only on relevant concepts; readers are referred to \cite{Lang1995-Manifolds} for a thorough treatment of geometry.

For the manifold PDMPs considered here, the natural state space is the tangent bundle, generalising the position-velocity product space in the Euclidean case. Accounting for this bundle structure is instrumental in developing a consistent mathematical formulation and subsequently proving invariance. The state spaces of both manifold PDMPs and sticky manifold PDMPs are constructed geometrically as pullback bundles, independently of the specific PDMP dynamics employed. We therefore review this construction before establishing the invariance of the manifold PDMP (Algorithm~\ref{alg;ManifoldBPS}) in Proposition~\ref{prop;pullback} and the sticky manifold PDMP (Algorithm~\ref{alg;TheAlgorithm}) in Theorem~\ref{thm;NewTheorem}. Here, we introduce only the geometric concepts needed for these constructions; readers are referred to \cite{Lang1995-Manifolds} for a thorough treatment.

A $p$-dimensional \textit{manifold} $\mathcal{M}$ is a space that looks locally like Euclidean space $\mathbb{R}^p$ through prescribed transformations called \textit{local coordinate charts} $\{\varphi_i\}$.
%Such a consistently defined set of coordinate charts is called an \textit{atlas} of the manifold $\mathcal{M}$.
We say $\mathcal{M}$ is \textit{smooth} when all transition maps $\varphi_i\circ\varphi_j^{-1}$ are smooth in the usual sense wherever they are defined.
%This abstract setup enables an intrinsic treatment of geometric structure. Alternatively, a manifold can be \textit{embedded} in Euclidean space $\mathbb{R}^d$ through an embedding $\varphi:\mathcal{M}\to\mathbb{R}^{d}$. For $d\ge p$, such an embedding always exists \citep{Whitney1936}, although a convenient explicit embedding need not be available. For an embedded manifold $\mathcal{M}\subset\mathbb{R}^d$, the difference in dimension, $d-p$, is called the \textit{codimension}.
In this work, we focus on \textit{embedded} manifolds $\mathcal{M}\subset\mathbb{R}^d$. The difference $d-p$ is called the \textit{codimension}. For any smooth manifold $\mathcal{M}$, such an embedding exists for sufficiently large $d$. %\citep{Whitney1936}.

%To define dynamics on manifolds, we need a suitable notion of velocity or momentum. Velocities are represented by \textit{tangent vectors}, and their distributions are thus naturally defined on the collection of them.
At a point $\xi\in\mathcal{M}$, the \textit{tangent space} $T_\xi(\mathcal{M})$ consists of all \textit{tangent vectors} $\eta$ that can be realised by smooth curves through $\xi$; $\eta$ belongs to $T_\xi(\mathcal{M})$ if there exists a smooth curve $\gamma:(-\epsilon,\epsilon)\to\mathcal{M}$ such that $\gamma(0)=\xi$ and $\dot{\gamma}(0)=\eta$.
The space $T_{\xi} (\mathcal{M})$ forms a $p$-dimensional vector space, and the collection of all tangent spaces forms the \textit{tangent bundle}, $T(\mathcal{M}):=\sqcup_{\xi\in\mathcal{M}}T_\xi(\mathcal{M})=\{(\xi,\eta)\mid\eta\in T_\xi(\mathcal{M})\}$.
For any bundle $E$ or a subset of it, we denote its \textit{fibre} at $x$ by $E_x:=\{v\mid (x,v)\in E\}$. 
Tangent vectors represent the natural generalisation of velocity in Euclidean space to Riemannian manifolds, and the state spaces for manifold PDMPs and sticky manifold PDMPs will all be formulated as bundles. 
%Its dual is the cotangent bundle $T^*(\mathcal{M}):=\sqcup_{\xi\in\mathcal{M}}T_\xi^*(\mathcal{M})$, which carries a canonical symplectic structure \cite[Section~V.7.2]{Lang1995-Manifolds}; see also \citet{betancourt2017geometric,Barp+2018} in the context of Monte Carlo methods.

In this work, two canonical constructions are repeatedly used: \textit{pullback} and \textit{lifting}.
\textit{Pullback} and its dual, \textit{pushforward}, are universal constructions for many mathematical objects.
The probability measure $\pi_*$ satisfying $\pi_*(A)=\pi(\phi^{-1}(A))$ is called the \textit{pushforward} of $\pi$ along a mapping $\phi:\mathcal{M}\to\mathcal{N}$.
When $\phi$ is invertible, $\pi$ is 
the \textit{pullback} of $\pi_*$.
A tangent bundle can also be pulled back along $\phi$ by defining
\[\phi^*(T(\mathcal{N})):=\{(\xi,\eta)\mid\xi\in\mathcal{M},\eta\in T_{\phi(\xi)}(\mathcal{N})\}.\]
When $\phi$ is the global chart, $\phi^*(T(\mathcal{N}))$ is the state space of manifold PDMPs. %, while the blow-down map provides that of sticky manifold PDMPs (Section~\ref{subsec-blowup}).
We define a pullback of a PDMP in Section~\ref{subsec;pullback-PDMP}.
%While pullback is a universal construction, many design choices arise in the form of lifting.
A \textit{lift} of a probability measure $\pi$ on $\mathcal{M}$ to its tangent bundle is any distribution defined on $T(\mathcal{M})$ with position marginal $\pi$. When $\pi$ is the target distribution of a PDMP, the choice of lift corresponds to the choice of velocity distribution.
%This corresponds to the choice of velocity distribution in the Euclidean case.
%We will face two lifting problems associated with the blow-up procedure; we define a lifting of target measures and PDMP dynamics to the blown-up manifold in Section~\ref{subsec-blowup}.
In Section~\ref{subsec-blowup}, we discuss a lifting of manifold PDMP dynamics.

% A \textit{Riemannian metric} on $\mathcal{M}$ is a family of inner products $g_\xi$ acting on elements of $T_\xi(\mathcal{M})$ that varies smoothly in $\xi$. A pair $(\mathcal{M},g)$ is called a \textit{Riemannian manifold}.
% In local coordinates $x=\varphi(\xi)$ of the tangent bundle $T(\mathcal{M})$, a Riemannian metric $g$ is represented by a symmetric positive definite matrix $G_x$, $g_x(u,w)= u^\top G_xw$.
% A Riemannian metric $g$ also induces a canonical volume measure on a manifold $\mathcal{M}$. In local coordinates $x$, it is given by $\mathcal{H}^p_{g}(dx)=\sqrt{\det G_x}\,dx$ with the matrix representation \citep[Section~XI.4]{Lang1995-Manifolds}. Intrinsically, this coincides with the $p$-dimensional Hausdorff measure seen as a metric space with Riemannian distance $\operatorname{dist}_g$, hence the notation. For a precise definition of the Hausdorff measure, see Appendix~\ref{appendix;subsec-Hausdorff}.

\subsection{Pulling back PDMPs through coordinate charts}\label{subsec;pullback-PDMP}

In this subsection, we consider the pullback of PDMP $(X_t,V_t)$ through a coordinate chart $\varphi:\mathcal{M}\to\mathcal{U}\subset\mathbb{R}^p$, defined by
\begin{equation}\label{eq;PullbackPDMP}
    \varphi^*Z_t:=(\varphi^{-1}(X_t),D\varphi^{-1}(X_t)V_t),
\end{equation}
where the position process is pulled back by $\varphi$, while the velocity is pulled back by the derivative $D\varphi$.
When $\pi$ is an invariant measure of $X_t$, this pullback PDMP dynamics leaves the pullback distribution $\pi^*:=(\varphi^{-1})_*\pi$ invariant, and $\pi^*$ is described through the Jacobian.
We defer the proof to Appendix~\ref{appendix;sec-ManifoldPDMP}.

\begin{proposition}[Invariant distribution of pullback PDMPs]\label{prop;pullback}
    Let $(\mathcal M,g)$ be a $p$-dimensional Riemannian manifold, $\mathcal{U}\subset\mathbb{R}^p$ an open set, and $\varphi:\mathcal{M}\to\mathcal{U}$ be a diffeomorphism. Consider a PDMP $(X_t,V_t)$ on $\mathcal{U}\times\mathbb{R}^p$ with an invariant distribution $\pi(x)\,dx\nu_x(dv)$. Then, the pullback PDMP process \eqref{eq;PullbackPDMP} on $T(\mathcal{M})$ has an invariant distribution
    \begin{equation}\label{eq;inversely-weighted-target}
        \pi(\varphi(\xi))J_\varphi^g(\xi)\mathcal{H}^p_g(d\xi)\nu_\xi^\varphi(d\eta),\qquad(\xi,\eta)\in T(\mathcal{M})
    \end{equation}
    where $J_\varphi^g$ is the Jacobian of $\varphi$ computed with the Riemannian metric $g$, $\mathcal{H}_g^p$ is the Hausdorff measure on $(\mathcal{M},g)$, and
    $\nu^\varphi_\xi:=(D\varphi(\xi)^{-1})_*\nu_{\varphi(\xi)}$ is the pullback distribution.
\end{proposition}

\begin{remark}[The choice of $g$]
    The Jacobian factor in \eqref{eq;inversely-weighted-target} disappears when $g$ is the pullback of the Euclidean metric $g_{\mathrm{Euclid}}$ through $\varphi$, i.e., $g=\varphi^*g_{\mathrm{Euclid}}$, or, in other words, when $\varphi$ is isometric; see Section~\ref{subsec-illustration-hypersphere} for an example.
    %Nevertheless, we retain the freedom to choose $g$, since we set $g$ to be the metric pulled back from the ambient Euclidean space through the embedding $i:\mathcal{M}\to\mathbb{R}^{d}$ when it is used as a subroutine within the sticky manifold PDMP algorithm (Algorithm~\ref{alg;TheAlgorithm}).
    However, the target distribution \eqref{eq;6manifoldtarget} of sticky manifold PDMPs needs to be specified with respect to the induced metric $g=i^*g_{\mathrm{Euclid}}$ through the embedding $i:\mathcal{M}\to\mathbb{R}^{d}$, which differs from $\varphi^*g_{\text{Euclid}}$ in general.
\end{remark}

Theorem~\ref{thm;pullback-generator} shows that the pulled-back PDMP follows the geodesics of the induced metric $\varphi^*g_{\text{Euclid}}$. The choice of parametrisation is therefore crucial and may result in both improvement and worsening of the geometry of the transformed target, as well as the resulting computational complexity of simulating the continuous time process. %See \cite{Bierkens-Kamatani-Roberts2024} for further analysis in the Euclidean setting.

\subsection{Illustration: two charts of the unit hypersphere}\label{subsec-illustration-hypersphere}

\begin{figure}
    \centering
    \includegraphics[width=0.9\linewidth]{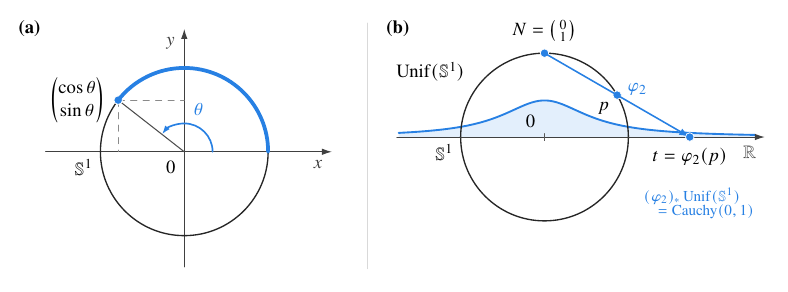}
    \caption{Two parametrisations for the one-dimensional unit hypersphere. (a) represents the arc length parametrisation $\varphi_1$. In this case, the Jacobian is constant, $J^g_{\varphi_1}=1$ and $\varphi_1$ is a Riemannian isometry. (b) represents the stereographic projection, pushing-forward the uniform distribution to the standard Cauchy distribution.}
    \label{fig:UnitSphere}
\end{figure}

We illustrate how different coordinate charts give rise to different weighted targets, using the one-dimensional unit hypersphere $\mathbb{S}^1$ as an example. It has two natural coordinate maps, $\varphi_1:\mathbb{S}^1\setminus\{(1,0)\}\to(0,2\pi)$ and $\varphi_2:\mathbb{S}^1\setminus\{(0,1)\}\to\mathbb{R}$, given by
\[\varphi_1(x,y)=\begin{cases}\arccos x&y\ge0,\\2\pi-\arccos x&y<0.\end{cases}\qquad\varphi_2(x,y)=\frac{x}{1-y},\]
respectively (Figure~\ref{fig:UnitSphere}).
The Jacobians with respect to the metric $g$ on $\mathbb{S}^1$ induced by the ambient Euclidean metric are
\[J^g_{\varphi_1^{-1}}(x)=1,\qquad J^g_{\varphi_2^{-1}}(x)=\frac{2}{1+x^2},\]
respectively. Thus, to sample from the uniform measure on $\mathbb{S}^1$, i.e., $\pi^*(x)=(2\pi)^{-1}$, the required weighted proxy targets are
\[\pi_1(x)=\frac{1}{2\pi}\quad(x\in(0,2\pi)),\qquad\pi_2(x)=\frac{1}{\pi}\frac{1}{1+x^2}\quad(x\in\mathbb{R}),\]
calculated via Equation~\eqref{eq;weighted-target} for $\varphi_1$ and $\varphi_2$ respectively.
The simplicity of $\pi_1$ derives from the constant Jacobian $1$, that means $\varphi_1$ is a Riemannian isometry. In this case, the weight is absorbed by the reference measure $\mathcal{H}^p_g$ in Equation~\eqref{eq;inversely-weighted-target}. By contrast, the weighted target $\pi_2$ is nontrivial, being the standard Cauchy distribution. This correspondence is utilised to efficiently sample from heavy-tailed distributions on $\mathbb{R}^d$; see \cite{Yang2024,Bell2024}. %Note that in practice there is no global chart for $\mathbb{S}^d$ and so principled sampling would require the construction of an atlas (Remark \ref{remark;GeneralManifolds}).

\begin{remark}[Higher-dimensional unit spheres]\label{remark-high-dimensional-hypersphere}
    We consider sampling on the two-dimensional unit sphere in Sections~\ref{subsec;ConcentricSpheres} and~\ref{subsec;QuantumStateTomography}.
    This sphere $\mathbb{S}^2$ or any higher-dimensional sphere differs from $\mathbb{S}^1$ in that no isometric coordinate chart $\varphi$ is available due to the non-trivial sectional curvature. Thus, isometric charts like $\varphi_1$ are not available in higher dimensions, while the stereographic projection $\varphi_2$ can be naturally generalised. We can, however, simulate the geodesic flow directly as constrained dynamics through the canonical embedding $\mathbb{S}^2\to\mathbb{R}^3$ \citep[e.g.,][]{Yang2024},
    \[\begin{cases}
        x_t=\cos(\|v_0\|t)x_0+\sin(\|v_0\|t)v_0/\|v_0\|\\
        v_t=-\|v_0\|\sin(\|v_0\|t)x_0+\cos(\|v_0\|t)v_0.
    \end{cases}\]
    % This follows a standard approach for manifold sampling when an explicit coordinate treatment is inconvenient \citep{Byrne2013,Yang2024,Bell2024}.
\end{remark}

\section{The sticky manifold PDMP}
\label{sec;3StickyManifold}

We now introduce the sticky manifold PDMP sampler. Throughout this Section we focus on a single embedded Riemannian manifold $\mathcal{M}\subset\mathbb{R}^d$ of codimension one that is closed in $\mathbb{R}^d$.
We also assume that $\mathcal{M}$ is \textit{orientable}: a global unit normal vector field $\boldsymbol{n}:\mathcal{M}\to\mathbb{R}^d$ exists and can be evaluated at each point $x\in\mathcal{M}$. We define the orthogonal decomposition of the velocity as
\[v^\perp:=\langle v,\boldsymbol{n}(x)\rangle\boldsymbol{n}(x),\qquad v^\tpar:=v-v^\perp,\]
at any point $x\in\mathcal{M}$.
We do not impose strong assumptions on the choice of PDMP sampler for sampling in the ambient space or on the submanifold, allowing any standard PDMP sampler mentioned in Section~\ref{sec;1Introducion} and manifold PDMP sampler in Section~\ref{sec;2ManifoldPDMPs} to be incorporated as subroutines.

\subsection{Algorithmic description}\label{subsec;theAlgorithm}

We present the algorithm in a slightly more general form than the one outlined in the Introduction (Section~\ref{subsec;Introduction-contribution}), allowing stochastic velocity transformations upon sticking and unsticking. These transformations are specified by the kernels $Q_{\mathrm{stick}}$ and $Q_{\mathrm{unstick}}$, respectively. The construction in the Introduction corresponds to taking these kernels to be Delta measures associated with deterministic velocity transformations. More generally, we allow the sticking kernel $Q_{\mathrm{stick}}$ to transform the tangential velocity component to meet the requirements of the chosen manifold dynamics while leaving the normal component unchanged. During the manifold dynamics, this normal component is parallel-transported along $\mathcal M$. The unsticking kernel $Q_{\mathrm{unstick}}$ uses the current tangential velocity and the stored normal component to generate the velocity with which the ambient dynamics resume. The same unsticking rate \eqref{eq;6Manifoldthawingrate} continues to be valid when both kernels preserve the norm of the normal component and satisfy the velocity compatibility condition \eqref{eq;VelocityCompatibilityCondition}. This condition requires $Q_{\mathrm{stick}}$ to map the incoming ambient velocity distribution to the prescribed joint distribution of the manifold velocity and the stored normal component, and $Q_{\mathrm{unstick}}$ vice versa.
%The algorithm requires a sticking kernel $Q_{\text{stick}}$ and an unsticking kernel $Q_{\text{unstick}}$ to transform the velocity upon these events.
% When these kernels meet the velocity compatibility condition of \eqref{eq;VelocityCompatibilityCondition}, the unsticking rate function $\Lambda$ is given by \eqref{eq;6Manifoldthawingrate}, which we recall here as
% \[\Lambda(x) = \|v^\perp_\sigma\|\frac{\omega}{1-\omega}\frac{\pi_{\mathbb{R}^d}(x)}{\pi_{\mathcal{M}}(x)},\tag{\ref{eq;6Manifoldthawingrate}}\]
% where $v^\perp_\sigma:=\langle v_\sigma,\boldsymbol{n}(x_\sigma)\rangle\boldsymbol{n}(x_\sigma)$ is the normal velocity at the $\mathcal{M}$-hitting time $\sigma$, and $\omega$ is the mixture weight on $\mathbb{R}^d$. The rate function \eqref{eq;6Manifoldthawingrate} suggests a longer sojourn when either (a) the normal hitting velocity was small, (b) the prior ambient mixture weight $\omega$ is small, or (c) the density $\pi_{\mathcal{M}}$ is large relative to $\pi_{\mathbb{R}^d}$, thereby balancing the incoming probability flux with the outgoing unsticking flux.
Given a PDMP sampler with position marginal $e^{-U(x)}\pi_{\mathbb{R}^d}(x)\,dx$ and a manifold PDMP sampler with $e^{-U(x)}\pi_{\mathcal{M}}(x)\mathcal{H}^{d-1}_{\mathcal{M}}(dx)$, the sticky manifold PDMP algorithm can be summarised as follows:

\begin{algorithm}[H]
\caption{Sticky manifold PDMP.}
\label{alg;TheAlgorithm}
\begin{algorithmic}[1]
\State Initialise $z_0=(x_0,v_0) \in \mathbb{R}^{d}\times\mathcal{V}$
\While{$t<T$}
    \State Run a standard PDMP until $\sigma = \inf\{s>t\mid x_s \in \mathcal{M}\}$
    \State Sample a new velocity $v_{\sigma}\sim Q_{\text{stick}}((x_{\sigma-},v_{\sigma-}),-)$
    \State Start a manifold PDMP in $\mathcal{M}$ from $(x_\sigma,v_{\sigma}^\tpar)$ until $\tau$, the first Poisson arrival with rate $\Lambda$
    \State Parallel-transport the normal component by setting $v_{\tau-}^\perp\gets\langle v_{\sigma},\boldsymbol{n}(x_{\sigma})\rangle \boldsymbol{n}(x_{\tau-})$
    \State Recover the ambient velocity by sampling $v_{\tau}\sim Q_{\text{unstick}}((x_{\tau-},v_{\tau-}),-)$
\EndWhile
\end{algorithmic}
\end{algorithm}

%The ambient normal component $v^\perp$ is parallel-transported along $\mathcal{M}$ while $(x,v^\tpar)$ develops within the manifold $\mathcal{M}$.
In the remainder of this subsection, we list examples of kernels $Q_{\text{stick}},Q_{\text{unstick}}$ in a few important situations.

\begin{example}[SM-PDMP with the identity sticking and side-flipping unsticking kernels]\label{example;SM-identity-sticking}
    When both ambient and manifold velocities are standard Gaussian,
    we can choose the sticking kernel to be the identity kernel $Q_{\mathrm{stick}}((x,v),-)=\delta_{(x,v)}$ and the unsticking kernel to be a side-flipping kernel $Q_{\mathrm{unstick}}((x^\pm,v),-)=\delta_{(x^\mp,v)}$, where $x^+$ and $x^-$ represent the same point of $\mathcal{M}$ on the two boundary faces. The side flip changes only the boundary-face label, leaving the physical position and velocity unchanged. Relative to the new face, the retained normal velocity points outwards, allowing the ambient dynamics to resume in the blown-up state space (Section~\ref{subsec-blowup}). This choice reduces Algorithm~\ref{alg;TheAlgorithm} to the one outlined in Section~\ref{subsec;Introduction-contribution}. These kernels satisfy condition~\eqref{eq;VelocityCompatibilityCondition}, since the ambient standard Gaussian velocity decomposes into independent standard Gaussian tangential and normal components, and side flipping preserves the ambient velocity distribution. %The algorithm has its simplest form in this Gaussian case; it stores the normal component $v^\perp_\sigma$ upon sticking and recovers the normal component $v_{\tau}\gets v_{\tau}^\tpar + v_\tau^\perp$ based on that record when leaving.
\end{example}

\begin{example}[SM-PDMP with unit velocities]\label{example;SM-UnitVelocity}
    When the invariant velocity distributions are the uniform measures on $\mathcal{V}=\mathbb{S}^{d-1}$ and $\mathbb{S}^{d-2}$, a modest transformation satisfying \eqref{eq;VelocityCompatibilityCondition} is a deterministic sticking map $\psi_x:\mathbb{S}^{d-1}\to\mathbb{S}^{d-2}\times(-1,1)$ defined by
    \[\psi_x(v)=\left(\frac{v^\tpar}{\|v^\tpar\|},\langle\boldsymbol{n}(x),v\rangle\right),\]
    where $v^\tpar$ depends on the position $x$.
    Upon unsticking, the inverse transformation is applied:
    \[\psi^{-1}_x(u^\tpar,u^\perp)=\sqrt{1-(u^\perp)^2}\;u^\tpar+u^\perp\boldsymbol{n}(x).\]
    %Overall, we can straightforwardly check that this transformation $\psi$ satisfies condition~\eqref{eq;PullbackCondition}, a specialisation of the velocity compatibility condition~\eqref{eq;VelocityCompatibilityCondition}.
    This sticking transformation $\psi_x$ pushes forward $\operatorname{Unif}(\mathbb{S}^{d-1})$ to the direct product of $\operatorname{Unif}(\mathbb{S}^{d-2})$ and the distribution of $2B-1$ for $B\sim\operatorname{Beta}((d-1)/2,(d-1)/2)$. This verifies condition~\eqref{eq;PullbackCondition}, and hence the velocity compatibility condition~\eqref{eq;VelocityCompatibilityCondition}.
    Alternatively, we may resample the normal component from a flux-weighted distribution upon unsticking. This approach is used with the tangential part of $\psi_x$ in \citet[Section~4.2]{Chevallier2023}, who consider coordinate-aligned linear subspaces. When combined with the Rao--Blackwellised unsticking rate (Corollary~\ref{cor;Rao-Blackwellized-rate}), this approach avoids storing the normal component but discards the memory of the incoming normal speed, potentially reducing persistence across model transitions.
\end{example}

\begin{example}[SM-Zigzag with velocity refreshment]\label{example;SM-Zigzag}
    For the zig-zag sampler, velocity refreshment upon sticking becomes crucial, even for non-linear submanifolds and tilted lines (Sections~\ref{subsec-GMM-illustration} and \ref{subsec;LinearSubspaces}).
    Since the tangent spaces of these submanifolds are no longer aligned with the coordinate axes, the new velocity needs to be transformed upon sticking, unlike the coordinate-aligned case of \citet[Section~4.1]{Chevallier2023}.
    A straightforward way to ensure velocity compatibility \eqref{eq;VelocityCompatibilityCondition} is to completely refresh the tangential velocity by sampling a new manifold velocity from $\operatorname{Unif}(\{\pm1\})$ upon sticking, while resuming the ambient dynamics upon unsticking with the stored normal velocity.
    This specification is used in the numerical examples of Section~\ref{subsec;LinearSubspaces}.
\end{example}

More generally, partial stochastic refreshments can be incorporated as long as the more general flux balance condition \eqref{eq;GeneralBalance} is met, being related to the discussion in \citet[Section~2.7]{Bierkens2023b}. Previous works suggest that this may lead to faster mixing when designed effectively \citep{Michel2020,Wu2020,Tran-Kleppe2025}. 
Note that Gaussian velocity distributions do not guarantee that the velocity of the manifold PDMP induced by its coordinate chart is an isotropic Gaussian distribution (see Section~\ref{subsec-illustration-hypersphere}). We discuss the velocity compatibility condition under such a setting further in Section~\ref{subsec;illustration-blowup-cone}.

\subsection{Blow-up construction of the state space}\label{subsec-blowup}

In this technical section we mathematically define a non-reversible Markov process that connects the PDMP dynamics in the ambient space with the dynamics on $\mathcal{M}$. We establish $\Pi$-invariance of the resulting process in Section~\ref{subsec;invarianceProof}. We construct the state space, $E$, as a pullback along the blow-down map, followed by a lifting of manifold dynamics. In $E$, states on the manifold $\mathcal{M}$ correspond to \textit{frozen boundaries} $\mathfrak{F}$, whose state is represented by $(x,v^\tpar,v^\perp)$ with $(x,v^\tpar)\in T(\mathcal{M})$ and $v^\perp$ perpendicular to $\mathcal{M}$. Importantly, the state of the manifold PDMP $(x,v^\tpar)\in T(\mathcal{M})$ is lifted by adding the ambient normal velocity $v^\perp$, that is parallel-transported along the submanifold while the process develops within $\mathfrak{F}$. This construction enables us to: \textit{i)} Distinguish the two sides of the manifold, denoted by $\mathcal{M}^+$ and $\mathcal{M}^-$, thereby deciding from which side the process stuck to $\mathcal{M}$ and from which side it should leave; \textit{ii)} Provide the normal velocity when the process restarts in the ambient space. This augmented state $(x,v^\tpar,v^\perp)\in\mathfrak{F}$ provides the complete information needed to define a Markov process representing the sticky manifold PDMP algorithm.

\begin{figure}
    \centering
    \includegraphics[width=\linewidth]{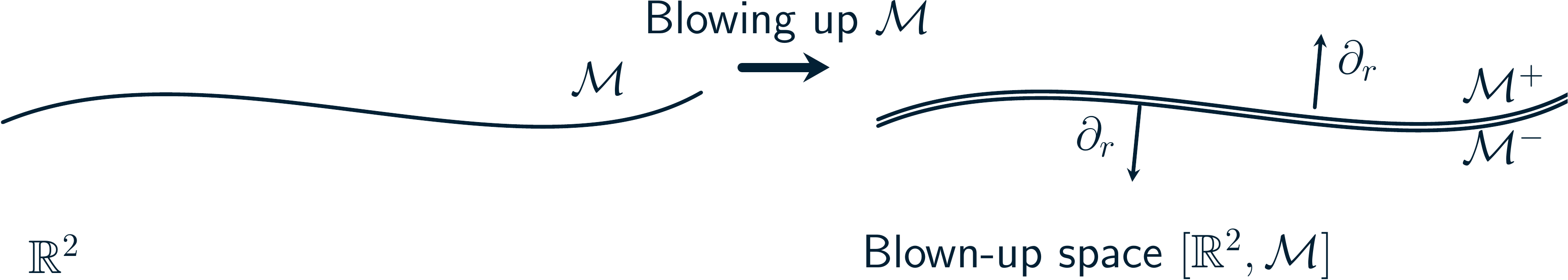}
    \caption{Radial real blow-up of a smooth embedded submanifold $\mathcal{M}\subset\mathbb{R}^{2}$. The blow-up replaces $\mathcal{M}$ by two boundary faces, $\mathcal{M}^{+}$ and $\mathcal{M}^{-}$, representing the two normal directions. There exists a natural vector field $\partial_{r}$ induced from our construction, representing the outward normal direction on each face.}
    \label{fig:RadialBlowup}
\end{figure}

%Formally, this construction is the geometric \textit{blow-up} that generalises the construction in \cite{Bierkens2023a} from the case of the coordinate-aligned linear space $\{y=0\}$, where two copies of the origin $0$ are introduced as $0^-$ and $0^+$, to an arbitrary manifold $\mathcal{M}$.
The radial blow-up of $\mathcal{M}$ in $\mathbb{R}^d$ gives an enlarged space $[\R^d;\cM]$ equipped with a smooth surjection $\beta:[\R^d;\cM]\to\R^d$, called the \textit{blow-down} map. As illustrated in Figure~\ref{fig:RadialBlowup}, the blown-up space $[\mathbb{R}^d;\mathcal{M}]$ has a boundary that consists of two copies of the original manifold $\beta^{-1}(\cM)=\cM^+\sqcup\cM^-$. Apart from this boundary, the interior is identical to $\mathbb{R}^d\setminus\mathcal{M}$, and the blow-down map $\beta$ is an identity map on it. Note that this construction generalises the approach in \cite{Bierkens2023a} for the coordinate-aligned linear space $\{y=0\}$, where two copies of the origin $0$ are introduced as $0^-$ and $0^+$. The precise geometric construction of $[\R^d;\cM]$ and $\beta$, and the unit vector field $\partial_r$ pointing outwards follows the standard radial real blow-up procedure used in differential geometry; see Appendix~\ref{appendix-subsec-blowup} for full details. It replaces the submanifold with its spherical normal bundle, thereby retaining the direction of approach.
The state space of the sticky manifold PDMP is then defined as the pullback bundle $E:=\beta^*(T(\mathbb{R}^d))$ along the blow-down map $\beta$.
The frozen boundary is defined by the misalignment of the velocity $v$ with the vector field $\partial_r$
\[\mathfrak{F}:=\{(\xi,v)\in E\mid \xi\in\cM^{\pm}, \brac{v,\partial_r(\xi)}<0\}.\]
In words, a state is confined in $\mathcal{M}^\pm$ when the normal velocity points inwards, the opposite direction to $\partial_r$.
Here $\mathfrak F$ represents the stuck states over $\mathcal M$, retaining both the tangential velocity $v^\tpar$ and the ambient normal velocity $v^\perp$. For a point $x=\beta(\xi)\in\mathcal{M}$ in the manifold, we write $\xi=x^+$ or $\xi=x^-$ depending on which face the position belongs to in the blown-up space.
We similarly define the ambient boundary as the subsets of $\Gamma:=\partial T(\mathbb{R}^d\setminus\mathcal{M})$ comprising states that satisfy $\langle v,\partial_r(x)\rangle<0$ and $\langle v,\partial_r(x)\rangle>0$ by $\Gamma_{\text{in}}$ and $\Gamma_{\text{out}}$, respectively. Although $\Gamma_{\text{in}}$ coincides with $\mathfrak{F}$ as a set, we distinguish the two for clarity.
The sticking kernel $Q_{\mathrm{stick}}$ is defined from $\Gamma^{\mathrm{in}}$ to $\mathfrak F$ and $Q_{\mathrm{unstick}}$ from $\mathfrak F$ to $\Gamma^{\mathrm{out}}$.
% For example, in the two-sided case, deterministic kernels can be defined by
% \begin{equation}\label{eq;UnstickingKernel}
%     Q_{\mathrm{stick}}f(x,v)=f(x,\phi_x(v)),\qquad Q_{\mathrm{unstick}}f(x,v)=f(x',\phi_x^{-1}(v)),\qquad x\ne x'\in\beta^{-1}(\beta(x)),
% \end{equation}
% for a test function $f:E\to\mathbb{R}$, where $\phi_x$ is a suitable invertible velocity transformation. Unsticking moves the process to the opposite boundary face.

Lastly, we need to lift the manifold dynamics on $T(\mathcal{M})$ to the frozen boundary $\mathfrak{F}$. We choose the \textit{horizontal lift} for this purpose: for a state $(\xi,v^\tpar,v^\perp)\in\mathfrak{F}$, the normal velocity $v^\perp$ is parallel-transported in the normal bundle along $\mathcal{M}$, while $(\xi,v^\tpar)$ evolves according to the manifold dynamics.
Given two PDMP generators $\mathcal L_{\R^d}$ and $\mathcal L_{\mathcal M}$ in $\R^d$ and $\cM$ respectively, and denoting the horizontal lift by $\mathcal{L}^{\mathrm{lift}}_{\mathcal{M}}$, the generator of the sticky manifold PDMP on the state space $E$ is then defined by
\begin{equation}\label{eq;generator1}
    \mathcal L f=\begin{cases}
        \mathcal L_{\mathbb R^d}f,&\text{on }E\setminus\mathfrak{F},\\
        \mathcal L_{\mathcal M}^{\text{lift}}f+\Lambda(Q_{\mathrm{unstick}}f-f),&\text{on }\mathfrak{F}.
    \end{cases}
\end{equation}
The state-dependent unsticking rate $\Lambda$ determines when the process leaves $\mathcal{M}$ via the kernel $Q_{\mathrm{unstick}}$.
Note that there is no jump term in \eqref{eq;generator1} that corresponds to the sticking move; the sticking move is encoded in the boundary condition $Q_{\mathrm{stick}}f=\lim_{t\searrow0}f\circ\phi_{-t}$, where $\phi_t$ is the ambient flow, satisfied by the test functions $f$ in the generator domain $\mathcal D(\mathcal L)$; see \citet[Theorem~5.5]{Davis1984} and \citet[Theorem~26.14]{Davis1993}.
See Appendix~\ref{appendix-subsec-lifting-manifold-dynamics} for mathematical details and the well-definedness result.

\subsection{Invariance}\label{subsec;invarianceProof}

Lifting the position target $\pi$ in \eqref{eq;6manifoldtarget} to the state space $E$ constructed in Section~\ref{subsec-blowup}, we establish the invariance of the possibly unnormalised finite measure $\Pi=\omega\Pi_{E\setminus\mathfrak{F}}+(1-\omega)\Pi_{\mathfrak{F}}$, where
\[\Pi_{E\setminus\mathfrak{F}}(dxdv)=\pi_{\mathbb{R}^d}(x)\,dx\,\nu_{E\setminus\mathfrak{F},x}(dv),\quad(x,v)\in E\setminus\mathfrak{F},\]
\[\Pi_{\mathfrak{F}}(dxdv)=\pi_{\mathcal{M}}(x)\mathcal{H}_{\mathcal{M}}^{d-1}(dx)\nu_{\mathfrak{F},x}(dv),\quad (x,v)\in\mathfrak{F}.\]
Here, $\mathcal{H}^{d-1}_{\mathcal{M}}$ is the Hausdorff measure induced from the ambient metric or, equivalently, the Riemannian volume measure, and $\nu_{E\setminus\mathfrak{F}}$ and $\nu_{\mathfrak{F}}$ are probability kernels representing conditional velocity distributions.
Throughout this subsection, we absorb the common factor $e^{-U}$ in \eqref{eq;6manifoldtarget} into the component densities and, with a slight abuse of notation, continue to denote the resulting possibly unnormalised densities by $\pi_{\mathbb{R}^d}$ and $\pi_{\mathcal{M}}$ as the result only depends on the ratio between them. All proofs are deferred to Appendix~\ref{appendix-subsec-proofoftheorem}.

\begin{assumption}\label{assumption;MainText}
    We assume that the ambient PDMP and the lifted manifold PDMP leave $\Pi_{E\setminus\mathfrak F}$ and $\Pi_{\mathfrak F}$ invariant, respectively,
    and that the embedded Riemannian manifold $\mathcal{M}\subset\mathbb{R}^d$ is orientable, $d-1$ dimensional, and closed in $\mathbb{R}^d$.
    Additionally, we work under the technical conditions ensuring well-definedness, regularity of $\Pi_{E\setminus\mathfrak F}$ and $\Pi_{\mathfrak F}$, and integrability of the flow and the overall jump intensity, outlined in Assumption \ref{NewAssumption}.
\end{assumption}

Define the restriction of the ambient velocity distribution on the inward and outward boundaries $\Gamma_{\mathrm{in}},\Gamma_{\mathrm{out}}$ by
\[\nu_x^{\mathrm{in/out}}(B):=\sum_{\xi\in\beta^{-1}(x)}\nu_{E\setminus\mathfrak{F},\xi}(B_\xi),\]
respectively, for a Borel measurable set $B$ in $\Gamma_{\mathrm{in/out},x}:=\{(\xi,v)\in\Gamma_{\mathrm{in/out}}\mid\beta(\xi)=x\}$.
We use the standard notation $\mu Q(d\zeta):=\int Q(z,d\zeta)\mu(dz)$ for a measure $\mu$ and a probability kernel $Q$.

\begin{theorem}[Invariance of sticky manifold PDMP]\label{thm;NewTheorem}
    Under Assumption~\ref{assumption;MainText},
    suppose that the two probability kernels $Q_{\mathrm{stick}},Q_{\mathrm{unstick}}$ preserve the position $\beta(x)$ and the norm of the normal velocity component $\|v^\perp\|$. Additionally, assume that
    \begin{equation}\label{eq;VelocityCompatibilityCondition}
        \nu^{\mathrm{in}}_xQ_{\mathrm{stick},x}=\nu_{\mathfrak{F},x},\qquad \nu_{\mathfrak{F},x}Q_{\mathrm{unstick},x}=\nu_{x}^\mathrm{out},
    \end{equation}
    for $\mathcal{H}^{d-1}_{\mathcal{M}}$-almost all $x\in\mathcal{M}$.
    Then, the sticky manifold PDMP with unsticking rate \eqref{eq;6Manifoldthawingrate}
    %, i.e.,\[\Lambda(x)=\|v_\sigma^\perp\|\frac{\omega}{1-\omega}\frac{\pi_{\mathbb{R}^d}(x)}{\pi_{\mathcal{M}}(x)}\]
    leaves $\Pi$ invariant.
\end{theorem}

\begin{remark}[Velocity pushforward condition for deterministic kernels]
    Suppose further that the kernels are deterministic and given by
    \[Q_{\text{stick},x}(v,-)=\delta_{\psi_x(v)},\qquad Q_{\text{unstick},x}(\eta,-)=\delta_{R_x\circ\psi_x^{-1}(\eta)},\]
    where $\psi_x:\Gamma_{\mathrm{in},x}\to\mathfrak{F}_x$ is a smooth diffeomorphism and $R_x:\Gamma_{\mathrm{in},x}\to\Gamma_{\mathrm{out},x}$ is a side-flipping map that takes $(y,v)\in\Gamma_{\mathrm{in},x}$ to the state $(y',v)\in\Gamma_{\mathrm{out},x}$ with the same velocity $v$ but opposite position of the blown-up front face $y'\in\beta^{-1}(x)\setminus\{y\}$.
    Condition~\eqref{eq;VelocityCompatibilityCondition} then reads
    \begin{equation}\label{eq;PullbackCondition}
        (\psi_x)_*\nu_{x}^\mathrm{in}=\nu_{\mathfrak{F},x},\qquad (R_x\circ\psi_x^{-1})_*\nu_{\mathfrak{F},x}=\nu_{x}^\mathrm{out}.
    \end{equation}
    When $(R_x)_*\nu_{x}^\mathrm{in}=\nu_{x}^\mathrm{out}$, these conditions reduce to $(\psi_x)_*\nu_{x}^\mathrm{in}=\nu_{\mathfrak F,x}$.
\end{remark}

\begin{remark}[General rate function beyond the velocity compatibility condition]\label{remark;GeneralResult}
    The proof of Theorem~\ref{thm;NewTheorem} reveals a more general picture.
    In full generality, the unsticking rate involves velocity distributions as well, while the velocity compatibility~\eqref{eq;VelocityCompatibilityCondition} is a sufficient condition for the velocity factor to disappear from the rate function.
    Define the boundary measures
    \[J_{\text{in/out}}(dz):=1_{\Gamma_{\text{in/out}}}(z)\|v^\perp\|\pi_{\mathbb{R}^d}(x)\mathcal{H}^{d-1}_{\mathcal{M}}(dx)\nu_{E\setminus\mathfrak{F}}(x,dv).\]
    $J_{\text{in/out}}$ contains additional factors of the normal velocity norm $\|v^\perp\|$, ambient position density $\pi_{\mathbb{R}^d}$, and volume measure $\mathcal{H}^{d-1}_{\mathcal{M}}$.
    For $J_{\text{in/out}}$, we assume the following balance condition:
    \begin{equation}\label{eq;GeneralBalance}
        J_{\mathrm{in}}Q_{\mathrm{stick}}Q_{\mathrm{unstick}}=J_{\mathrm{out}},\qquad \omega J_{\mathrm{in}}Q_{\mathrm{stick}}=(1-\omega)\Lambda\Pi_{\mathfrak{F}}.
    \end{equation}
    This is a generalisation of \eqref{eq;VelocityCompatibilityCondition}. Indeed, the assumptions of Theorem~\ref{thm;NewTheorem} imply
    \[(\|v^\perp\|\nu_x^{\mathrm{in}})Q_{\textrm{stick},x}=\|v^\perp\|\nu_{\mathfrak{F},x},\qquad(\|v^\perp\|\nu_{\mathfrak{F},x})Q_{\textrm{unstick},x}=\|v^\perp\|\nu_{x}^{\mathrm{out}}.\]
    This can be interpreted as a fibre-wise balance condition that, in turn, implies \eqref{eq;GeneralBalance} after applying the coarea formula.
    Under \eqref{eq;GeneralBalance} and absolute continuity $J_{\mathrm{in}}Q_{\mathrm{stick}}\ll\Pi_{\mathfrak{F}}$, the rate given by the Radon--Nikodym derivative
    \begin{equation}\label{eq;GeneralRate}
        \Lambda(\zeta)=\frac{\omega}{1-\omega}\frac{d(J_{\mathrm{in}}Q_{\mathrm{stick}})}{d\Pi_{\mathfrak F}}(\zeta),\qquad\Pi_{\mathfrak F}\text{-a.e.}
    \end{equation}
    makes the sticky manifold PDMP $\Pi$-invariant.
\end{remark}

\subsection{Rao--Blackwellisation and extensions}

We establish a two-stage Rao--Blackwellisation within our sticky manifold framework that preserves the target position marginal. First, we average over the retained normal velocity in the rate, obtaining a rate analogous to that in reversible-jump BPS \cite[Proposition~2]{Chevallier2023}. When this averaged rate is constant, we further replace exponential sticking times by their expectations, extending the construction of \cite{Chin-Nishimura2026} for the sticky zig-zag sampler. Together, these operations remove both sources of randomness in the sticking time. Our numerical experiments demonstrate improvements in estimation accuracy and qualitatively different behaviours in the model-index autocorrelation function (Section~\ref{sec;4NumericalExamples}).

\begin{corollary}[Rao--Blackwellized unsticking rate]\label{cor;Rao-Blackwellized-rate}
    Assume the conditions of Theorem~\ref{thm;NewTheorem} with the standard Gaussian ambient velocity, and that the rate \eqref{eq;6Manifoldthawingrate} can be written as $\Lambda=c\|v^\perp\|$ for a measurable function $c(x)>0$. Then, replacing this rate with the velocity-averaged one $\Lambda_{\mathrm{RB}}=c\sqrt{2/\pi}$ still leaves the position marginal of $\Pi$ invariant, but now with reweighted velocity distribution $\nu_{\mathfrak{F},x}^{\mathrm{RB}}(dv)=\sqrt{{\pi}/{2}}\|v^\perp\|\nu_{\mathfrak F,x}(dv)$.
\end{corollary}

%We can push this approach further to average over the exponential clock to obtain deterministic sticking times. We continue to work under Gaussian ambient velocity, although this principle applies to general velocity distributions.

\begin{corollary}[Rao--Blackwellized sticking time]\label{cor;deterministic-sticking}
    Under the assumptions of Corollary~\ref{cor;Rao-Blackwellized-rate}, suppose further that $c$ is constant. Replace the exponential unsticking time with the deterministic residence time $T=\Lambda_{\mathrm{RB}}^{-1}=c^{-1}\sqrt{\pi/2}$. The resulting process has an invariant distribution whose position marginal distribution is identical to that of the original case.
\end{corollary}

Later, we also consider the case where the ambient space is a region $U\subset\mathbb{R}^d$ and $\mathcal{M}=\partial U$ is the boundary of $U$; see examples in Section~\ref{subsec;QuantumStateTomography} and~\ref{subsec;BayesianPCA}. In this case, the unsticking rate has to be altered accordingly.

\begin{corollary}[One-sided case]\label{cor-one-sided-case}
    Under Assumption~\ref{assumption;MainText}, assume that the ambient dynamics are confined within an open region $U\subset\mathbb{R}^d$ with boundary $\mathcal{M}=\partial U$, and that the ambient velocity distribution is everywhere spherically symmetric.
    Further suppose that $Q_{\mathrm{stick}}$ and $Q_{\mathrm{unstick}}$ preserve the position $x$ together with the norm of the normal velocity component $\|v^\perp\|$ and satisfy the compatibility condition
    \[\nu_x^{\mathrm{in}}Q_{\mathrm{stick},x}=\frac{1}{2}\nu_{\mathfrak{F},x},\qquad\frac12\nu_{\mathfrak{F},x}Q_{\mathrm{unstick},x}=\nu^{\mathrm{out}}_x.\]
    Then, half of the rate \eqref{eq;6Manifoldthawingrate} 
    \[\Lambda(x)=\frac{\|v_\sigma^\perp\|}{2}\frac{\omega}{1-\omega}\frac{\pi_{\mathbb{R}^d}(x)}{\pi_{\mathcal{M}}(x)}%\frac{d\nu_{E\setminus\mathfrak{F}}(x,-)}{d\nu_{\mathfrak{F}}(x,-)}(v)
    \]
    ensures the correct position marginal of the invariant distribution.
\end{corollary}

\begin{remark}[Nested strata]\label{remark-nested-sequence}
    The preceding construction extends directly to a finite nested sequence of submanifolds $\mathcal{M}_0\subset\mathcal{M}_1\subset\cdots\subset\mathcal{M}_n$ by a repeated application of the Gauss--Green argument (Lemma~\ref{lemma;ExtendedGaussGreen}), provided that each adjacent inclusion is of codimension one, transitions occur only between adjacent strata, and the corresponding assumptions and balance conditions hold at every interface. Note that a minor modification arises in the mixture weights, with the pair $\omega,1-\omega$ replaced by a sequence of weights $\omega_0,\omega_1,\cdots,\omega_n$ that sum to one.
\end{remark}

\begin{remark}[Submanifolds overlapping over a null set]\label{remark-overlapping-manifolds}
    In the survival extrapolation example (Section~\ref{subsec;survival-extrapolation}), we consider three codimension-one manifolds $\mathcal{M}_{\text W},\mathcal{M}_\Gamma,\mathcal{M}_{\text A}$ that intersect at $\{\lambda=\sigma=1\}$. Due to this overlap, the union $\mathcal{M}_{\text W}\cup\mathcal{M}_\Gamma\cup\mathcal{M}_{\text A}$ is not an embedded manifold. We nevertheless regard the three submodels as distinct labelled components within the disjoint union $\widetilde{M}:=\cup_{i\in\{\text{W},\Gamma,\text{A}\}}\{i\}\times\mathcal{M}_i$; a trajectory evolving in one of the components is not allowed to transition directly to the other components. Although $\widetilde{M}$ is no longer an embedded submanifold, our theorem also extends to this case by applying the invariance argument component-wise, given that the ambient trajectory hits two or more hypersurfaces simultaneously with probability zero.
\end{remark}

\subsection{Illustration: two charts of the quadratic cone and blow-up}\label{subsec;illustration-blowup-cone}

The punctured cone, defined by
\begin{equation}\label{eq;cone}
    C = \left\{(x,y,z)^\top \in \mathbb{R}^3 \,\middle|\, x^2+y^2=z^2,z>0\right\},
\end{equation}
is perhaps the simplest example that naturally produces a singularity. For $C$, we consider two coordinate charts: the projection coordinate chart and the polar coordinate chart.
The \textit{projection coordinate} $\varphi:C\to\mathbb{R}^2\setminus\{0\}$, defined by $\varphi(x,y,z)=(x,y)$, has constant Jacobian, $J^g_\varphi=2^{-1/2}$, and as such does not affect the weighted target \eqref{eq;weighted-target}. This simplifies the algorithm when the sole aim is sampling on $C$. A different perspective arises, however, when the aim is to also sample from the apex $0$ of $C$.

\begin{example}[Cone apex as the null matrix]\label{exp;ConicalParametrisation}
    In Section~\ref{subsec;BayesianPCA}, we deal with the space of the $d\times d$ positive semi-definite symmetric matrices $\operatorname{Sym}_d(\mathbb{R})_+$. When $d=2$, it is a solid cone with three strata $\mathcal{M}_i=\{\operatorname{rank}(G)=i\}$ of rank $i=0,1,2$ matrices. In this context, $C$ corresponds to the stratum of rank $1$ positive symmetric matrices $\mathcal{M}_1$ and its apex to the null matrix singleton $\mathcal{M}_0=\{0\}$ through the isometry $\Phi:\mathbb{R}^3\to\operatorname{Sym}_2(\mathbb{R})$ given by
    \[\Phi(x,y,z)=\frac{1}{\sqrt{2}}\begin{pmatrix}z+x&y\\y&z-x\end{pmatrix}.\]
    Since the two strata $\mathcal{M}_1,\mathcal{M}_0$ differ in dimension by two, the sampler defined by the projection coordinate chart is unable to sample from the apex. In contrast, the polar coordinate allows the apex to be hit with positive probability.
\end{example}

\begin{figure}
    \centering
    \includegraphics[trim={0 1cm 0 0},width=0.9\linewidth]{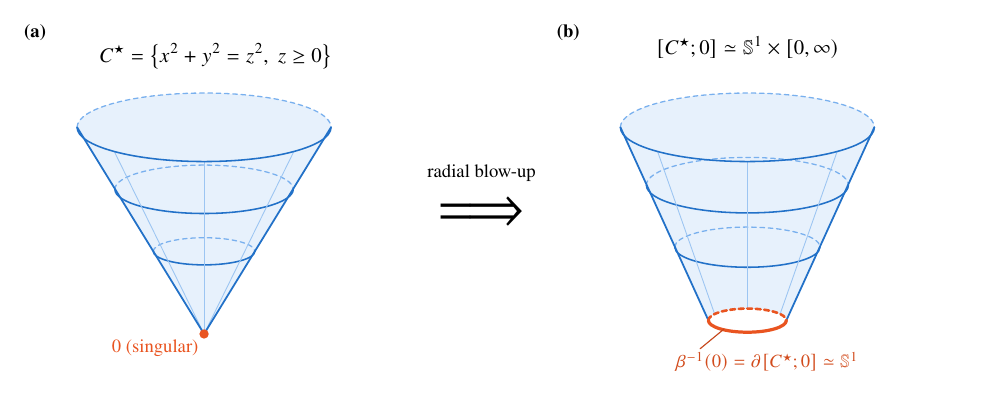}
    \caption{Blow-up of a half cone, or equivalently the space of all symmetric semi-definite rank-one matrices. (a) Singular cone $C^\star$ has its apex as a point of singularity. (b) Represents the blown-up manifold with its front face given by $\beta^{-1}(0)\simeq\mathbb{S}^1$.}
    \label{fig:ConeRadialBlowup}
\end{figure}

\textit{Polar coordinates} $(r,\theta)$ on $C$ are defined by the inverse of a parametrisation $\beta$
\[\beta:(0,\infty)\times\mathbb{S}^1\ni(r,\theta)\mapsto(r\cos\theta,r\sin\theta,r)\in C.\]
Note that the cone with apex $C^\star:=C\cup\{0\}=\{x^2+y^2=z^2,z\ge0\}$ is not a smooth manifold, although its punctured version $C$ is a two-dimensional smooth manifold on which $\beta^{-1}:C\to(0,\infty)\times\mathbb{S}^1$ is defined and smooth.
%In geometry, radial blow-up has been used to resolve such a singularity as the apex of $C^\star$; see \cite{Grieser2001,Melrose2008}, and Appendix~\ref{appendix-subsec-blowup} for the abstract construction.
Blowing up $C^\star$ results in a manifold with boundary $[C^\star;0]:=[0,\infty)\times\mathbb{S}^1$, on which the polar parametrisation $\beta$ naturally extends. This extension $\overline{\beta}:[C^\star;0]\to C^\star$ is the blow-down map (Figure~\ref{fig:ConeRadialBlowup}).
Importantly, the point $0\in C^\star$ is now replaced with a codimension one front face $\{0\}\times\mathbb{S}^1=\partial[C^\star;0]$. This allows a manifold PDMP sampler on $[C^\star;0]$ with the polar coordinate $(r,\theta)$ to stick to this boundary with a non-zero hitting probability. %This is achieved by extending the polar coordinate $r\searrow0$ with the direction of approach $\theta$ retained. 

\begin{remark}[Jacobian of the blow-up coordinates]\label{remark;divergingJacobian}
    The Jacobian of the chart is $J_\beta^g(r,\theta)=\sqrt{2}r$ for the induced metric $g$ of $C$; the weighted target \eqref{eq;weighted-target} is thus $\pi(r,\theta)=\pi^*(\beta(r,\theta))\cdot\sqrt{2}r$. Therefore, for naïve PDMP dynamics on the blown-up space to reach $r=0$ with positive probability, $\pi^*$ must diverge sufficiently rapidly as $r\searrow0$ to compensate for the vanishing Jacobian.
\end{remark}

\begin{example}[Higher-dimensional rank cones]
\label{exp;HigherDimensionalRankCones}
    We can generalise the above construction to higher dimensions using lifting. First, the polar coordinate chart generalises to a spectral parametrisation over $\operatorname{Sym}_d(\mathbb{R})_+$. Write
    \[\mathcal M_k:=\{G\in\operatorname{Sym}_d(\mathbb R)_+\mid\operatorname{rank}(G)=k\},\qquad k=0,\cdots,d.
    \]
    The dimension of $\mathcal M_k$ is $k(2d-k+1)/2$, so that the dimensions of
    successive rank strata differ by $d-k+1$. We define the spectral lift by
    \[\widetilde{\mathcal M}_k=C_k\times\operatorname{St}(d,d),\qquad\beta_k(r,Q)=Q\operatorname{diag}(r_1,\ldots,r_k,0,\ldots,0)Q^\top,\qquad C_k=\{r_1>\cdots>r_k>0\},\]
    with $C_0$ a singleton and $\beta_0(Q)=0$.
    The columns of $Q\in\operatorname{St}(d,d)=O(d)$ form an orthonormal eigenbasis of $\beta_k(r,Q)$. The map $\beta_k$ is a smooth fibre bundle over the subset of $\mathcal{M}_k$ with distinct positive eigenvalues, with fibre $\{\pm1\}^k\times O(d-k)$.
    Extending $r_k$ to zero produces the sticking boundary
    \[\{r_1>\cdots>r_{k-1}>0,\ r_k=0\}\times O(d)\simeq\widetilde{\mathcal M}_{k-1}.
    \]
    Thus, every adjacent rank transition becomes codimension-one in the lifted space.
\end{example}

\section{Numerical examples}
\label{sec;4NumericalExamples}

\subsection{Overview}

In this Section, we first validate our methodology on two toy examples where $\mathcal{M}$ corresponds to linear subspaces of $\mathbb{R}^d$ (Section~\ref{subsec;LinearSubspaces}) and to hyperspheres (Section~\ref{subsec;ConcentricSpheres}). The former illustrates how our rate \eqref{eq;6Manifoldthawingrate} accounts for changing incoming flux as the slope of the line varies; the latter demonstrates the effectiveness of Rao--Blackwellisation while highlighting a limitation of our method: infrequent manifold visits may slow down between-model mixing. We then apply the sampling approach to three examples of applied statistical interest arising in quantum state tomography (Section~\ref{subsec;QuantumStateTomography}), Bayesian principal component analysis (Section~\ref{subsec;BayesianPCA}), and survival analysis (Section~\ref{subsec;survival-extrapolation}). The quantum state tomography example highlights the computational value of model transitions that require no additional likelihood evaluations. Bayesian principal component analysis provides a higher-dimensional setting for joint inference on rank and covariance. Finally, the survival analysis example demonstrates how a single construction accommodates multiple parametric families without the need to develop several bespoke reversible-jump proposals.

In these examples, we mainly use forward event-chain Monte Carlo (FECMC) as the default PDMP dynamics, given its favourable performance \citep{Michel2020,Shiba-Kamatani2026}, although our sticky manifold construction is agnostic to the choice of within-model dynamics. Our main comparator is the celebrated reversible jump algorithm \citep{Green1995}, defining between-model proposals that are bespoke to each application. For within-model sampling, we consider two alternatives, Hamiltonian Monte Carlo proposals (RJ-HMC) and standard FECMC dynamics (RJ-FECMC, Section~\ref{subsec-GMM-illustration}), on the same parametrisation. In the former case, the algorithm alternates between within and across model proposals, while in the latter reversible jump proposals are triggered with a time-homogeneous Poisson rate $\lambda>0$, both tuned in pilot runs to yield a comparable number of model transitions to the sticky manifold algorithm (SM-FECMC). RJ-FECMC serves as a bridge between SM-FECMC and RJ-HMC that helps distinguish the effects of within-model dynamics from those of the model-transition mechanism. To yield competitive performance, we tuned the integration step $T$ and step size $\epsilon$ of the HMC dynamics based on the effective sample size per CPU time, while we did not allow any tuning or preconditioning for FECMC dynamics in SM-FECMC and RJ-FECMC. For a given sampler, its statistical efficiency is measured through the root mean squared error (RMSE) over 100 independent runs of the same budget. The ground truth is either computed by numerical quadrature or approximated by a long run, whose error or Monte Carlo standard error is verified to be sufficiently smaller than the RMSE. Further details on implementation are in Appendix~\ref{appendix-ExperimentalDetails}.

\subsection{Varying hypersphere radii}\label{subsec;ConcentricSpheres}

\begin{figure}[!th]
    \centering
    \includegraphics[width=\linewidth]{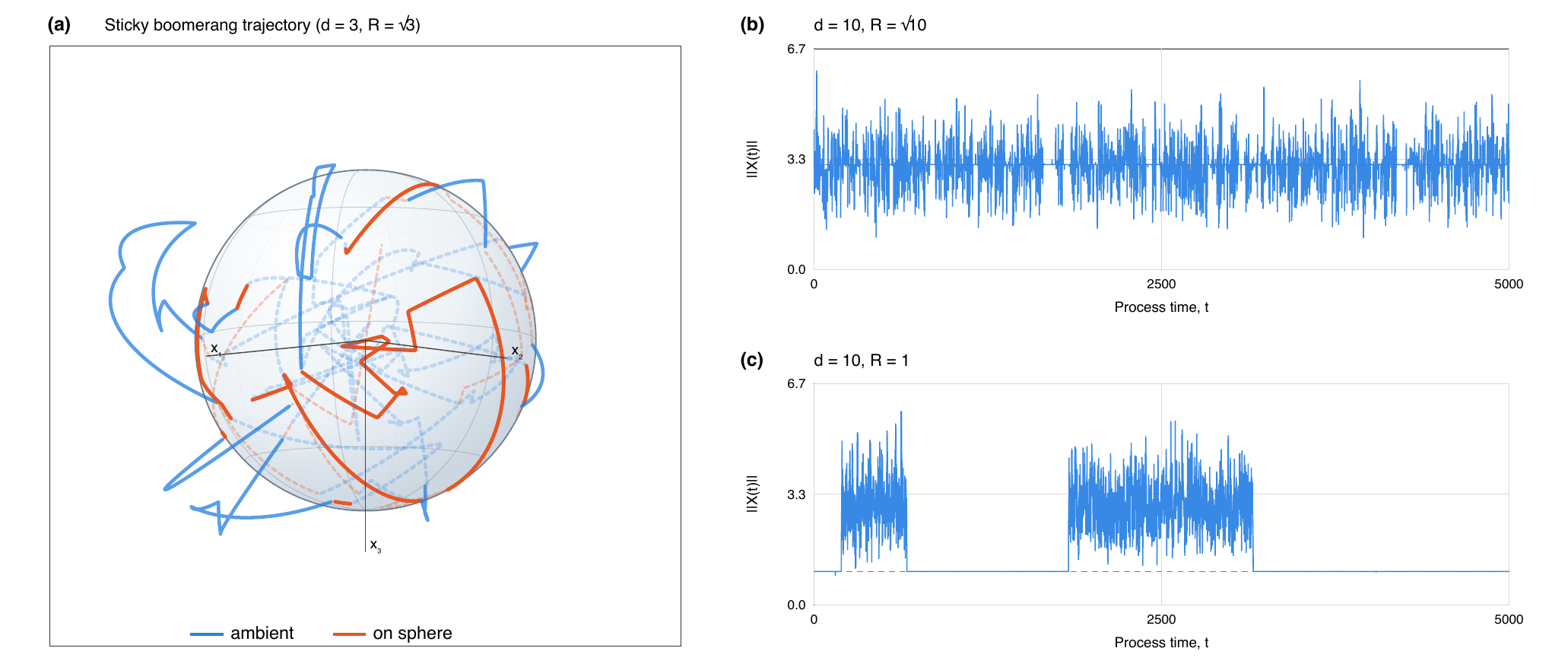}
    \caption{(a) Trajectory of the sticky boomerang sampler for $d=3$ and $R=\sqrt{d}$; blue segments evolve in the ambient space while orange segments on the sphere along its geodesic flow (see Remark~\ref{remark-high-dimensional-hypersphere}), with back-facing portions shown faint and dashed. (b), (c) Radius traces for $d=10$ with (b) $R=\sqrt{d}$ and (c) $R=1$. Both targets assign mass $1/2$ to the sphere. The smaller sphere is reached rarely and produces longer residence episodes.}
    \label{fig:6PDMPSphere2}
\end{figure}

In this section we sample from an equal mixture distribution of the $d$-dimensional standard Gaussian $\pi_{\mathbb{R}^d}$ and its restriction $\pi_{\mathcal{M}}$ to the hypersphere $\mathcal{M}_R:=\{\|x\|=R\}\subset\mathbb{R}^d$. Using Gaussian velocity (Example~\ref{example;SM-identity-sticking}), the rate is $\Lambda_R=\|v^\perp\|\chi_d(R)$, where $\chi_d$ is the chi density function with $d$ degrees of freedom.
Figure~\ref{fig:6PDMPSphere2} (b) and (c) show the sample paths of $\|x\|$ for $d = 10$, and $R= \sqrt{d}$ and $1$, respectively, in consideration that the mass concentrates around $\|x\| = \sqrt{d}$ as $d$ becomes large. This highlights that, when $\mathcal{M}$ is close to the high-density region of the target, sticking times are short to account for shorter excursions away from $\mathcal{M}$, and vice versa. Thus, when the density ratio $\pi_{\mathbb{R}^d}/\pi_{\mathcal{M}}$ is small and manifold visits are rare, the mixing of the marginal model-index process can be limited by the longer sojourn, even if the mixture weights of the target distribution are equal $\omega=1-\omega=1/2$. This deterioration is analogous to slow mixing in the reversible-jump methods with poorly tuned proposal distributions.

Table~\ref{tab:four-variants-fecmc} compares four variants of the sticky manifold PDMP sampler for $R = \sqrt{d}, 1$ across varying dimensions. In particular these variants are \textit{i)} The standard sampler; \textit{ii)} The standard sampler where in addition $v^\perp$ is refreshed at refreshment events; \textit{iii)} The standard sampler with the averaged unsticking rate (Corollary \ref{cor;Rao-Blackwellized-rate}); \textit{iv)} The previous sampler with deterministic sticking time (Corollary \ref{cor;deterministic-sticking}). For each sampler, RMSE changes only moderately as $d$ increases when $R = \sqrt{d}$. When $R = 1$, however, RMSE increases rapidly with no model transitions for $d = 20, 50$. In both cases the Rao--Blackwellisation procedures improve sampler efficiency, most notably for the fixed unsticking time.

\begin{table}[!thb]
\centering
\caption{RMSE for estimating the ambient mixture weight $\omega=1/2$, mean number of trans-dimensional jumps, and the empirical variance of sticking times. Each condition uses 1000 independent stationary-started runs of time horizon $T=10^4$.}
\label{tab:four-variants-fecmc}\vspace{1mm}
\normalsize\setlength{\tabcolsep}{5pt}
\resizebox{\linewidth}{!}{%
\begin{tabular}{rcccccccccccc}
\toprule
 & \multicolumn{3}{c}{FECMC} & \multicolumn{3}{c}{+ normal refreshment} & \multicolumn{3}{c}{+ averaged rate} & \multicolumn{3}{c}{+ fixed unsticking time} \\
\cmidrule(lr){2-4}\cmidrule(lr){5-7}\cmidrule(lr){8-10}\cmidrule(lr){11-13}
$d$ & RMSE & jumps & \shortstack{Sticking time\\variance} & RMSE & jumps & \shortstack{Sticking time\\variance} & RMSE & jumps & \shortstack{Sticking time\\variance} & RMSE & jumps & \shortstack{Sticking time\\variance} \\
\midrule
\addlinespace[0.5em]
\multicolumn{13}{l}{$R=\sqrt{d}$} \\
3 & $0.0242$ & 4263.6 & $37.83$ & $0.0096$ & 4260.4 & $7.45$ & $0.0089$ & 4258.0 & $5.51$ & $\mathbf{0.0067}$ & 4262.6 & $0.00$ \\
5 & $0.0222$ & 4356.7 & $38.98$ & $0.0098$ & 4357.5 & $7.13$ & $0.0093$ & 4351.4 & $5.27$ & $\mathbf{0.0078}$ & 4356.8 & $0.00$ \\
10 & $0.0212$ & 4439.7 & $39.19$ & $0.0110$ & 4429.2 & $6.94$ & $0.0103$ & 4425.6 & $5.10$ & $\mathbf{0.0089}$ & 4427.0 & $0.00$ \\
20 & $0.0227$ & 4457.0 & $36.78$ & $0.0126$ & 4470.1 & $6.81$ & $0.0117$ & 4465.3 & $5.02$ & $\mathbf{0.0104}$ & 4463.1 & $0.00$ \\
50 & $0.0247$ & 4486.8 & $52.61$ & $0.0150$ & 4482.2 & $6.77$ & $0.0140$ & 4491.1 & $4.96$ & $\mathbf{0.0129}$ & 4488.7 & $0.00$ \\
\addlinespace[0.5em]
\multicolumn{13}{l}{$R=1$} \\
3 & $0.0279$ & 3854.1 & $60.04$ & $0.0124$ & 3860.0 & $8.84$ & $0.0113$ & 3863.5 & $6.71$ & $\mathbf{0.0102}$ & 3866.1 & $0.00$ \\
5 & $0.0401$ & 1284.5 & $367.90$ & $0.0252$ & 1283.7 & $67.79$ & $0.0242$ & 1287.8 & $59.91$ & $\mathbf{0.0221}$ & 1287.2 & $0.00$ \\
10 & $0.2349$ & 12.4 & $6.40\times10^{5}$ & $0.2156$ & 12.5 & $5.05\times10^{5}$ & $0.1954$ & 12.5 & $5.24\times10^{5}$ & $\mathbf{0.1743}$ & 12.7 & $0.00$ \\
20 & $0.5000$ & 0.0 & \textemdash & $0.5000$ & 0.0 & \textemdash & $0.5000$ & 0.0 & \textemdash & $0.5000$ & 0.0 & \textemdash \\
50 & $0.5000$ & 0.0 & \textemdash & $0.5000$ & 0.0 & \textemdash & $0.5000$ & 0.0 & \textemdash & $0.5000$ & 0.0 & \textemdash \\
\bottomrule
\end{tabular}}
\end{table}

\subsection{The Bloch sphere in quantum state tomography}\label{subsec;QuantumStateTomography}

%We place the same spherical geometry as in the previous section in the context of quantum state tomography.
Quantum state tomography estimates a quantum state from measurements on identically prepared systems \citep{Blume-Kohout2010}. Experiments often aim to prepare a pure quantum state, but noise can produce a mixed state. The state of a single qubit is represented by a Bloch vector $r=(r_x,r_y,r_z)\in B^3:=\{\|r\|\le1\}$, with pure states on the boundary $\mathbb{S}^2=\partial B^3$ and mixed states in the interior. To quantify uncertainty between the two models, we therefore assign positive prior mass to both the sphere and the interior, yielding a mixed-dimensional prior for Bayesian model averaging \citep{Ferrie2014}.
Specifically, we use an equal mixture of uniform distributions on $B^3$ and $\mathbb{S}^2$.

A measurement in the direction $a\in\mathbb{S}^2$ yields $Y_a\in\{\pm1\}$ with likelihood
\[
\operatorname{P}[Y_a=y\,|\,r]=\frac{1+ya^\top r}{2},\qquad y\in\{-1,+1\}.
\]
Synthetic data are generated from $r_\alpha=\alpha(1,1,1)^\top/\sqrt{3}$, with $\alpha\in\{1,.99\}$ and $N=100$ independent measurements in uniformly random directions.
We use Rao--Blackwellized sticky manifold FECMC (RB-SM-FECMC) with fixed sticking times (Corollary~\ref{cor;deterministic-sticking}), and the unsticking rate is halved to account for one-sidedness (see Corollary~\ref{cor-one-sided-case}).
For RJ-HMC, we use $T=3$ and $\epsilon\in\{0.08,0.09\}$, guided by pilot searches maximising the ESS per CPU second, and geodesic Monte Carlo \citep{Byrne2013} on $\mathbb{S}^2$.
The reversible-jump proposal is tailored to the spherical geometry \citep{Brooks2003}: it proposes a radius $R\in(0,1)$ when entering the ball and radially projects onto $\mathbb S^2$ in the reverse direction. Its frequency is calibrated separately for each $(\alpha,\epsilon)$ to match RB-SM-FECMC's model-transition count per CPU second.
Each algorithm is initialised from an exact posterior sample from a rejection sampler and run for one second 100 times. We report RMSEs of four quantities: the posterior pure state probability $p_{\mathbb{S}^2}:=\operatorname{P}[\|r\|=1\,|\,y]$, two conditional posterior means $\mu_{\mathbb{S}^2}:=\operatorname{E}[r\,|\,\|r\|=1,y],\;\mu_{B^3}:=\operatorname{E}[r\,|\,\|r\|<1,y]$, and the posterior mean $\operatorname{E}[r\,|\,y]$.

\begin{table}[!ht]
\centering
\caption{RMSE from 100 paired runs under two budgets: (a) 1 process CPU second and (b) $10^6$ potential or gradient evaluations. Both panels use the same data and initial states. $N_U$ and $N_{\nabla U}$ count evaluations of the potential $U$ and its gradient, respectively, including directional derivatives in the latter. Bold values indicate the smallest RMSE.}
\label{tab:quantum_1cpu_second}
\vspace{1mm}
\fontsize{8.5}{10.5}\selectfont
\setlength{\tabcolsep}{4pt}
\renewcommand{\arraystretch}{1.25}
\resizebox{\linewidth}{!}{%
\begin{tabular}{@{}clrrrrrrr@{}}
\toprule
\multirow{2}{*}{$\alpha$} & \multirow{2}{*}{Sampler} & \multicolumn{4}{c}{RMSE} & \multirow{2}{*}{Transitions} & \multirow{2}{*}{$N_U$} & \multirow{2}{*}{$N_{\nabla U}$} \\
\cmidrule(lr){3-6}
 & & $p_{\mathbb S^2}$ & $\mu_{\mathbb S^2}$ & $\mu_{B^3}$ & $\operatorname{E}[r\mid y]$ & & & \\
\midrule
\multicolumn{9}{@{}l}{\textbf{(a) 1 process CPU second}} \\
\addlinespace[2pt]
\multirow{4}{*}{1.00} & RB-SM-FECMC & $\mathbf{4.91}\!\times\!10^{-4}$ & $3.56\!\times\!10^{-4}$ & $8.34\!\times\!10^{-4}$ & $3.30\!\times\!10^{-4}$ & $1.69\!\times\!10^{5}$ & $0$ & $5.80\!\times\!10^{5}$ \\
 & RJ-FECMC & $9.90\!\times\!10^{-4}$ & $6.21\!\times\!10^{-4}$ & $1.12\!\times\!10^{-3}$ & $6.28\!\times\!10^{-4}$ & $1.69\!\times\!10^{5}$ & $7.55\!\times\!10^{5}$ & $1.71\!\times\!10^{5}$ \\
 & RJ-HMC ($\epsilon=0.08$) & $6.40\!\times\!10^{-4}$ & $2.60\!\times\!10^{-4}$ & $\mathbf{5.77}\!\times\!10^{-4}$ & $2.46\!\times\!10^{-4}$ & $1.71\!\times\!10^{5}$ & $1.36\!\times\!10^{6}$ & $2.26\!\times\!10^{6}$ \\
 & RJ-HMC ($\epsilon=0.09$) & $6.25\!\times\!10^{-4}$ & $\mathbf{2.36}\!\times\!10^{-4}$ & $6.54\!\times\!10^{-4}$ & $\mathbf{2.25}\!\times\!10^{-4}$ & $1.71\!\times\!10^{5}$ & $1.36\!\times\!10^{6}$ & $2.25\!\times\!10^{6}$ \\
\midrule
\multirow{4}{*}{0.99} & RB-SM-FECMC & $\mathbf{4.87}\!\times\!10^{-4}$ & $4.33\!\times\!10^{-4}$ & $6.84\!\times\!10^{-4}$ & $3.48\!\times\!10^{-4}$ & $1.61\!\times\!10^{5}$ & $0$ & $6.18\!\times\!10^{5}$ \\
 & RJ-FECMC & $1.59\!\times\!10^{-3}$ & $5.26\!\times\!10^{-4}$ & $9.58\!\times\!10^{-4}$ & $5.37\!\times\!10^{-4}$ & $1.59\!\times\!10^{5}$ & $5.99\!\times\!10^{5}$ & $2.77\!\times\!10^{5}$ \\
 & RJ-HMC ($\epsilon=0.08$) & $7.91\!\times\!10^{-4}$ & $2.38\!\times\!10^{-4}$ & $\mathbf{5.13}\!\times\!10^{-4}$ & $2.39\!\times\!10^{-4}$ & $1.62\!\times\!10^{5}$ & $1.32\!\times\!10^{6}$ & $2.38\!\times\!10^{6}$ \\
 & RJ-HMC ($\epsilon=0.09$) & $8.32\!\times\!10^{-4}$ & $\mathbf{2.22}\!\times\!10^{-4}$ & $5.22\!\times\!10^{-4}$ & $\mathbf{2.36}\!\times\!10^{-4}$ & $1.64\!\times\!10^{5}$ & $1.32\!\times\!10^{6}$ & $2.38\!\times\!10^{6}$ \\
\midrule[0.8pt]
\multicolumn{9}{@{}l}{\textbf{(b) $10^6$ combined potential/first-derivative evaluations}} \\
\addlinespace[2pt]
\multirow{4}{*}{1.00} & RB-SM-FECMC & $\mathbf{3.61}\!\times\!10^{-4}$ & $\mathbf{2.92}\!\times\!10^{-4}$ & $\mathbf{6.05}\!\times\!10^{-4}$ & $\mathbf{2.58}\!\times\!10^{-4}$ & $2.92\!\times\!10^{5}$ & $0$ & $1.00\!\times\!10^{6}$ \\
 & RJ-FECMC & $9.16\!\times\!10^{-4}$ & $5.82\!\times\!10^{-4}$ & $1.07\!\times\!10^{-3}$ & $5.95\!\times\!10^{-4}$ & $1.83\!\times\!10^{5}$ & $8.16\!\times\!10^{5}$ & $1.84\!\times\!10^{5}$ \\
 & RJ-HMC ($\epsilon=0.08$) & $1.15\!\times\!10^{-3}$ & $4.98\!\times\!10^{-4}$ & $1.20\!\times\!10^{-3}$ & $4.64\!\times\!10^{-4}$ & $4.73\!\times\!10^{4}$ & $3.77\!\times\!10^{5}$ & $6.23\!\times\!10^{5}$ \\
 & RJ-HMC ($\epsilon=0.09$) & $1.11\!\times\!10^{-3}$ & $4.40\!\times\!10^{-4}$ & $1.16\!\times\!10^{-3}$ & $4.02\!\times\!10^{-4}$ & $4.74\!\times\!10^{4}$ & $3.76\!\times\!10^{5}$ & $6.24\!\times\!10^{5}$ \\
\midrule
\multirow{4}{*}{0.99} & RB-SM-FECMC & $\mathbf{4.10}\!\times\!10^{-4}$ & $\mathbf{3.17}\!\times\!10^{-4}$ & $\mathbf{5.44}\!\times\!10^{-4}$ & $\mathbf{2.58}\!\times\!10^{-4}$ & $2.61\!\times\!10^{5}$ & $0$ & $1.00\!\times\!10^{6}$ \\
 & RJ-FECMC & $1.48\!\times\!10^{-3}$ & $4.93\!\times\!10^{-4}$ & $8.99\!\times\!10^{-4}$ & $5.10\!\times\!10^{-4}$ & $1.81\!\times\!10^{5}$ & $6.84\!\times\!10^{5}$ & $3.16\!\times\!10^{5}$ \\
 & RJ-HMC ($\epsilon=0.08$) & $1.41\!\times\!10^{-3}$ & $5.34\!\times\!10^{-4}$ & $1.06\!\times\!10^{-3}$ & $5.01\!\times\!10^{-4}$ & $4.39\!\times\!10^{4}$ & $3.56\!\times\!10^{5}$ & $6.44\!\times\!10^{5}$ \\
 & RJ-HMC ($\epsilon=0.09$) & $1.43\!\times\!10^{-3}$ & $4.76\!\times\!10^{-4}$ & $9.98\!\times\!10^{-4}$ & $4.31\!\times\!10^{-4}$ & $4.44\!\times\!10^{4}$ & $3.56\!\times\!10^{5}$ & $6.44\!\times\!10^{5}$ \\
\bottomrule
\end{tabular}}
\end{table}

% \begin{figure}[!hbt]
%     \centering
%     \includegraphics[width=\linewidth]{Quantum_autocorrelation.pdf}
%     \caption{Model-indicator autocorrelation. Lags are converted to CPU seconds using the average CPU time needed for one unit of FECMC process time or one RJ-HMC sweep. Only RB-SM-FECMC with fixed sticking times (gray) plunges into the negative region, followed by damped oscillations.}
%     \label{fig:Quantum_autocorrelation}
% \end{figure}

\begin{figure}[!hbt]
    \centering
    \includegraphics[width=\linewidth]{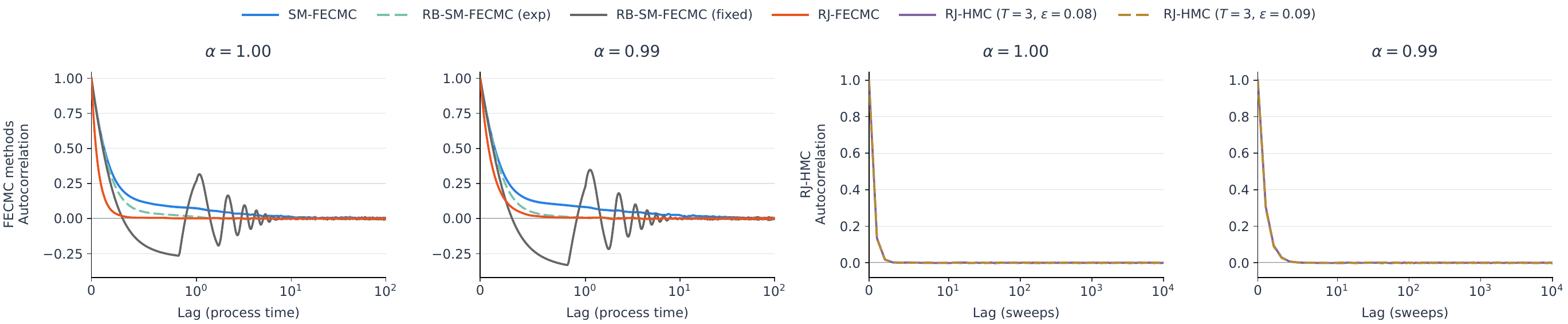}
    \caption{Model-indicator autocorrelation. Only RB-SM-FECMC with fixed sticking times (gray) plunges into the negative region, followed by damped oscillations. RB-SM-FECMC (exp) represents the version with the Rao--Blackwellised rate (Corollary~\ref{cor;Rao-Blackwellized-rate}), which leads to exponential sticking times.}
    \label{fig:Quantum_autocorrelation}
\end{figure}

At a budget of one CPU second, RB-SM-FECMC achieves smaller RMSEs than RJ-FECMC for all reported quantities across the two scenarios (Table~\ref{tab:quantum_1cpu_second}, panel (a)). It also estimates the posterior pure-state probability $p_{\mathbb{S}^2}$ more accurately than RJ-HMC, despite approximately matched model-transition counts. RJ-HMC, however, gives more accurate estimates of the conditional posterior means $\mu_{\mathbb{S}^2}$ and $\mu_{B^3}$, as well as the model-averaged posterior mean $\operatorname{E}[r\,|\, y]$.
The good mixing of the model indicator process under RB-SM-FECMC is reflected in the autocorrelation function (Figure~\ref{fig:Quantum_autocorrelation}), in which RB-SM-FECMC exhibits negative regions and damped oscillations, whereas these features are absent from the other algorithms. These negative correlations may contribute to the reduced RMSE through antithetic effects.

Computing the unsticking rate \eqref{eq;6Manifoldthawingrate} does not involve likelihood evaluations and therefore RB-SM-FECMC is the only sampler compared in panel (a) of Table~\ref{tab:quantum_1cpu_second} that requires no evaluations of $U$ during trajectory simulation, as recorded by $N_U=0$.
To distinguish efficiency per target evaluation from the remaining simulation costs in this three-dimensional example, panel (b) of Table~\ref{tab:quantum_1cpu_second} reports a complementary
result at a fixed budget of $N_U+N_{\nabla U}=10^6$. Each evaluation of $U$ or its full gradient $\nabla U$ counts as one epoch, including all evaluations required for event-time simulation. Under this cost measure, RB-SM-FECMC achieves the smallest observed RMSE for every reported quantity in all scenarios.

\subsection{Blown-up rank strata in Bayesian principal component analysis}\label{subsec;BayesianPCA}

In this example, we apply the blow-up construction in Section~\ref{subsec;illustration-blowup-cone} to Bayesian principal component analysis \citep{Bishop1998,Zhang+2004} with unknown signal rank. %Specifically, we consider 20-dimensional observations $y_i|x_i\sim N_{20}(x_i,\sigma^2I_{20})$ from a Gaussian source $x_i\sim N_{20}(0,G)$ with a known noise variance $\sigma^2>0$. 
Specifically, for $y_i \in \mathbb{R}^{20}$ and known $\sigma^2$,
\begin{equation*}
    y_i|x_i\sim N_{20}(x_i,\sigma^2I_{20}), \quad x_i\sim N_{20}(0,G), \quad i=1,\cdots,n.
\end{equation*}
%When $d=2$, the rank-one cone $\mathcal{M}_1$ in the space of $2\times2$ positive semi-definite symmetric matrices $\operatorname{Sym}_2(\mathbb{R})_+$ is a conical surface (Figure~\ref{fig:Cone}).
We generate $n=80$ observations with $\sigma=0.5$ in six scenarios, where $G$ has eigenvalues A: $(2,1.5,1,0.5,0,\ldots,0)$; B: $(1,1,1,0.1,\ldots,0.1)$; C: $(1,1,1,0.1,0.1,0.1,0,\ldots,0)$; D: $(1,1,1,1,1,0,\ldots,0)$; E: $(1,0.95,\ldots,0.05)$; O: $(0,\ldots,0)$.
We consider a mixed-dimensional prior on $G$ with equal weights on each rank stratum $\mathcal{M}_i$, in which $G=U_kA_kU_k^\top$ where $A_k\sim W_k(k+1,(k+1)^{-1}I_k)$ is Wishart-distributed and $U_k$ is uniformly distributed on the Stiefel manifold $\operatorname{St}(k,d)$.
The full-rank stratum $\mathcal{M}_{20}$ has dimension 210.
We report RMSEs for three quantities: the posterior rank-probability vector $(p_0,\ldots,p_{20})$, where $p_k=\operatorname{P}[\operatorname{rank}(G)=k\,|\, y]$, the posterior mean covariance $\operatorname{E}[G\,|\, y]$, and the posterior mean signal (or denoised signal) $\operatorname{E}[x\,|\, y]$. SM-FECMC uses Rao--Blackwellised rates (Corollary~\ref{cor;Rao-Blackwellized-rate}). HMC dynamics combine reflective dynamics on the constrained Euclidean regions $C_k$ with geodesic Monte Carlo \citep{Byrne2013} on the Stiefel manifold $O(20)$, and use integration step $T$ and step size $\epsilon$ that are guided by pilot searches maximising the ESS per CPU time, while we did not allow any tuning or preconditioning for FECMC dynamics. All three algorithms are run on the same spectral lift coordinates (Example~\ref{exp;HigherDimensionalRankCones}) for 3 CPU seconds over 100 times with shared initial states from a separate long run.

\begin{figure}[!thb]
    \centering
    \includegraphics[width=\linewidth]{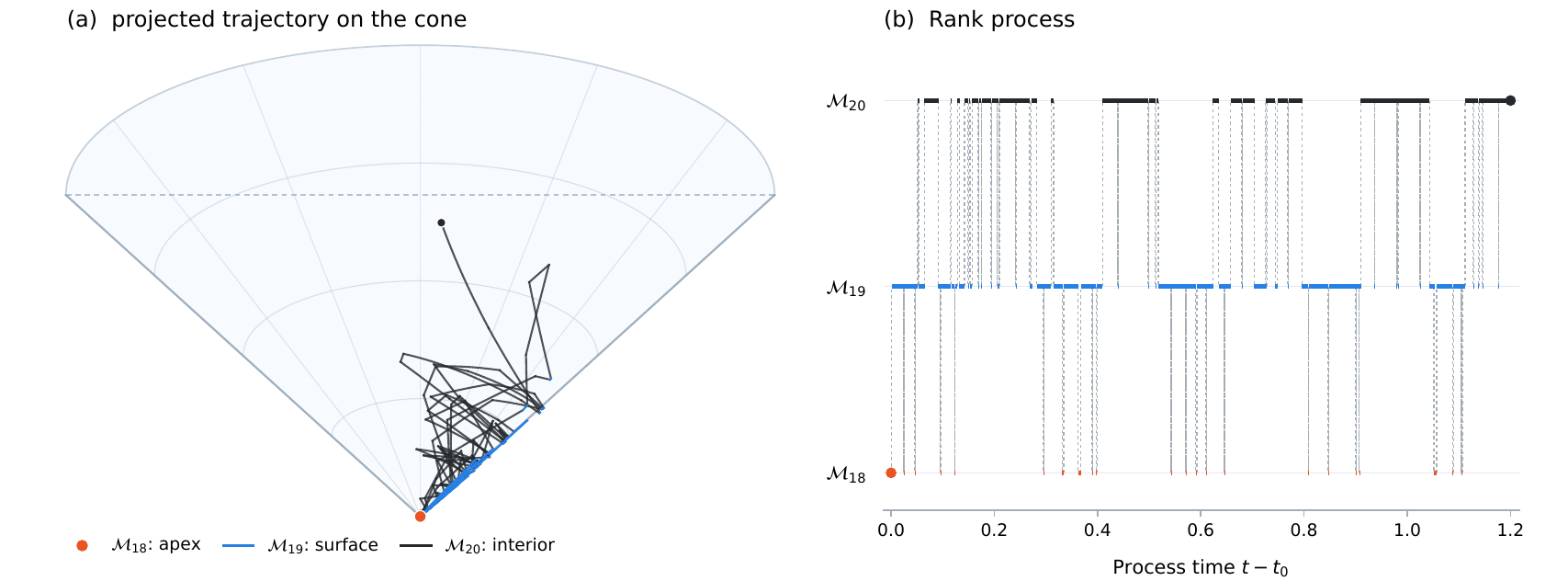}
    \caption{An illustrative SM-FECMC posterior trajectory in scenario E. (a) The trajectory of the Schur-complement $S_t:=D_t-B_t^\top A_t^{-1}B_t\in\operatorname{Sym}_2(\mathbb{R})_+$ via the conical parametrisation (Example~\ref{exp;ConicalParametrisation}), where $U^{-1}G_tU=(A_t\;B_t;B_t^\top\;D_t)$ for a $U\in O(20)$. (b) The corresponding rank process over a short time interval.}
    \label{fig:Cone}
\end{figure}

\begin{table}[!htb]
\centering
\caption{RMSEs for the six scenarios from 100 paired replicates with 3 CPU seconds. $k^\star$ denotes the rank with the maximum posterior probability $\widehat{p}_{k^*}$, which is computed from the reference run.}
\label{tab:BPCA}
\small
\setlength{\tabcolsep}{4pt}
\renewcommand{\arraystretch}{1.08}\vspace{1mm}
\begin{tabular}{@{}clrrrr@{}}
\toprule
Scenario & Sampler & \shortstack{Rank-vector\\RMSE} & \shortstack{Mean-$G$\\RMSE} & \shortstack{Denoising\\RMSE} & \shortstack{Mean rank\\transitions} \\
\midrule
\multirow{4}{*}{\shortstack{A\\{\footnotesize $(2,1.5,1,0.5,0,\ldots,0)$}\\{\footnotesize $k^\star=6,\ \widehat p_{k^\star}=0.459$}}} & SM-FECMC & $1.12\times 10^{-2}$ & $\mathbf{1.46\times 10^{-2}}$ & $\mathbf{4.95\times 10^{-3}}$ & $3.96\times 10^{4}$ \\
 & RJ-FECMC & $1.15\times 10^{-2}$ & $1.60\times 10^{-2}$ & $5.25\times 10^{-3}$ & $3.62\times 10^{4}$ \\
 & RJ-HMC ($T=3,\ \epsilon=0.04$) & $1.05\times 10^{-2}$ & $2.31\times 10^{-2}$ & $6.26\times 10^{-3}$ & $3.96\times 10^{4}$ \\
 & RJ-HMC ($T=3,\ \epsilon=0.05$) & $\mathbf{8.95\times 10^{-3}}$ & $2.11\times 10^{-2}$ & $5.81\times 10^{-3}$ & $4.02\times 10^{4}$ \\
\midrule
\multirow{4}{*}{\shortstack{B\\{\footnotesize $(1,1,1,0.1,\ldots,0.1)$}\\{\footnotesize $k^\star=9,\ \widehat p_{k^\star}=0.450$}}} & SM-FECMC & $\mathbf{1.27\times 10^{-2}}$ & $\mathbf{1.74\times 10^{-2}}$ & $\mathbf{9.63\times 10^{-3}}$ & $2.85\times 10^{4}$ \\
 & RJ-FECMC & $1.32\times 10^{-2}$ & $1.79\times 10^{-2}$ & $9.77\times 10^{-3}$ & $2.97\times 10^{4}$ \\
 & RJ-HMC ($T=3,\ \epsilon=0.03$) & $3.03\times 10^{-2}$ & $1.05\times 10^{-1}$ & $6.44\times 10^{-2}$ & $3.01\times 10^{4}$ \\
 % & RJ-HMC ($T=3,\ \epsilon=0.04$) & $4.13\times 10^{-2}$ & $1.23\times 10^{-1}$ & $7.44\times 10^{-2}$ & $3.22\times 10^{4}$ \\
 % & RJ-HMC ($T=3,\ \epsilon=0.05$) & $3.90\times 10^{-2}$ & $1.24\times 10^{-1}$ & $7.52\times 10^{-2}$ & $3.23\times 10^{4}$ \\
  & RJ-HMC ($T=10,\ \epsilon=0.007$) & $2.09\times 10^{-2}$ & $5.86\times 10^{-2}$ & $2.70\times 10^{-2}$ & $2.86\times 10^{4}$ \\
\midrule
\multirow{4}{*}{\shortstack{C\\{\footnotesize $(1,1,1,0.1,0.1,0.1,0,\ldots,0)$}\\{\footnotesize $k^\star=4,\ \widehat p_{k^\star}=0.484$}}} & SM-FECMC & $1.34\times 10^{-2}$ & $\mathbf{1.09\times 10^{-2}}$ & $4.92\times 10^{-3}$ & $3.52\times 10^{4}$ \\
 & RJ-FECMC & $1.35\times 10^{-2}$ & $1.13\times 10^{-2}$ & $5.40\times 10^{-3}$ & $3.82\times 10^{4}$ \\
 & RJ-HMC ($T=3,\ \epsilon=0.04$) & $9.03\times 10^{-3}$ & $1.77\times 10^{-2}$ & $5.33\times 10^{-3}$ & $3.73\times 10^{4}$ \\
 & RJ-HMC ($T=3,\ \epsilon=0.05$) & $\mathbf{8.56\times 10^{-3}}$ & $1.50\times 10^{-2}$ & $\mathbf{4.80\times 10^{-3}}$ & $3.71\times 10^{4}$ \\
\midrule
\multirow{4}{*}{\shortstack{D\\{\footnotesize $(1,1,1,1,1,0,\ldots,0)$}\\{\footnotesize $k^\star=7,\ \widehat p_{k^\star}=0.462$}}} & SM-FECMC & $1.07\times 10^{-2}$ & $\mathbf{1.79\times 10^{-2}}$ & $\mathbf{7.64\times 10^{-3}}$ & $5.85\times 10^{4}$ \\
 & RJ-FECMC & $1.23\times 10^{-2}$ & $2.05\times 10^{-2}$ & $8.36\times 10^{-3}$ & $5.64\times 10^{4}$ \\
 & RJ-HMC ($T=3,\ \epsilon=0.04$) & $\mathbf{9.34\times 10^{-3}}$ & $3.10\times 10^{-2}$ & $9.53\times 10^{-3}$ & $5.91\times 10^{4}$ \\
 & RJ-HMC ($T=3,\ \epsilon=0.05$) & $1.07\times 10^{-2}$ & $2.96\times 10^{-2}$ & $9.80\times 10^{-3}$ & $5.91\times 10^{4}$ \\
\midrule
\multirow{5}{*}{\shortstack{E\\{\footnotesize $(1,0.95,0.9,\ldots,0.05)$}\\{\footnotesize $k^\star=20,\ \widehat p_{k^\star}=0.606$}}} & SM-FECMC & $\mathbf{2.13\times 10^{-2}}$ & $\mathbf{5.15\times 10^{-2}}$ & $\mathbf{1.42\times 10^{-2}}$ & $6.45\times 10^{4}$ \\
 & RJ-FECMC & $2.94\times 10^{-2}$ & $6.15\times 10^{-2}$ & $1.70\times 10^{-2}$ & $6.91\times 10^{4}$ \\
  & RJ-HMC ($T=3,\ \epsilon=0.03$) & $1.65\times 10^{-1}$ & $1.05\times 10^{0}$ & $3.10\times 10^{-1}$ & $7.28\times 10^{4}$ \\
 % & RJ-HMC ($T=3,\ \epsilon=0.04$) & $3.71\times 10^{-1}$ & $1.85\times 10^{0}$ & $5.82\times 10^{-1}$ & $7.52\times 10^{4}$ \\
 % & RJ-HMC ($T=3,\ \epsilon=0.05$) & $4.09\times 10^{-1}$ & $1.99\times 10^{0}$ & $6.35\times 10^{-1}$ & $7.32\times 10^{4}$ \\
 & RJ-HMC ($T=10,\ \epsilon=0.007$) & $2.41\times 10^{-2}$ & $1.59\times 10^{-1}$ & $4.50\times 10^{-2}$ & $6.71\times 10^{4}$ \\
\midrule
\multirow{4}{*}{\shortstack{O\\{\footnotesize $(0,\ldots,0)$}\\{\footnotesize $k^\star=0,\ \widehat p_{k^\star}=0.816$}}} & SM-FECMC & $8.30\times 10^{-3}$ & $5.40\times 10^{-4}$ & $8.39\times 10^{-4}$ & $3.76\times 10^{4}$ \\
 & RJ-FECMC & $6.53\times 10^{-3}$ & $\mathbf{4.61\times 10^{-4}}$ & $\mathbf{7.13\times 10^{-4}}$ & $3.58\times 10^{4}$ \\
 & RJ-HMC ($T=2,\ \epsilon=0.1$) & $\mathbf{5.82\times 10^{-3}}$ & $5.84\times 10^{-4}$ & $8.82\times 10^{-4}$ & $3.92\times 10^{4}$ \\
 & RJ-HMC ($T=3,\ \epsilon=0.05$) & $9.75\times 10^{-3}$ & $8.28\times 10^{-4}$ & $1.26\times 10^{-3}$ & $3.77\times 10^{4}$ \\
\bottomrule
\end{tabular}
\end{table}

\begin{figure}[!ht]
    \centering
    \includegraphics[width=\linewidth]{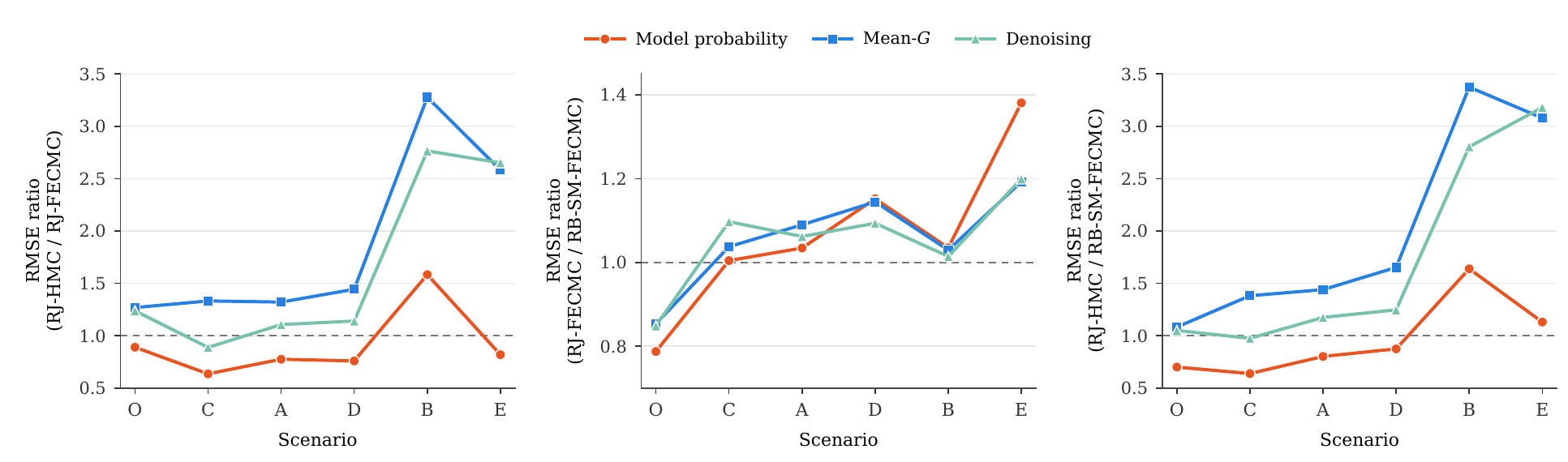}
    \caption{Relative RMSEs across the six PPCA scenarios. Left: RJ-HMC relative to RJ-FECMC; middle: RJ-FECMC relative to SM-FECMC; right: RJ-HMC relative to SM-FECMC. For each scenario and quantity, the RJ-HMC ratio uses the smallest observed RMSE among all configurations.}
    \label{fig:RelativeRMSEsMain}
\end{figure}

Table~\ref{tab:BPCA} shows that either SM-FECMC or RJ-FECMC performs best for the posterior mean of $G$ across all scenarios, while RJ-HMC with $(T,\epsilon)=(3,0.05)$ performs best in scenario C for the posterior mean of $x$. Interestingly, across scenarios A--E, the RMSE ratio of RJ-HMC against either SM-FECMC or RJ-FECMC tends to increase as the posterior modal rank $k^\star$ increases (Figure~\ref{fig:RelativeRMSEsMain}). Moreover, the average acceptance probability of HMC proposals with the best-performing tuning also decreases accordingly (Appendix~\ref{appendix-subsec;AdditionalResultPCA}). This may reflect the increasing number of order constraints on the eigenvalues, near whose boundaries the characteristic length scales of the posterior, and hence the acceptance rates, differ significantly (Appendix~\ref{appendix-subsubsec;PriorBCA}), requiring small step sizes that may be suboptimal for other parts of the posterior and for dynamics on $O(20)$. In fact, the best-performing RJ-HMC in scenarios B and E uses $T=10$ integration steps with a smaller step size $\epsilon$, while in the other scenarios it uses $T=3$. %a mechanism commonly reported for HMC \citep[Section~5.4.2.2]{Neal2011}.
In contrast, the PDMP dynamics adapt to such anisotropy and are relatively robust to multiscale structure without further tuning or preconditioning, consistent with previous results \citep{Bierkens-Kamatani-Roberts2024,Lu-Luo2026}.

\begin{figure}[!htb]
\centering
\includegraphics[width=\textwidth]{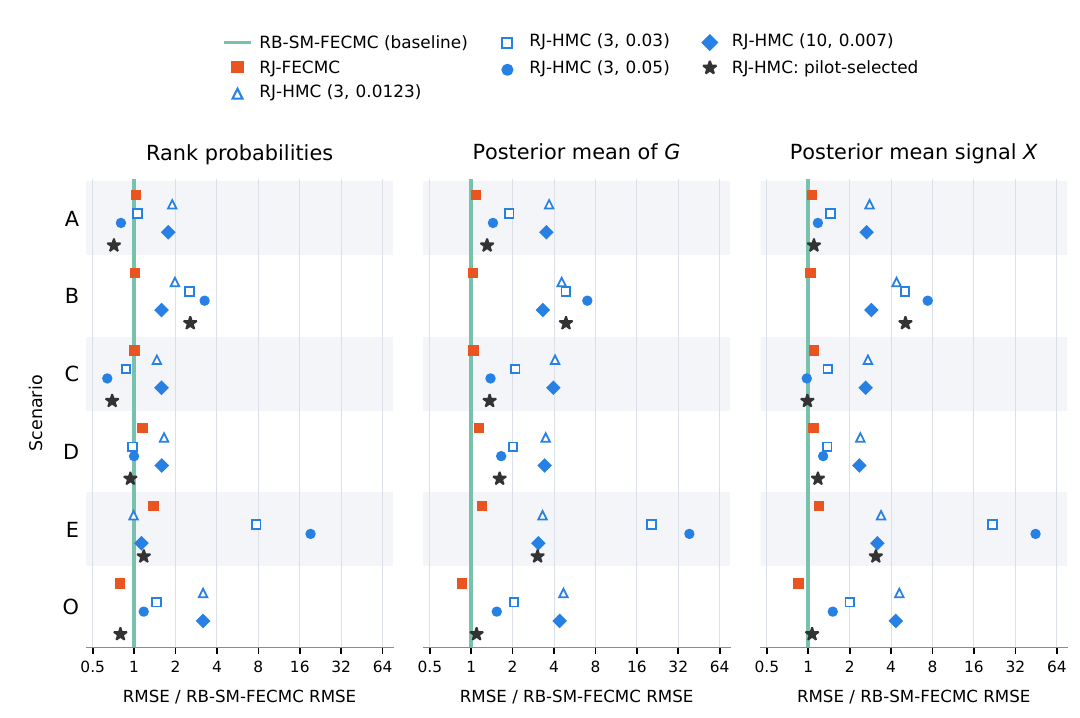}
\caption{Sensitivity of RJ-HMC to hyperparameters across six scenarios. Blue symbols use four common configurations, while stars use scenario-specific settings selected by pilot runs based on ESS-per-CPU, which took about 2 hours for each scenario on average; see Appendix~\ref{appendx-subsec;TuningHMC} for details.}
\label{fig:TuningSensitivityPCA}
\end{figure}

This robustness to scale differences provides a further reason to use PDMP dynamics for trans-dimensional sampling, in which posterior geometry may vary across strata, whereas RJ-HMC may benefit from stratum-wise tuning or preconditioning. For example, in scenario C, the average acceptance probability of RJ-HMC with $(T,\epsilon)=(3,0.05)$ ranges from 69.9\% in $\mathcal{M}_3$ to 7.4\% in $\mathcal{M}_6$ (Table~\ref{tab;AcceptanceProbability}). Additionally, RJ-HMC is sensitive to step-size tuning; Figure~\ref{fig:TuningSensitivityPCA} shows that no single setting is uniformly competitive with SM-FECMC and RJ-FECMC, whereas the baseline SM-FECMC dynamics required no tuning.

Comparison with RJ-FECMC highlights a computational advantage of the rejection-free nature of the sticky manifold framework.  In addition to the cost of evaluating acceptance probabilities, rejected reversible-jump proposals may incur computational overhead on interrupted event-time simulation. Additional results suggest that SM-FECMC achieves longer trajectories under the same CPU budget by avoiding these costs (Appendix~\ref{appendix-subsec;AdditionalResultPCA}). %while the sticky-manifold framework keeps the number of wasted interventions to the minimum.
%Since RJ-FECMC and SM-FECMC share the continuous-time dynamics, the time consumed in between-stratum transitions is reflected in the time horizon as stochastic processes reached within the 3 CPU seconds budget. SM-FECMC proceeds the process time more in scenarios D and E, while less in scenarios B, C, O, as opposed to RJ-FECMC. This may be derived from, and is reflected in the better rank-probability estimates in scenarios D and E. See Appendix~\ref{appendix-subsec;AdditionalResultPCA} for further details.

\subsection{Model-averaged survival extrapolation}\label{subsec;survival-extrapolation}

In the following we consider an example based on the application of survival models to time-to-event data arising from clinical trials. Concretely, we consider time-to-event data $y \in \mathbb{R}_{+}^N$ administratively censored at $y_+$ and where the estimand of interest is mean survival over a lifetime time-horizon $(0,y_\infty)$, requiring survival curves to be extrapolated over the interval $(y_+,y_\infty)$. This task commonly arises in the cost-effectiveness evaluation of novel medical interventions in publicly funded healthcare systems, and precludes the use of standard non-parametric methods (e.g. Kaplan--Meier estimates) as they lack a mechanism to extrapolate beyond final observation times. See \cite{Baio2020,Hardcastle2025} for more general introductions to this problem setting. 

A common solution is to assume that $y$ are generated by a parametric survival distribution e.g. Weibull or log-Normal \citep{Latimer2011}, however extrapolations are then sensitive to the choice of survival model. A solution proposed in the literature is the use of Bayesian model averaging to both account for this uncertainty and to incorporate prior information into extrapolations through priors on the choice of underlying model \citep{negrin2017bayesian,gallacher2021extrapolating,Hardcastle2024}.

\begin{figure}[!ht]
    \centering
    \includegraphics[width=\linewidth]{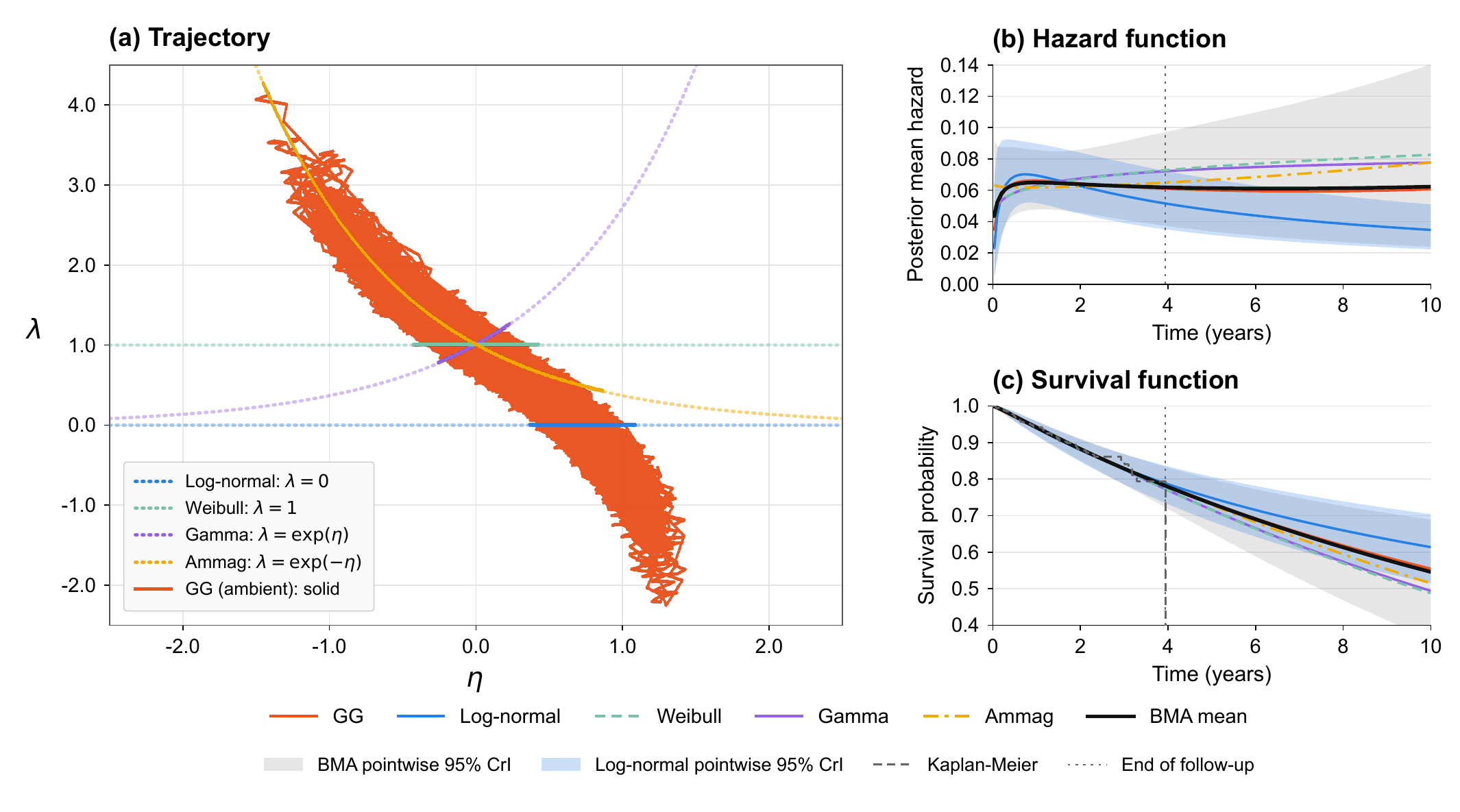}
    \caption{(a) Full SM-FECMC trajectory for the joint AFT model with uniform prior model probabilities on the $(\eta,\lambda)$ plane with $\eta=\log\sigma$. (b,c) Posterior mean hazard and survival functions for the control arm fitted separately under the informative model prior.}
    \label{fig:SurvivalFullTrajectory}
\end{figure}

We propose a novel computational approach to this problem using the sticky manifold PDMP sampler and the generalised gamma family of distributions \citep{Stacy1962}, a three-parameter distribution that contains the log-normal, Weibull, gamma, and ammag distributions as special cases. Using the parametrisation $(\mu,\sigma,\lambda)$ of \cite{Prentice1974} these cases correspond to the submanifolds $\mathcal{M}_{\text{LN}}=\{\lambda=0\}$, $\mathcal{M}_{\text{W}}:=\{\lambda=1\}$, $\mathcal{M}_{\Gamma}=\{\lambda=\sigma\}$, and $\mathcal{M}_{\text{A}}:=\{\lambda=1/\sigma\}$ (Figure~\ref{fig:SurvivalFullTrajectory}). We place prior mass on these submanifolds, and then sample from the resulting posterior using the sticky manifold PDMP (Remark~\ref{remark-overlapping-manifolds}), avoiding the design of multiple reversible jump steps, and facilitating model averaging over the parametric sub-families.

We consider the following prior over $\theta=(\alpha,\beta,\eta,\lambda)$ with $\eta:=\log\sigma$:
\[\pi(d\theta)= \omega_{GG}\pi_{\text{GG}}(\theta)\,d\theta+\sum_{i\in\{\text{LN},\text{W},\Gamma,\text{A}\}}\omega_{i}\pi_i(\theta)\mathcal{H}^3_{\mathcal{M}_i}(d\theta),\quad \sum_{i\in\{\text{GG},\text{LN},\text{W},\Gamma,\text{A}\}}\omega_i = 1,\]
where $\pi_{\mathrm{GG}}(\theta)\,d\theta$ assigns independent Gaussian priors to the elements of $\theta$ and $\pi_i\mathcal{H}^3_{\mathcal{M}_i}$ is its restriction onto the constrained manifold. In particular, the prior weights $\omega_i$ can be chosen to reflect prior information about the choice of submodel based on previous analyses.

We apply the model to data for 810 leukemia patients from the CLL-8 trial that informed NICE technology appraisal TA174 \citep{Williams+2017} evaluating the efficacy of the novel immunotherapy Rituximab. Final observation times occur at $y_+ = 4$, and we evaluate mean survival up to $y_\infty = 30$. We consider two versions of the model, the first with control and treatment arms related through an accelerated failure time assumption, and the second where independent models are fitted for control and treatment arms. This latter approach is common in health technology assessment as it avoids strong assumptions on the relationship between treatment arms. We implement the model with uniform prior weights, and further consider a second hypothetical scenario where previous analyses indicate that the log-Normal model is the most appropriate model for the control arm, with $\omega_{\mathrm{LN}} = 1/2$. This acts as a prior for control and treatment groups under the accelerated failure time assumption, but is only applied to the model for the control arm when the groups are modelled independently.

Running the SM-FECMC algorithm with $T = 10^5$ results in the trajectories shown in panel (a) of Figure~\ref{fig:SurvivalFullTrajectory}, with the sampler placing non-negligible posterior probability on all four parametric sub-families. Example hazard and survival functions for the control group when treatment arms are modelled separately with an informative prior are shown in panel (b) and (c) of Figure \ref{fig:SurvivalFullTrajectory}, with 95\% credible intervals under the model averaged posterior and the log-normal posterior shown in grey and blue respectively. Note the larger uncertainty intervals in the extrapolation period under the averaged model, accounting for the epistemic uncertainty that selecting a single parametric model does not address. Full posterior model probabilities are included in Appendix~\ref{appendix-ExperimentalDetails}. For uniform $\omega_i$ the ammag model is the most favoured model with 55\%, 54\% and 46\% posterior probabilities. All other models have $<20\%$ posterior probability with the generalised-gamma receiving the least posterior mass in all cases. With the informative prior, the log-normal model goes from $11\%$ posterior probability to $31\%$ posterior probability under the accelerated failure time model, and $12\%$ to $34\%$ posterior probability in the independent control arm model. The impact on mean survival is shown in Table \ref{tab:rmst-prior-sensitivity}. Note that model averaging increases the uncertainty in estimates of mean survival when compared to selecting a single parametric model. Further, the impact of the informative prior has a small impact under the accelerated failure time assumption, but reduces the expected difference in mean survival when the arms are modelled separately by about a year. Given the current NICE threshold of £35,000 per quality adjusted life year, this difference would potentially be large enough to alter the final funding decision for the treatment.

\begin{table}[!ht]
\centering
\caption{Estimates of 30-year restricted mean survival time (years). Entries are posterior means (95\% credible intervals).}
\label{tab:rmst-prior-sensitivity}
\setlength{\tabcolsep}{7pt}\vspace{1mm}
\resizebox{\linewidth}{!}{%
\begin{tabular}{lrrrrr}
\toprule
 & & & & \multicolumn{2}{c}{Model averaged} \\
\cmidrule(lr){5-6}
Model & Gen-Gamma & Log-normal & Ammag & Uniform prior & Informative prior \\
\midrule
\addlinespace[2pt]
\multicolumn{6}{l}{\textbf{Treatment arms modelled together}} \\
Control    & $15.63\;(8.82,\,20.70)$ & $18.16\;(15.74,\,20.48)$ & $13.08\;(8.49,\,18.02)$ & $13.85\;(8.81,\,19.48)$ & $14.82\;(9.02,\,19.96)$ \\
Treatment  & $17.89\;(11.56,\,22.05)$ & $19.85\;(17.47,\,22.13)$ & $16.04\;(11.33,\,20.41)$ & $16.59\;(11.65,\,21.22)$ & $17.33\;(11.93,\,21.63)$ \\
Difference & $2.26\;(-0.56,\,5.73)$ & $1.70\;(-0.55,\,3.89)$ & $2.97\;(-0.64,\,6.72)$ & $2.74\;(-0.61,\,6.44)$ & $2.51\;(-0.58,\,6.24)$ \\
\midrule
\addlinespace[2pt]
\multicolumn{6}{l}{\textbf{Treatment arms modelled separately}} \\
Control    & $14.81\;(7.86,\,20.53)$ & $17.19\;(14.17,\,20.08)$ & $12.74\;(8.35,\,17.95)$ & $13.35\;(8.45,\,19.14)$ & $14.33\;(8.70,\,19.55)$ \\
Treatment  & $18.25\;(11.31,\,22.98)$ & $20.10\;(17.17,\,22.73)$ & $16.55\;(11.37,\,21.34)$ & $16.98\;(11.53,\,21.90)$ & $16.98\;(11.53,\,21.90)$ \\
Difference & $3.45\;(-5.71,\,12.34)$ & $2.92\;(-1.20,\,6.93)$ & $3.80\;(-3.59,\,10.67)$ & $3.63\;(-4.42,\,11.01)$ & $2.65\;(-5.28,\,10.58)$ \\
\bottomrule
\end{tabular}}
\end{table}

\section{Discussion}

We have developed the trans-dimensional moves based on sticking and unsticking that are rejection-free, require no additional likelihood evaluation, and apply to both linear and non-linear submanifolds under the same principle. In numerical experiments, we demonstrated regimes in which the sticky manifold framework leads to better estimates with Rao--Blackwellisation than the reversible-jump counterparts, a high-dimensional example that highlights the potential of PDMP dynamics in varying posterior geometry, and an application where multiple non-linear submodel constraints can be handled without bespoke jump proposals. A broader goal is to give practitioners greater freedom to choose statistically-motivated parametrisations, rather than computationally-motivated ones. Realising this goal more generally will require practical strategies for choosing coordinates and dynamics and for simulating boundary hitting and event times.

The blow-up procedure (Section~\ref{subsec;illustration-blowup-cone}) is a geometric solution to high codimensions, as illustrated in the Bayesian PCA example (Section~\ref{subsec;BayesianPCA}). However, it may limit the range of priors that can be used with our sticky manifold construction, for regular priors on the associated radial blow-up coordinate translate into densities on the original space with a diverging Jacobian factor near the sticking boundaries (see Remarks~\ref{remark;divergingJacobian} and~\ref{remark;Pourzanjani}). If, instead, a prior does not diverge sufficiently rapidly in the original space, a naïve application of PDMP dynamics on the blow-up coordinates will have zero probability of hitting the front face, producing reducible dynamics. Additional procedures are therefore required to accommodate such priors within the blow-up procedure.

In our numerical experiments, RJ-FECMC was used primarily as a bridge to disentangle the effects of within-model PDMP dynamics from those of the model-transition mechanism, rather than to identify an optimal sampler. In practice, reversible-jump and sticky-manifold transitions could complement each other rather than serve as competing components in trans-dimensional PDMP sampling. Sticky-manifold transitions require no additional likelihood evaluations, while reversible-jump moves may facilitate between-model exploration, particularly when boundary hits are infrequent. Further improvements may come from adaptive tuning and preconditioning of the within-model dynamics. Such methods have been developed for both PDMP-based samplers \citep{Bertazzi2022,Chevallier-Raab2025,Chevallier+2026} and HMC \citep{Hoffman-Gelman2014,Bou-Rabee+2026}.

There are several immediate extensions to the example in Section~\ref{subsec;survival-extrapolation}, for example, using this approach for distributional selection in more complex survival models \citep[e.g.]{Hardcastle2024} or combining the model and sampling architecture with standard sticky PDMP dynamics to perform simultaneous distributional and variable selection.

\section*{Acknowledgements}

The authors thank Alexandros Beskos, Samuel Livingstone, and Gianluca Baio for insightful discussions. They thank the Isaac Newton Institute for Mathematical Sciences, Cambridge, for support and hospitality during the programme \textit{Monte Carlo sampling: beyond the diffusive regime}, where initial discussions on this work took place. They are also grateful to the Institute of Statistical Mathematics for its hospitality during LH’s research visits from 17 to 31 May 2025 and from 24 October to 7 November 2026, which were instrumental in the development of this work. HS is supported by JST BOOST, Japanese Grant Number JPMJBS2412. LH was supported by EPSRC grant EP/W523835/1 while a PhD student at University College London, and was further supported by Medical Research Council Career Development Award UKRI332. KK was supported by JST CREST, Grant Number JPMJCR2115.

\bibliography{TheBib}

\appendix

\section{Manifold PDMP}\label{appendix;sec-ManifoldPDMP}

\subsection{The Hausdorff measure and area formula}\label{appendix;subsec-Hausdorff}

For a Riemannian manifold $\mathcal M$, the volume measure induced from the Riemannian volume form \citep[see][Chapter~X]{Lang1995-Manifolds} agrees (on all Borel measurable sets) with the Hausdorff measure induced from the Riemannian distance, with the normalisation convention of \citet[2.10.2]{Federer1996}. % This MathOverflow post is interesting: https://mathoverflow.net/questions/87930/hausdorff-measure-and-the-volume-form
This fact is proven in \citealp[3.2.46]{Federer1996}. Throughout this paper, we refer to this measure as the Hausdorff measure, and we provide a definition following \citealp[Section~2.10]{Federer1996} and \citealp[Section~3.10~(iii)]{Bogachev2007}.

\begin{definition}[Hausdorff measure]
    Let $(\mathcal M,g)$ be a $q$-dimensional Riemannian manifold, $\dist_g$ the Riemannian distance, and $\alpha_q$ be the volume of the $q$-dimensional unit ball in $\R^q$. Define a function $\zeta:2^\mathcal M\to[0,\infty]$ by
    \[\zeta^q(S)=\frac{\alpha_q}{2^q}(\diam_g S)^q,\qquad \emptyset\ne S\subset \mathcal M,\]
    where $\diam_gS:=\sup\dist_g(S\times S)$.
    We define $\zeta^q(\emptyset)=0$. This $\zeta^q$ defines an outer measure in the sense of Carathéodory \citep[see][Definition~1.11.1]{Bogachev2007}, i.e., a monotone and $\sigma$-subadditive set function $\cH^q$ by
    \[\cH^q(A):=\lim_{\delta\searrow0}\cH^q_\delta(A)\qquad A\subset\mathcal M,\]
    \[\cH_\delta^q(A)=\inf\Brace{\sum_{i=1}^\infty\zeta(S_i)\,\middle|\,A\subset\bigcup_{i=1}^\infty S_i,\diam_g(S_i)\le\delta}.\]
    The resulting complete measure $\cH^q$, constructed from Carathéodory's extension theorem \citep[Theorem~1.11.4]{Bogachev2007}, is, in fact, a restriction of $\cH^q$ onto a complete $\sigma$-algebra and will be called \textit{Hausdorff measure} on the Riemannian manifold $(\mathcal M,g)$.
    We sometimes write $\cH^q$ as $\cH^q_g$ to clarify the Riemannian distance on which it depends.
\end{definition}

The Hausdorff measure $\cH^q$ defined above is the same as the Lebesgue measure when $\mathcal M=\R^q$, including the domain of the definition.

For subsequent reference, we give the area formula, which is an extension of the change of measures formula \citep{Federer1996,Diaconis2013}. For a Riemannian manifold $(\mathcal{M},g)$,
\begin{equation*}
\int_{A}g(f(x))J_mf(x)\lambda^m(\mathrm{d}x) = \int_{\mathcal{M}}g(y)|\{x\in A : f(x)= y\}|\mathcal{H}^m_g(\mathrm{d}y),
\end{equation*}
where $f:\mathbb R^m\to\mathcal{M}$ is Lipschitz continuous, $A\subset\mathbb R^m$, and $\abs{\cdot}$ denotes the cardinality of a set.

\subsection{Proof of Proposition~\ref{prop;pullback}}

\begin{proof}
    Defining the mapping $T_\varphi:T(\mathcal{M})\to\mathcal{U}\times\mathbb{R}^p$ as $T_\varphi(\xi,\eta):=(\varphi(\xi),D\varphi(\xi)\eta)$, the above pullback construction can be expressed as
    \[T_\varphi^{-1}(X_t,V_t)=(\varphi^{-1}(X_t),D\varphi^{-1}(X_t)V_t).\]
    Since $T_\varphi$ is again a diffeomorphism, actually a bundle isomorphism,
    we see that $T_\varphi^{-1}(X_t,V_t)$ is again a PDMP, whose transition semigroup $(P_t^\varphi)$ satisfies the intertwining relationship
    \[(P^\varphi_tF)=P_t(F\circ T^{-1}_\varphi)\circ T_\varphi,\]
    where $(P_t)$ is the semigroup of the original PDMP $(X_t,V_t)$.
    Using this conjugacy and the area formula \citep[Theorem~3.2.5, 3.2.46]{Federer1996}, we conclude that
    \[\mu^\varphi(d\xi d\eta):=\pi(\varphi(\xi))J_\varphi^g(\xi)\mathcal{H}^p_g(d\xi)\nu_\xi^\varphi(d\eta),\qquad(\xi,\eta)\in T(\mathcal{M})\]
    is invariant with respect to $P^\varphi_t$, where $\nu^\varphi_\xi:=(D\varphi(\xi)^{-1})_*\nu_{\varphi(\xi)}$ is the pullback probability measure. The calculation goes as follows
    \begin{align*}
        \int_{T(\mathcal{M})}P^\varphi_tF(\xi,\eta)\mu^\varphi(d\xi d\eta)&=\int_{T(\mathcal{M})}P_t(F\circ T^{-1}_\varphi)(T_\varphi(\xi,\eta))\mu^\varphi(d\xi d\eta)\\
        &=\int_{\mathcal{U}\times\mathbb{R}^p}P_t(F\circ T_\varphi^{-1})(x,v)\pi(x)\,dx\nu_x(dv)\\
        &=\int_{\mathcal{U}\times\mathbb{R}^p}F(T^{-1}_\varphi(x,v))\pi(x)\,dx\nu_x(dv)=\int_{T(\mathcal{M})}F(\xi,\eta)\mu^\varphi(d\xi d\eta),
    \end{align*}
    where we used the invariance of $(X_t,V_t)$ in the second-to-last equality.
\end{proof}

\subsection{Dynamics of manifold PDMPs}\label{appendix-subsec;pullback-generator}

\begin{theorem}[Pullback of PDMP generators]\label{thm;pullback-generator}
    Let $(\mathcal M,g)$ be a $p$-dimensional Riemannian manifold, $\mathcal{U}\subset\R^p$ an open set, and $\varphi:\mathcal{M}\to\mathcal{U}$ be a diffeomorphism. Assume that the PDMP $(X_t,V_t)$ on $\mathcal{U}\times\R^p$ has a strong generator of the form
    \[\mathcal{L}f(z)=\langle v,\nabla_xf(z)\rangle+\lambda(z)(Qf-f)(z),\qquad z=(x,v),\]
    for a function $f:\mathcal{U}\times\mathbb{R}^p\to\mathbb{R}$ in the domain.
    Then, the generator of the induced PDMP process $Z^\varphi_t:=(\varphi^{-1}(X_t),D\varphi^{-1}(X_t)V_t)$ on $T(\mathcal{M})$ is given by, for a test function $F:T(\mathcal{M})\to\mathbb{R}$,
    \begin{equation}\label{eq;pulledback-generator}
        \mathcal{L}^\varphi F(\zeta)=\mathcal{L}^\varphi_DF(\zeta)+\lambda^\varphi(\zeta)(Q^\varphi F-F)(\zeta),\qquad\zeta=(\xi,\eta),
    \end{equation}
    \[\mathcal{L}_D^\varphi=v^i\frac{d}{dx^i}-\Gamma^i_{jk}v^jv^k\frac{d}{dv^i},\]
    where $\mathcal{L}_D^\varphi$ is the metric spray of $g_\varphi:=\varphi^*g_E$, the pullback metric of the Euclidean metric $g_E$.
    Additionally, if $\mathcal{L}$ admits $C^1_c(\mathcal{U}\times\mathbb{R}^p)$ as a core, then $C^1_c(T(\mathcal{M}))$ is also a core of $\mathcal{L}^\varphi$.
\end{theorem}

\begin{proof}
    We discuss locally by fixing a chart for $\mathcal{M}$.
    Denote the generator of $Z=(X,V)$ on $\mathcal{U}\times\R^d$ by
    \[\mathcal{L}f(z)=\langle v,\nabla_xf(z)\rangle+\lambda(z)(Qf-f)(z),\qquad z=(x,v),\]
    for a function $f:\mathcal{U}\times\mathbb{R}^d\to\mathbb{R}$ in the domain.
    Then, defining the bundle isomorphism as $T_\varphi(\xi,\eta):=(\varphi(\xi),D\varphi(\xi)\eta)$, the process $T_\varphi^{-1}(X,V)$ on $T(\mathcal{M})$ is again a PDMP.
    Its generator can be written as
    \begin{equation}
        \mathcal{L}^\varphi F(\zeta)=\langle\eta,\nabla_\xi F(\zeta)\rangle-\langle\Gamma_\varphi(\zeta),\nabla_\eta F(\zeta)\rangle+\lambda^\varphi(\zeta)(Q^\varphi F-F)(\zeta),\qquad\zeta=(\xi,\eta),\;F\in C^1_c(T(\mathcal{M})),
    \end{equation}
    where $\lambda^\varphi:=\lambda\circ T_\varphi$, $Q^\varphi F:=Q(F\circ T_\varphi^{-1})\circ T_\varphi$, and
    \[\Gamma_\varphi(\zeta):=-D^2\varphi^{-1}(x)[v,v],\qquad (x,v)=T_\varphi(\xi,\eta)\]
    is the Christoffel acceleration term that pushes the PDMP flow along the geodesics of the metric induced through $\varphi$, i.e., $g_\varphi:=\varphi^*g_{\mathbb{R}^d}$.
    This construction works for a general manifold $\mathcal{M}$ with the deterministic part of $\mathcal{L}^\varphi$ replaced with the metric spray of $g_\varphi$.
    It is straightforward to check that the diagram commutes
    \[\mathcal{L}f\circ T_\varphi=\mathcal{L}^\varphi(f\circ T_\varphi)\]
    for any $f:\mathcal{U}\times\mathbb{R}^d\to\mathbb{R}$ in the domain of $\mathcal{L}$.
    This indicates that the precomposition map $T_\varphi^*(f)=f\circ T_\varphi$ defines an isometry between $C_c^1(\mathcal{U}\times\mathbb{R}^d)$ and $C_c^1(T(\mathcal{M}))$ with respect to the graph norm. Therefore, $C^1_c(T(\mathcal{M}))$ is a core of $\mathcal{L}^\varphi$, given that $C^1_c(\mathcal{U}\times\mathbb{R}^d)$ is a core of $\mathcal{L}$, the set of all smooth functions with compact supports \citep[Corollary~24]{Durmus+2021}.
\end{proof}

\section{Sticky manifold PDMP}

\subsection{Blow-up construction and tangent-normal splitting}\label{appendix-subsec-blowup}

For an embedded submanifold $\cM\mono\R^d$ with codimension at least one, merely taking the boundary results in $\partial(\R^d\setminus\cM)=\ov{\cM}$. To retain the information as to which side the process approached the submanifold $\cM$, we consider \textit{blowing-up} $\cM$. For a general treatment of blow-ups, especially as a technique of singularity resolution in geometry, see \citet[Section~1.5]{Melrose2008} or \citet[Section~2.3]{Grieser2001}. Here, we define the blow-up of $\cM$ in $\R^d$ as a set
\[[\R^d;\cM]:=(\R^d\setminus\cM)\sqcup S(N_{\cM}(\R^d)),\]
where $S(N_{\cM}(\R^d))=(N_{\cM}(\R^d)\setminus0_{\cM})/\R_{>0}$ is the unit sphere subbundle of the normal bundle $N_\cM(\R^d)$. We equip this set with a surjection $\beta:[\R^d;\cM]\to\R^d$ such that
\[\beta(y)=y\quad(y\in\mathbb{R}^d\setminus\cM),\qquad\beta(x,s)=x\quad((x,s)\in S(N_\cM(\R^d))),\]
called a \textit{blow-down map}. We define a manifold structure on $[\R^d;\cM]$ so that $\beta$ is smooth and $[\R^d;\cM]$ has boundary $\beta^{-1}(\cM)$, that is also called \textit{front face}. To do so, we use a restriction of the exponential map $\exp|_{N_\cM(\R^d)}:N_\cM(\R^d)\to\R^d$, that is diffeomorphic on an open neighbourhood $U$ of the zero section $0_\cM=\{(x,0)\}_{x\in\cM}\subset N_{\cM}(\R^d)$ \citep[Theorem~IV.5.3]{Lang1995-Manifolds}. A key observation is that this neighbourhood, after excluding the zero section $U\setminus0_\cM$, admits a polar coordinate expression in the fibre direction $0\ne v\in N_\cM(\R^d)$, i.e.,
\begin{equation}\label{eq;radial-blowup}
    (x,v)=(x,rs),\qquad x\in\cM,\;r>0,\;s\in S_x(N_{\cM}(\R^d)),
\end{equation}
where $S_x(N_{\cM}(\R^d))$ is the unit sphere in the fibre of $x$ in $N_{\cM}(\R^d)$. The angular part $s$ represents the direction of approach to $x\in0_{\cM}$, which corresponds to $\cM$ via $\beta$, and the radial part $r>0$ indicates the distance to $0_{\cM}\simeq\cM$.
Together with $x$, the triplet $(x,r,s)$ induces local coordinates on $U\setminus0_\mathcal{M}$, an open set in $N_\cM(\R^d)$. By extending this local coordinate to $r=0$, we obtain local coordinates of the blown up manifold $[\R^d;\cM]$ and the blow-down map $\beta(x,r,s)=\exp(x,rs)$, which is smooth since it is a smooth extension of the restricted exponential map $\exp|_{N_\cM(\R^d)}$.

As a result, each point in the front face $(x,s)\in \beta^{-1}(\cM)$ is distinguished by the direction of approach $s$ to $x\in\cM$.
When the submanifold $\cM$ is coorientable and of codimension $1$, the normal bundle is one dimensional and trivial: $N_{\cM}(\R^d)\simeq{\cM}\times\R$. In this case, fixing a coorientation, $\beta^{-1}(\cM)\simeq\cM\times\{\pm1\}$ and we see that the boundaries of $[\R^d;\cM]$ consist of two copies of $\cM$.

The state space $E$ of the sticky manifold PDMP is the pullback of the tangent bundle $T(\R^d)$ along the blow-down map $\beta$, denoted by
\[E:=\beta^*(T(\R^d))=\{(x,v)\mid x\in[\R^d;\cM],v\in T_{\beta(x)}(\R^d)\}.\]
Since the blown-up manifold $[\R^d;\cM]$ can be decomposed into the interior $\R^d\setminus\cM$ and the boundary, which we denote by $\cM_+\sqcup\cM_-$ assuming $\cM$ is of codimension one and coorientable, we have a decomposition, after an appropriate identification,
\[E=T(\R^d\setminus\cM)\sqcup T_{\beta(\cM_+)}(\R^d)\sqcup T_{\beta(\cM_-)}(\R^d),\]
where $T_{\cM}(\R^d)=\{(x,v)\in T(\R^d)\mid x\in\cM\}$ is a restriction of $T(\R^d)$. This vector bundle admits a canonical splitting induced by the Euclidean structure \citep[see also][Section~IV.3]{Lang1995-Manifolds},
\[T_{\cM}(\R^d)\simeq T(\cM)\oplus N_{\cM}(\R^d),\]
where $\oplus$ is the direct sum of vector bundles \citep[Section~III.4]{Lang1995-Manifolds}. In other words, on a boundary point $x\in\cM_+$, the velocity $v\in \beta^*(T(\R^d))_x= T_{\beta(x)}(\R^d)$ can be decomposed into the tangent component $v_{\cM_+}$ and the normal component $v_{\cM_+}^\perp$.

Next we define the frozen boundary $\mathfrak{F}$ in $E$.
The normal component $v_{\cM_+}^\perp$ is \textit{pointing outwards from $\cM$} when it is positively aligned with $\partial_r$, derived from the local coordinate \eqref{eq;radial-blowup}, i.e. $\brac{v_{\cM_+}^\perp,\partial_r}>0$. The opposite direction is said to be \textit{pointing inwards}. The states in the subset
\[\mathfrak{F}^\pm:=\{(x,v)\in \beta^*(T(\R^d))\mid x\in\cM_{\pm}, \brac{v^\perp_{\cM_\pm},\partial_r}<0\}\]
are considered to be `stuck' to $\cM$; $\mathfrak{F}:=\mathfrak{F}^+\cup\mathfrak{F}^-$ is called the `frozen boundary' \cite[see also][Section~2.1]{Bierkens2023a}.
The pair $(x,v_{\cM_+})$ is naturally identified with a point in $T(\cM)$, describing dynamics confined in $\cM$. The remaining normal component $v_{\cM_+}^\perp$ is considered to be an auxiliary variable that retains the old velocity. The base variable $x\in\cM_{\pm}$ retains the information of which side the process entered the submanifold $\cM$. At a thawing event, $x\in\cM_+$ is mapped to the corresponding point in the opposite boundary component $\cM_-$, possibly with an appropriate modification of velocity, so that the normal component $v_{\cM_+}^\perp$ now points outwards. The process will then leave $\cM$, ensuring the whole procedure will yield a well-defined Markov process. %Overall, the state space $E$ can be decomposed into three parts; $T(\R^d\setminus\cM)$, the frozen boundary $\mathfrak{F}$, and the states immediately before leaving the submanifold, denoted by $G:=\beta^*(T(\R^d))_{\cM^\pm}\setminus\mathfrak{F}$ in Section~\ref{subsec-blowup}, hence yielding Equation~\eqref{eq;state-space-E}. %The rest of the boundary states $\beta^*(T(\R)^d)_{\cM_\pm}\setminus F$ possess a velocity pointing outwards from $\cM$ and are just before leaving the submanifold $\cM$.

The above construction generalises naturally to the cases that the codimension of $\cM$ is larger than one. In such cases, a subtle difference arises in that the gauge dimension of $s\in S(N_\cM(\R^d))$ becomes non-trivial. Motion along these fibres is invisible under the blow-down map, although it may influence the direction of subsequent departures from the front face.
As in our probabilistic PCA example (Section~\ref{subsec;BayesianPCA}), we specify this gauge dynamics as $\dot{\theta}\sim N(0,1)$ in the rank-one stratum.
%We also remark that when the codimension is zero our construction reduces the problem of sampling in a restricted domain \citep{Bierkens2018,Bierkens2023b,Chevallier+2024}.

\subsection{Lifting manifold dynamics to the frozen boundary}\label{appendix-subsec-lifting-manifold-dynamics}

To align it with the blown-up state space $E$, we need to horizontally lift the manifold PDMP generator along $\overline{\beta}:\mathfrak{F}\to T(\mathcal{M})$, which is defined by
\[\overline{\beta}(x,v)=(x,P_{x}v),\]
where $P_x:T_x([\mathbb{R}^d;\mathcal{M}])\to T_x(\mathcal{M})$ denotes the orthogonal projection. Although other ways of lifting are possible, we do not consider them in this work.

We can alternatively describe this locally. Let the deterministic part of the generator on $\mathcal{M}$ be given by
\begin{equation*}
    \mathcal{L}_D=\eta^i\frac{\d}{\d\xi^i}-\Gamma^i_{jk}(\xi)\eta^j\eta^k\frac{\d}{\d\eta^i},
\end{equation*}
for local coordinates $(\xi,\eta)\in T(\mathcal{M})$, and $\Gamma$ being the Christoffel symbols of the connection induced by, say, a chosen coordinate chart $\varphi$.
Then, the lifted generator is given by the same form
\begin{equation*}
    \mathcal{L}_D^{\text{lift}}=(v^\tpar)^i\frac{\d}{\d x^i}-\Gamma^i_{jk}(x)(v^\tpar)^j(v^\tpar)^k\frac{\d}{\d (v^\tpar)^i},
\end{equation*}
but now for a global parametrisation $(x,v^\tpar,a)\in\mathfrak{F}$ induced by the decomposition
\[v=v^\tpar+v^\perp\in T_x([\mathbb{R}^d;\mathcal{M}])=T_x(\mathcal{M})\oplus N_x(\mathcal{M}),\]
with $a=\|v^\perp\|$ when $(x,v)\in\mathfrak{F}^+$ and $a=-\|v^\perp\|$ when $(x,v)\in\mathfrak{F}^-$.
In particular, it follows that $\mathcal{L}^{\text{lift}}_D(\|v^\perp\|)=0$.
In other words, the normal component $v^\perp$ is parallel-transported along $\mathcal{M}^\pm$, with the norm $\norm{v^\perp}$ preserved.

Equivalently, $\mathcal{L}^{\text{lift}}_{\mathcal{M}}$ can be characterised by the following properties, which will be used in the subsequent proof. First, the coordinate chart $(x,v^\tpar,v^\perp)$ on $\mathfrak{F}$ has an inverse mapping $\Psi:T(\mathcal{M})\times(\mathbb{R}\setminus\{0\})\to\mathfrak{F}$ defined by
\begin{equation}\label{eq;Phi}
    \Psi(x,v,a)=\begin{cases}
    (x,v-a\partial_r)\in\mathfrak{F}^+&a>0,\\
    (x,v+a\partial_r)\in\mathfrak{F}^-&a<0.
\end{cases}
\end{equation}
which is a diffeomorphism. This can be understood as a kind of trivialisation of the `tangled' tangent bundle $\mathfrak{F}$. Then,
% the following diagram commutes
% \[\xymatrix{
%     \mathfrak{F}\ar[r]^-{\mathcal{L}^\text{lift}_{\mathcal{M}}}&\mathfrak{F}\\
%     T(\mathcal{M})\ar[r]^-{\mathcal{L}_a}\ar[u]^-{\Psi_a}&T(\mathcal{M})\ar[u]_-{\Psi_a}
% }\]
% for all $a\in\mathbb{R}$ and $\Psi_a(x,v):=\Psi(x,v,a)$, or equivalently,
\begin{equation}\label{eq;IntertwiningRelation}
    (\mathcal{L}^{\text{lift}}_{\mathcal{M}}f)\circ\Psi(x,v,a)=\mathcal{L}_{\mathcal{M}}(f\circ\Psi(\cdot,\cdot,a))(x,v).
\end{equation}

The lifted dynamics preserve the invariance property. We give a proof in the Gaussian velocity case.

\begin{proposition}[Invariance of the lifted manifold dynamics]
    \label{prop;invarianceOfLiftedDynamics}
    Assume invariant manifold dynamics before lifting.
    Under Assumption~\ref{NewAssumption} and Gaussian velocity $\nu_{E\setminus\mathfrak{F}}(x,-)=\nu_{\mathfrak{F}}(x,-)=N_d(0,I_d)$,
    \[\int_{\mathfrak{F}}\mathcal{L}_{\mathcal{M}}^{\mathrm{lift}}f\,d\Pi_{\mathfrak{F}}=0,\]
    for every $f\in\mathcal{D}(\mathcal{L}_{\mathcal{M}}^{\mathrm{lift}})$.
\end{proposition}
\begin{proof}
    The general strategy is that $\mathfrak{F}$ has a natural foliation with respect to $a\in\mathbb{R}\setminus\{0\}$ and we employ Fubini's theorem after trivialising the foliation induced by the conserved value: the normal velocity. This is done by $\Psi_a(x,v):=\Psi(x,v,a)$ defined in \eqref{eq;Phi}.
    Recall that the diffeomorphism $\Psi:T(\mathcal{M})\times(\mathbb{R}\setminus\{0\})\to\mathfrak{F}$ satisfies \eqref{eq;IntertwiningRelation}.
    The fact that $\nu=\gamma^d$ is the $d$-dimensional standard Gaussian distribution simplifies the proof since an orthogonal transformation of a standard Gaussian is again a standard Gaussian and any one-dimensional projection is again standard Gaussian independent of the rest, rendering a product disintegration
    \[\Psi^*_x\gamma^d=\gamma^{d-1}\otimes\gamma^1\]
    on each fibre $T_x([\mathbb{R}^d;\mathcal{M}])$ identical over all $x\in\mathcal{M}^\pm$.
    Therefore, using Fubini's theorem after a change of variables along $\Psi$,
    \begin{align*}
        \int_{\mathfrak{F}}\mathcal{L}_{\mathcal{M}}^{\text{lift}}f\,d\Pi_{\mathfrak{F}}&=\int_{\mathbb{R}}\int_{T(\mathcal{M})}(\mathcal{L}_{\mathcal{M}}^{\text{lift}}f)\circ\Psi(x,v^\tpar,a)\,\pi_{\mathcal{M}}(x)\mathcal{H}^{d-1}_{\mathcal{M}}(dx)\gamma^{d-1}(dv^\tpar)\gamma^1(da)\\
        &=\int_{\mathbb{R}}\paren{\int_{T(\mathcal{M})}\mathcal{L}_{\mathcal{M}}(f\circ\Psi(\cdot,\cdot,a))(x,v^\tpar)\,\pi_{\mathcal{M}}(x)\mathcal{H}^{d-1}_{\mathcal{M}}(dx)\gamma^{d-1}(dv^\tpar)}\gamma^1(da),
    \end{align*}
    where we used the intertwining relation \eqref{eq;IntertwiningRelation} in the last equation. The term inside the parentheses equals zero since $f\circ\Psi(\cdot,\cdot,a)\in\mathcal{D}(\mathcal{L}_{\mathcal{M}})$ for almost all $a\in\mathbb{R}$.
\end{proof}

The proof extends to the general case in which there exists a general disintegration $\Psi^*\Pi_\mathfrak{F}(dxdv^\tpar da)=\Pi^a_{\mathfrak{F}}(dxdv^\tpar)\nu^\perp_x(da)$ and $\Pi^a_{\mathfrak{F}}$ is invariant for almost all $a$.

\subsection{Assumptions for the theorems}

\begin{assumption}\label{NewAssumption}
    For ambient PDMP dynamics in $\mathbb{R}^d$ and manifold PDMP dynamics in an orientable closed submanifold $\mathcal{M}\subset\mathbb{R}^d$ of codimension one, we assume the following conditions:
    \begin{enumerate}[label={[A\arabic*]}, font=\upshape]
        \item\label{item:well-definedness1} Both dynamics are velocity-jump PDMPs that have deterministic flows generated by a smooth second-order vector field $\mathfrak{X}_{(x,v)}=(v,F(x,v))$ \citep[Section~IV.3]{Lang1995-Manifolds}, satisfying the standard condition (24.8) of \cite{Davis1993}.
        \item\label{item:well-definedness2} The overall number of jumps is finite: $\operatorname{E}_z[N_t]<\infty$ for all $z\in E$ and $t>0$.
        \item\label{item:well-definedness3} The unsticking rate $\Lambda$ is locally integrable along the flow.
        \item\label{item:well-definedness4} Almost surely, the velocity is not tangent to $\mathcal{M}$ whenever the position lies within $\mathcal{M}$, including the initial state.
        \item\label{item:regularity1} The ambient velocity distribution $\nu_{E\setminus\mathfrak{F}}(x,dv)$ has density, denoted by $\nu_x(v)$, with respect to a reference measure $\rho_x(dv)$ on the fibre $\mathcal V_x(\mathbb R^d)$ of a smooth fibre bundle $\mathcal V(\mathbb R^d)\subset T(\mathbb R^d)$, whose total space is an embedded submanifold without boundary. The deterministic vector field $\mathcal X$ is tangent to $\mathcal V(\mathbb R^d)$.
        Further, the joint density $m(z):=\pi_{\mathbb{R}^d}(x)\nu_x(v)$ is of class $C^1(\mathcal{V}(\mathbb{R}^d))$, and the joint measure
        \[\Omega(dx,dv):=dx\rho_x(dv)\]
        defines a smooth positive density on $\mathcal{V}(\mathbb{R}^d)$ \citep[Section~XI.4, p.304]{Lang1995-Manifolds}.
        % \item\label{item:regularity2} $\widetilde{\Pi}_{E\setminus\mathfrak{F}}\ll\Pi_{\mathfrak{F}}$, where $\widetilde{\Pi}_{E\setminus\mathfrak{F}}(d\zeta):=\pi_{\mathbb{R}^d}(\xi)\mathcal{H}^{d-1}_{\mathcal{M}}(d\xi)\nu_{E\setminus\mathfrak{F}}(\xi,d\eta)$ is the restriction to the boundary.
        \item\label{item:regularity2} For $\mathcal{H}^{d-1}_\mathcal{M}$-almost all $\xi\in\mathcal{M}$, $\pi_{\mathcal{M}}(\xi)=0$ implies $\pi_{\mathbb{R}^d}(\xi)=0$.
        \item\label{item:assumption-tails1} The velocity and the acceleration component $F$ of the vector field $\mathfrak{X}$ have $\Pi$-integrable norm:
        \[\int_E\bigg(\|v\|+\|F(z)\|\bigg)\,\Pi(dz)<\infty.\]
        \item\label{item:assumption-tails2} The overall jump intensity is $\Pi$-integrable:
        \[\int_{E}\lambda(z)\Pi(dz)<\infty.\]
    \end{enumerate}
\end{assumption}

The last two integrability assumptions, \ref{item:assumption-tails1} and \ref{item:assumption-tails2}, are necessary in a functional analytic argument of Lemma~\ref{lem;cutoff}, while they are not used in the main local arguments that focus on test functions with compact support.
The regularity assumption \ref{item:regularity1} is required to perform integration by parts in Lemma~\ref{lemma;ambientInvariance}. The Gaussian case corresponds to $\mathcal{V}_x(\mathbb{R}^d)=T_x(\mathbb{R}^d)$, while the two primary non-Gaussian examples of the fibre subbundle $\mathcal{V}(\mathbb{R}^d)$ are the unit sphere bundle $\mathcal{V}_x(\mathbb{R}^d)=\mathbb{S}^{d-1}$ and the discrete bundle $\mathcal{V}_x(\mathbb{R}^d)=\{\pm1\}^d$. \ref{item:regularity2} is required to obtain the unsticking rate function $\Lambda$.
The first four assumptions, \ref{item:well-definedness1}-\ref{item:well-definedness4}, provide the well-definedness of the sticky manifold PDMP.

\begin{proposition}[Well-definedness of sticky manifold PDMP]
\label{prop;well-definedness}
    Assume \ref{item:well-definedness1}, \ref{item:well-definedness2}, \ref{item:well-definedness3}, and \ref{item:well-definedness4} in Assumption~\ref{NewAssumption}. Then the process $(X_t,V_t)$ defined by the sticky manifold PDMP algorithm (Algorithm~\ref{alg;TheAlgorithm}) is a well-defined PDMP and has \eqref{eq;generator1} as its extended generator.
    Additionally, the domain $\mathcal{D}(\mathcal{L})$ is composed of measurable functions $f:E\to\mathbb{R}$ that satisfy the following conditions
    \begin{enumerate}
        \item Along the deterministic flow $\phi_t$, $t\mapsto f(\phi_t(z))$ is absolutely continuous on each bounded time interval for any $z\in E$.
        \item the cumulative jump increment is locally integrable, i.e., 
        \[\operatorname{E}_z\left[\sum_{t=0}^{\sigma_n}|\Delta f(X_t,V_t)|\right]<\infty,\]
        for an increasing sequence of stopping times $\{\sigma_n\}$.
        \item Boundary condition:
        \[f_{\mathrm{in}}(z)=Q_{\text{stick}}f(z)=\int_{\mathfrak{F}} Q_{\text{stick}}(z,d\zeta)f(\zeta),\qquad z\in\Gamma_{\mathrm{in}},\]
        where $f_{\mathrm{in}}(z):=\lim_{t\nearrow0}f\circ\phi_t(z)$.
    \end{enumerate}
\end{proposition}
\begin{proof}
    By considering the state space $E$ as the direct sum of two components $E_1:=E\setminus\mathfrak{F},E_2:=\mathfrak{F}$,
    we can apply Theorem~26.14 of \cite{Davis1993} after checking Conditions 1 to 4 in (24.8) of \cite{Davis1993}. We check these four conditions in turn:
    \begin{enumerate}
        \item The regularity of the vector field $\mathfrak{X}$ on each component, together with the non-explosion of its flow, is assumed in \ref{item:well-definedness1}.
        \item On $\mathfrak{F}$, the intensity of the manifold PDMP jump $\lambda_{\mathfrak{F}}$ is assumed to be locally integrable along the flow in \ref{item:well-definedness1}. Together with \ref{item:well-definedness3}, we obtain the locally integrable rate of the sticky manifold PDMP on $\mathfrak{F}$.
        \item Our construction of sticking and unsticking mechanisms, denoted here also by $Q_{\text{stick}}, Q_{\text{unstick}}$, ensures $Q_{\text{stick}}(z,\{z\})=Q_{\text{unstick}}(z,\{z\})=0$.
        \item The overall number of jumps, including forced jumps, is finite given \ref{item:well-definedness2}. This is because a sticking event only takes place after an unsticking event.
    \end{enumerate}
    By Theorem~26.14 of \cite{Davis1993}, we conclude that $\mathcal{D}(\mathcal{L})$ is as such.
\end{proof}

\begin{corollary}[measure-separating property of the extended generator domain]
\label{cor-measure-separation}
    Denote the subset of $\mathcal{D}(\mathcal{L})$ of bounded functions $f$ satisfying $\mathcal{L}f\in L^1(\Pi)$ by $\mathcal{D}_0$. Then, under the same assumptions of Proposition~\ref{prop;well-definedness} in addition to \ref{item:assumption-tails1} and \ref{item:assumption-tails2}, $\mathcal{D}_0$ separates probability measures on $E$: for two probability measures $\mu,\nu$, the following condition implies $\mu=\nu$
    \begin{equation}\label{eq;MeasureSeparation}
        \int_E f\,d\mu=\int_Ef\,d\nu,\qquad f\in\mathcal{D}_0.
    \end{equation}
\end{corollary}
\begin{proof}
    We show that \eqref{eq;MeasureSeparation} implies that the same equality holds for an arbitrary $g\in C^1_c(E)$.
    This will conclude the measure separating property for $\mathcal{D}_0$ since $C^1_c(E)$ is measure separating \citep[Example~2.12.10]{Bogachev2007}.
    This will be concluded from the bounded convergence theorem by taking a bounded sequence $\{f_n\}\subset\mathcal{D}_0$ that converges to $g$ pointwise.
    $f_n$ assumes the same value as $g$ on $\mathfrak{F}$, while it converges to the value of $g_{\text{in}}:=Q_{\text{stick}}g$ at $\mathfrak{F}$ from within $E\setminus\mathfrak{F}$.
    To construct $f_n$, we first define a flow-box local coordinate along $\mathfrak{X}$ on $E\setminus\mathfrak{F}$.
    For a point $z\in E\setminus\mathfrak{F}$, define
    \[\tau(z):=\inf\{t>0\mid\phi_t(z)\in\Gamma_{\mathrm{in}}\},\]
    and $b(z):=\phi_{\tau(z)}(z)$ when $\tau(z)<\infty$ where $\phi_t$ is the flow of $\mathfrak{X}$. Then, using a standard smooth cut-off function $0\le\chi\le1$ with $\chi(0)=1$ and $\operatorname{supp}\chi\subset[0,1]$, define $f_n$ as
    \[f_n(z)=(1-\chi(n\tau(z)))g(z)+\chi(n\tau(z))g_{\mathrm{in}}(b(z))\]
    when $z\in E\setminus\mathfrak{F}$ satisfies $\tau(z)<\infty$, and $f_n(z)=g(z)$ otherwise. We can check $f_n\in\mathcal{D}_0$ as follows.
    First, $f_n\in\mathcal{D}(\mathcal{L})$ since $(f_n)_{\mathrm{in}}=g_{\mathrm{in}}=Q_{\mathrm{stick}}f$ and $f_n$ is absolutely continuous along the flow:
    \[\mathfrak{X}f_n(z)=n\chi'(n\tau(z))g(z)+(1-\chi(n\tau(z)))\mathfrak{X}g-n\chi'(n\tau(z))g_{\mathrm{in}}(b(z))\]
    on $z\in E\setminus\mathfrak{F}$ such that $\tau(z)<\infty$.
    Second, $\mathcal{L}f_n$ is integrable since
    \begin{align*}
        \int_E\|\mathcal{L}f_n\|\,d\Pi&\le\int_E\|\mathfrak{X}g\|\,d\Pi+2n\|\chi'\|_\infty\|g\|_\infty\Pi(E\setminus\mathfrak{F})+2\|g\|_\infty\int_E\lambda\,d\Pi<\infty,
    \end{align*}
    given $\norm{\mathfrak{X}_z}$ is $\Pi$-integrable from \ref{item:assumption-tails1} and $\lambda$ is $\Pi$-integrable from \ref{item:assumption-tails2}. Finally, $\|f_n\|_\infty\le\|g\|_\infty$ implies that the sequence $\{f_n\}$ is uniformly bounded.
\end{proof}

\subsection{Proof of Theorem~\ref{thm;NewTheorem}}\label{appendix-subsec-proofoftheorem}

We will see that the essential condition is the equation of measures \eqref{eq;RateMatchingEquation}, from which the unsticking rate is obtained under the absolute continuity condition~\ref{item:regularity2}. Recall the definition of the inward and outward weighted boundary measure flux:
\[J_{\text{in/out}}(dz)=1_{\Gamma_{\text{in/out}}}(z)|\langle v,\partial_r(x)\rangle|\pi_{\mathbb{R}^d}(x)\mathcal{H}^{d-1}_{\mathcal{M}}(dx)\nu_{E\setminus\mathfrak{F}}(x,dv).\]

\begin{proof}
    Denote the subset of $\mathcal{D}(\mathcal{L})$ of bounded functions $f$ satisfying $\mathcal{L}f\in L^1(\Pi)$ by $\mathcal{D}_0$. We prove $\int_E\mathcal{L}f\,d\Pi=0$ for all $f\in\mathcal{D}_0$. This implies invariance when $\mathcal{D}_0$ separates probability measures on $E$ %, as discussed in the proof of \citet[Theorem~1]{Chevallier+2024} and 
    \citep[Proposition~34.7]{Davis1993}, which is indeed the case in our context (Corollary~\ref{cor-measure-separation}).
    This approach, establishing $\int_E\mathcal{L}f\,d\Pi=0$ for $f$ in a measure-separating domain that is larger than the strong generator domain, is also taken in \citet[Theorem~34.31]{Davis1993} and \citet[Section~3]{Chevallier+2024}. The latter work is closely related to our proof; we here extend it by providing a flow-wise Gauss--Green formula (Lemma~\ref{lemma;ExtendedGaussGreen}). This can be of independent interest since it accommodates a wide range of PDMP samplers with non-linear flows, such as the boomerang sampler and manifold PDMPs discussed in Section~\ref{sec;2ManifoldPDMPs}.
    
    For every $f\in\mathcal{D}_0$, using the generator expression~\eqref{eq;generator1}, we have a decomposition
    \begin{equation}\label{eq;invariance-decomposition}
        \int_E\mathcal{L}f\,d\Pi=\omega\int_{E\setminus\mathfrak{F}}\mathcal{L}_{\mathbb{R}^d}f\,d\Pi_{E\setminus\mathfrak{F}}+(1-\omega)\int_{\mathfrak{F}}\mathcal{L}_{\mathcal{M}}^{\text{lift}}f\,d\Pi_{\mathfrak{F}}+(1-\omega)\int_{\mathfrak{F}}\Lambda(Q_{\mathrm{unstick}}f-f)\,d\Pi_{\mathfrak{F}}.
    \end{equation}
    The second term equals zero by the assumed invariance of the lifted manifold dynamics.
    To match the first term with the third, the Gauss--Green divergence theorem \citep[Theorem~XIII.3.2]{Lang1995-Manifolds} would imply
    \begin{equation}\label{eq;Gauss-Green}
      \int_{E\setminus\mathfrak{F}}\mathcal{L}_{\mathbb{R}^d}f\,d\Pi_{E\setminus\mathfrak{F}}=\int_{\Gamma_{\mathrm{in}}}f_{\mathrm{in}}\,dJ_{\mathrm{in}}-\int_{\Gamma_{\mathrm{out}}}f\,dJ_{\mathrm{out}}
    \end{equation}
    where $f_{\mathrm{in}}(z):=\lim_{t\searrow0}f(\phi_{-t}(z))$ on $z\in\Gamma_{\mathrm{in}}$ for the ambient flow $\phi$.
    % \[\int_{E\setminus\mathfrak{F}}\mathcal{L}_{\mathbb{R}^d}f(z)m(z)\,dz=-\int_{\mathfrak{F}}(Qf-f)(\zeta)m(\zeta)|\langle\eta,\partial_r(\xi)\rangle|\,d\zeta,\]
    % where $z=(x,v)$, $\zeta=(\xi,\eta)\in\mathfrak{F}$, and $m(z):=\pi_{\mathbb{R}^d}(x)\gamma^d(v)$ with Gaussian density $\gamma^d$.
    However, the standard result only covers the case $f|_{E\setminus\mathfrak{F}}\in C_c^1(T(\mathbb{R}^d))$. We therefore apply Lemma~\ref{lemma;ExtendedGaussGreen} to extend the result to an arbitrary $f\in\mathcal{D}_0$ with compact support, and an approximation argument to the general case $f\in\mathcal{D}_0$ (Lemma~\ref{lem;cutoff}).
    Using the boundary condition $f_{\mathrm{in}}=Q_{\mathrm{stick}}f$, \eqref{eq;invariance-decomposition} now reads
    \[\int_E\mathcal{L}f\,d\Pi=\omega\int_{\mathfrak{F}}Q_{\mathrm{stick}}f\,dJ_{\mathrm{in}}-\omega\int_{\Gamma_{\mathrm{out}}}f\,dJ_{\mathrm{out}}+(1-\omega)\int_{\mathfrak{F}}\Lambda(Q_{\mathrm{unstick}}f-f)\,d\Pi_{\mathfrak{F}}.\]
    The assumptions on the kernels ensure the following equalities:
    \begin{equation}\label{eq;RateMatchingEquation}
        \omega J_{\mathrm{in}}Q_{\mathrm{stick}}=(1-\omega)\Lambda\Pi_{\mathfrak{F}},\qquad(1-\omega)(\Lambda\Pi_{\mathfrak{F}})Q_{\mathrm{unstick}}=\omega J_{\mathrm{out}}.
    \end{equation}
    % \[(1-\omega)\Lambda(\zeta)\Pi_{\mathfrak{F}}(d\zeta)=\omega\,|\langle\eta,\partial_r(\xi)\rangle|\,\widetilde{\Pi}_{E\setminus\mathfrak{F}}(d\zeta),\]
    % where $\widetilde{\Pi}_{E\setminus\mathfrak{F}}(d\zeta):=\pi_{\mathbb{R}^d}(\xi)\mathcal{H}^{d-1}_{\mathcal{M}}(d\xi)\nu_{E\setminus\mathfrak{F}}(\xi,d\eta)$ is the corresponding surface restriction on $\mathfrak{F}$.
    This balance condition \eqref{eq;RateMatchingEquation} concludes \eqref{eq;invariance-decomposition} equals zero, and rate $\Lambda$ is the Radon--Nikodym derivative
    \begin{equation}\label{eq;RadonNikodym}
        \Lambda(\zeta)=\frac{\omega}{1-\omega}\frac{d(J_{\mathrm{in}}Q_{\mathrm{stick}})}{d\Pi_{\mathfrak{F}}}(\zeta)=|\langle\eta,\partial_r(\xi)\rangle|\frac{\omega}{1-\omega}\frac{\pi_{\mathbb{R}^d}(\xi)}{\pi_{\mathcal{M}}(\xi)}.
    \end{equation}
    % which reduces to \eqref{eq;6Manifoldthawingrate} when the velocity distributions are Gaussian $\nu_{E\setminus\mathfrak{F}}(x,-)=\nu_{\mathfrak{F}}(x,-)=N_d(0,\sigma^2)$.
\end{proof}

\subsection{Lemmas for the theorems}

Behind the proof, there are two key techniques. The first is disintegration over foliations, similar to the approach in \citealt[Theorem~2]{Chevallier+2024}, where the authors prove an integration by parts formula along the linear foliation $x+\mathbb{R}v$ parametrised by $v\in\mathbb{R}$. This approach naturally aligns with the martingale approach of \cite{Davis1984,Davis1993}, as the domain of the extended generator is described by conditions on the foliation induced by the deterministic flow; see \citealt[Theorem~5.5]{Davis1984} and \citealt[Theorem~26.14]{Davis1993}. Accordingly, the divergence theorem is replaced by the fundamental theorem of calculus along each one-dimensional leaf. The second key technique is related to the blow-up construction of the state space: the trivialisation of the bundle $\mathfrak{F}$ through $\Psi$ given in \eqref{eq;Phi}. This connects the two foliations, one in the ambient space and the other in the manifold.

The following lemma is the core of the proof of Theorem~\ref{thm;NewTheorem}.
To compute the first term in \eqref{eq;invariance-decomposition}, we could use the standard Gauss--Green divergence theorem and the Leibniz rule in $\mathbb{R}^d\setminus\mathcal{M}$ to derive the result if $f$ were smooth. However, this time $f$ is only assumed to be absolutely continuous along the flow, or we have to establish the Feller property and identify functional-analytic arguments about a core. Therefore, we consider foliating the state space along the flow $\phi_t$ of the vector field $\mathfrak{X}$ corresponding to $\mathcal{L}_D$ and use the fundamental theorem of calculus in each one-dimensional leaf. This approach is similar to \citealt[Theorem~4]{Chevallier+2024}, but we first construct a tubular neighbourhood and coordinate chart induced by $\mathfrak{X}$ to straighten the flow, a similar idea to the straightening-out theorem, also known as the flow-box theorem \citep[Proposition~IV.2.13]{Lang1995-Manifolds}. The resulting proof accommodates PDMPs with a non-linear flow such as the boomerang sampler and randomised Hamiltonian Monte Carlo.

\begin{lemma}[flowwise Gauss--Green formula on $E\setminus\mathfrak{F}$]\label{lemma;ExtendedGaussGreen}
    Under Assumption~\ref{NewAssumption}, it holds that
    % \[\int_{E\setminus\mathfrak{F}}\mathcal{L}_{\mathbb{R}^d}f(z)m(z)\,dz=-\int_{\mathfrak{F}}(Qf-f)(\zeta)m(\zeta)|\langle\eta,\partial_r(\xi)\rangle|\,d\zeta,\]
    \[\int_{E\setminus\mathfrak{F}}\mathcal{L}_{\mathbb{R}^d}f\,d\Pi_{E\setminus\mathfrak{F}}=\int_{\Gamma_{\mathrm{in}}}f_{\mathrm{in}}\,dJ_{\mathrm{in}}-\int_{\Gamma_{\mathrm{out}}}f\,dJ_{\mathrm{out}}\]
    for every bounded compactly supported Borel function that has integrability $\mathfrak{X}f\in L^1(\Pi_{E\setminus\mathfrak{F}})$, absolute continuity along the ambient flow $\phi$, and limits $f_{\mathrm{in}}(z):=\lim_{t\searrow0}f(\phi_{-t}(z))$ on $z\in\Gamma_{\mathrm{in}}$.
    In particular, it holds for every compactly supported $f\in\mathcal{D}_0$.
\end{lemma}
\begin{remark}[Key differences to existing theorems]
    Similar Gauss--Green type formulae are also established in \citet[Proposition~2.7]{Bierkens2023b} and \citet[Proposition~10]{Chevallier+2024}. These two results rely on complementary sets of assumptions, in that neither set of assumptions completely subsumes the other. Among several technical conditions, \citet[Proposition~2.7]{Bierkens2023b} assumes $C_c^1$-regularity of a test function $f$, while allowing a potentially curved flow of the form $\dot{x}=s(x)v$ with speed function $s$. By contrast, \citet[Proposition~10]{Chevallier+2024} only imposes a substantially weaker regularity condition, namely absolute continuity along the foliation, but it is confined to linear dynamics. Our theorem may be viewed as a common generalisation of the two, simultaneously accommodating non-linear second-order flows, including boomerang and Hamiltonian dynamics, and dealing directly with non-differentiable test functions in the domain of the extended generator.
\end{remark}
\begin{proof}
    We mainly consider the standard Gaussian case; the non-Gaussian case can be similarly deduced by replacing $T(\mathbb{R}^d)$ by $\mathcal{V}(\mathbb{R}^d)$, $dz$ by the density $\Omega(dz)$, and $\operatorname{div}$ by $\operatorname{div}_\Omega$.
    We then have $\Pi_{E\setminus\mathfrak{F}}(dz)=m(z)dz$ for $m(z)=\pi_{\mathbb{R}^d}\gamma^d(v)$ with Gaussian density $\gamma^d$. From the ambient invariance, the left-hand side reduces to
    \begin{equation}\label{eq;jump}
        \int_{E\setminus\mathfrak{F}}\mathcal{L}_{\mathbb{R}^d}f(z)m(z)\,dz=\int_{E\setminus\mathfrak{F}}(m(z)\mathcal{L}_Df(z)+f(z)\operatorname{div}(m\mathfrak{X})(z))\,dz,
    \end{equation}
    by Lemma~\ref{lemma;ambientInvariance}.
    The last integrand is actually the Lie derivative $\mathcal{L}_{\mathfrak{X}}(f\Omega_m)$ for the volume form $\Omega_m$ with density $m$ when $f$ is smooth.
    Therefore we could apply the Gauss--Green divergence theorem to derive the boundary term if $f$ were assumed to be smooth.
    Now that all terms in the right-hand side are local and do not involve jump terms, we proceed with the calculation locally by foliating $E\setminus\mathfrak{F}$.

    There exists a countable and locally finite open covering $(U_i)_{i\in I}$ of the frozen boundary $\mathfrak{F}$, and a corresponding family of open sets $(V_i)_{i\in I}$, open in $\mathbb{R}^{2d}$, with a diffeomorphism
    \begin{equation}\label{eq;Phi_i}
        \Phi_i:(-\epsilon_i,\epsilon_i)\times U_i\to V_i\overset{\text{open}}{\subset}\mathbb{R}^{2d}
    \end{equation}
    defined by $\Phi_i(t,\zeta)=\phi_t(\zeta)$ on a relatively open set $U_i$ in $\mathfrak{F}$, since the flow $\Phi(t,\zeta)=\phi_t(\zeta)$ has a full-rank differential at each point $(0,\zeta)$ satisfying $\zeta\in\mathfrak{F}$, similar to the argument of the flow-box theorem \citep[Lang~IV.2.13]{Lang1995-Manifolds}.
    Note that, by construction, every $(x,v)\in\mathfrak{F}$ satisfies $v^\perp\ne0$; the set $\{v=0\}$ can be disregarded since it has Lebesgue measure zero.
    We first show how to pull back an integral on $V_i$ to $(-\epsilon_i,\epsilon_i)\times U_i$.

    We first calculate the Jacobian of $\Phi_i$. On a point $(0,\zeta),\;\zeta\in U_i$, its derivative is given by
    \[D(\Phi_i)_{(0,\zeta)}(s,\alpha)=s\mathfrak{X}(\zeta)+\alpha,\qquad \alpha\in T_\zeta(\mathfrak{F}),\]
    where $\alpha\in T_\zeta(\mathfrak{F})$ is identified with a point in $\alpha\in T_\zeta(E)\simeq\mathbb{R}^{2d}$. The subfibre $T_\zeta(\mathfrak{F})$ forms a $2d-1$ dimensional subspace in $T_\zeta(E)$,
    Therefore, the Jacobian of the linear mapping $D(\Phi_i)_{(0,\zeta)}$ is the absolute value of the component of $\mathfrak{X}$ orthogonal to $T_\zeta(\mathfrak{F})$, i.e., $|\langle\eta,\partial_r(\xi)\rangle|$, since $\mathfrak{X}$ is a second-order vector field.
    Since $\Phi_i{(t,\zeta)}=\phi_t(\Phi_i(0,\zeta))$, we obtain the Jacobian at $(t,\zeta)\in(-\epsilon_i,\epsilon_i)\times U_i$ as
    \[J_{\Phi_i}(t,\zeta)=|\det D\phi_t(\zeta)||\langle\eta,\partial_r(\xi)\rangle|.\]
    
    Given the Jacobian, we can pull back an integral on $V_i$ through the change of variable formula.
    For any integrable function $F(\zeta)$, application of the Fubini theorem gives
    \begin{align*}
        \int_{V_i}F\,d\Pi_{E\setminus\mathfrak{F}}&=\int_{-\epsilon_i}^{\epsilon_i}\int_{U_i}F(\phi_t(\zeta))m(\phi_t(\zeta))J_{\Phi_i}(t,\zeta)\,dtd\zeta\\
        &=\int_{U_i}|\langle\eta,\partial_r(\xi)\rangle|\,d\zeta\int^{\epsilon_i}_{-\epsilon_i}F(\phi_t(\zeta))m(\phi_t(\zeta))J_{\phi_-(\zeta)}(t)\,dt,
    \end{align*}
    where $J_{\phi_-(\zeta)}(t):=|\det D\phi_t(\zeta)|$.
    %where $I_\zeta:=\Phi((-\epsilon,\epsilon),\zeta)$ and an integral with respect to $d\zeta'$ should be read as a line integral along $\zeta':=\phi_t(\zeta)$ where $t\in(-\epsilon,\epsilon)$.
    Observe that the family $\{\Phi_i((-\epsilon,\epsilon),\zeta)\}_{\zeta\in\mathfrak{F}}$ provides a one-dimensional foliation of $V_i$.
    Substituting $F(\zeta)=\mathcal{L}_Df(\zeta)+f(\zeta)\operatorname{div}(m\mathfrak{X})(\zeta)/m(\zeta)$, we continue the calculation of \eqref{eq;jump} as
    \begin{align}
        &=\int_{U_i}|\langle\eta,\partial_r(\xi)\rangle|\,d\zeta\int_{-\epsilon_i}^{\epsilon_i}\bigg(m(\phi_t(\zeta))\mathcal{L}_Df(\phi_t(\zeta))+f(\phi_t(\zeta))\operatorname{div}(m\mathfrak{X})(\phi_t(\zeta))\bigg)J_{\phi_-(\zeta)}(t)\,dt\\
        &=\int_{U_i}|\langle\eta,\partial_r(\xi)\rangle|\,d\zeta\int_{-\epsilon_i}^{\epsilon_i}\frac{d}{dt}\bigg(f(\phi_t(\zeta))m(\phi_t(\zeta))J_{\phi_-(\zeta)}(t)\bigg)\,dt,\label{eq;leafIntegral}
    \end{align}
    where we used $\mathcal{L}_Df(\phi_t(\zeta))=\frac{d}{dt}f(\phi_t(\zeta))$ and that
    \[\frac{d}{dt}\bigg(m(\phi_t(z))|\det D\phi_t(z)|\bigg)=\operatorname{div}(m\mathfrak X)(\phi_t(z))|\det D\phi_t(z)|,\]
    which is a $\phi_t$ pullback of the relationship $\mathcal{L}_{\mathfrak{X}}\Omega_m=\operatorname{div}(m\mathfrak{X})\Omega$, a result that follows from the Cartan formula \cite[Proposition~V.5.3]{Lang1995-Manifolds} where $\Omega=dz$ is the canonical volume form \cite[see also][Theorem~9.10.6]{Bogachev2007}.
    Since $f(\phi_t(\zeta))$ is absolutely continuous only on $(-\epsilon,0)\cup(0,\epsilon)$, using the fundamental theorem of calculus to the integrand $G(t,\zeta):=f(\phi_t(\zeta))m(\phi_t(\zeta))|\operatorname{det} D\phi_t(\zeta)|$, we have \eqref{eq;leafIntegral} equals
    \begin{equation}\label{eq;FundamentalTheoremOfCalculus}
    \int_{U_i}\bigg(G(\epsilon,\zeta)-G(-\epsilon,\zeta)-G(0+,\zeta)+G(0-,\zeta)\bigg)|\langle\eta,\partial_r(\xi)\rangle|\,d\zeta.
    \end{equation}
    In the last step, we will get rid of these $G(\pm\epsilon,\zeta)$.

    Finally, we get back to the global integral \eqref{eq;jump}. We first extend $(V_i)$, the open covering of $\mathfrak{F}$, to an open covering of $\operatorname{supp}(f)$, still denoted by $(V_i)$. Through a partition of unity subordinated to $(V_i)$, we have
    \begin{align*}
        &\int_{E\setminus\mathfrak{F}}\bigg(m(z)\mathcal{L}_Df(z)+f(z)\operatorname{div}(m\mathfrak{X})(z)\bigg)\,dz\\
        &=\sum_{i\in I}\int_{V_i}\bigg(m(z)\mathcal{L}_D(\chi_if)(z)+\chi_if(z)\operatorname{div}(m\mathfrak{X})(z)\bigg)\,dz\\
        &=\sum_{i\in I}\int_{U_i}\bigg(\chi_i(\phi_
        {0-}(\zeta))f(\phi_{0-}(\zeta))-\chi_i(\phi_
        {0+}(\zeta))f(\phi_{0+}(\zeta))\bigg)m(\zeta)|\langle\eta,\partial_r(\xi)\rangle|\,d\zeta\\
        &=\int_{\Gamma_{\mathrm{in}}}f_{\mathrm{in}}(\zeta)\,m(\zeta)|\langle\eta,\partial_r(\xi)\rangle|\,d\zeta-\int_{\Gamma_{\mathrm{out}}}f(\zeta)\,m(\zeta)|\langle\eta,\partial_r(\xi)\rangle|\,d\zeta
    \end{align*}
\end{proof}

\begin{lemma}[Implication of the ambient invariance]\label{lemma;ambientInvariance}
    Define $m(z):=\pi_{\mathbb{R}^d}(x)\nu^d_x(v)$. Then, under Assumption~\ref{NewAssumption},
    \[\int_{T(\mathbb{R}^d)}\mathcal{L}_{J}f(z)m(z)\,dz=\int_{T(\mathbb{R}^d)}f(z)\operatorname{div}(m\mathfrak{X})(z)\,dz,\]
    for every bounded, compactly supported Borel measurable function $f$.
\end{lemma}
\begin{proof}
    This can be directly checked for many standard PDMPs, including randomised Hamiltonian Monte Carlo, bouncy particle sampler, Zig-Zag sampler, boomerang sampler, etc.
    Nonetheless, we provide a proof in this generic setting.
    Before considering an arbitrary bounded compactly supported Borel function $f$, we first show the relationship for differentiable $f\in C_c^1(T(\mathbb{R}^d))$. The regularity assumptions of \ref{item:well-definedness1} and \ref{item:regularity1} enable us to use the Gauss divergence theorem \citep[Theorem~XIII.3.2]{Lang1995-Manifolds} %by considering a coordinate expression $(v,F(x,v))$ for the second-order vector field $\mathfrak{X}$ \ref{item:assumption-tails2}
    as follows
    \begin{align*}
        \int_{T(\mathbb{R}^d)}\mathcal{L}_{J}fm\,d\Omega&=
        -\int_{T(\mathbb{R}^d)}\mathcal{L}_Dfm\,d\Omega
        % \\&=-\int_{T(\mathbb{R}^d)}\left(v^i\frac{\partial f}{\partial x^i}(z)+F^i(x,v)\frac{\partial f}{\partial v^i}(z)\right)m(z)\,dz\\
        % &=\int_{T(\mathbb{R}^d)}\left(f(z)\frac{\partial (mv^i)}{\partial x^i}(z)+f(z)\frac{\partial(mF^i(x,v))}{\partial v^i}(z)\right)\,dz&\\
        =\int_{T(\mathbb{R}^d)}f\operatorname{div}_\Omega(m\mathfrak{X})\,d\Omega,
    \end{align*}
    where we used the assumed ambient invariance in the first equality.
    To extend the result for any bounded measurable function $f$, we first realise that both the leftmost and rightmost sides define the same linear functional on $C_c^1(T(\mathbb{R}^d))$. This equality first extends to $C_c(T(\mathbb{R}^d))$ by a standard approximation argument under the integrability assumption of \ref{item:assumption-tails2}.
    Then, they have to coincide as a Borel measure by the uniqueness of the Riesz representation \cite[Theorem~7.11.3]{Bogachev2007}.
\end{proof}

\begin{lemma}[Approximation argument for the flowwise Gauss--Green formula]
    \label{lem;cutoff}
    Assume that, under Assumption~\ref{NewAssumption},
    \begin{equation}\label{eq;lemma;cutoff}
        \int_{E\setminus\mathfrak{F}}\mathcal{L}_{\mathbb{R}^d}f\,d\Pi_{E\setminus\mathfrak{F}}=\int_{\Gamma_{\mathrm{in}}}f_{\mathrm{in}}\,dJ_{\mathrm{in}}-\int_{\Gamma_{\mathrm{out}}}f\,dJ_{\mathrm{out}}
    \end{equation}
    has been proven for every bounded, compactly supported Borel function $f$ satisfying the condition in Lemma~\ref{lemma;ExtendedGaussGreen}. Further suppose the flux balance condition \eqref{eq;GeneralBalance}
    \[J_{\mathrm{in}}Q_{\mathrm{stick}}Q_{\mathrm{unstick}}=J_{\mathrm{out}},\qquad \omega J_{\mathrm{in}}Q_{\mathrm{stick}}=(1-\omega)\Lambda\Pi_{\mathfrak{F}}.\]
    Then, the same identity holds for every $f\in\mathcal{D}_0$.
\end{lemma}
\begin{proof}
    Take a sequence of non-negative cutoff functions $\{\rho_n\}\subset C^\infty_c(E)$ such that $\rho_n\nearrow1\;\text{a.s.}$, for example,
    \[\rho_n(x,v):=\rho\left(\frac{\|\beta(x)\|^2+\|v\|^2}{n^2}\right)\]
    for a smooth decreasing function $\rho:\mathbb{R}_+\to\mathbb{R}_+$ satisfying $\rho([0,1])=\{1\}$ and $\rho([4,\infty))=\{0\}$.
    For any $f\in\mathcal{D}_0$, we can then check that $f_n:=\rho_nf$ satisfies the required conditions of Lemma~\ref{lemma;ExtendedGaussGreen}. $f_n$ also has compact support since
    \[\operatorname{supp}\rho_n\subset\{(x,v)\in E\mid\|\beta(x)\|^2+\|v\|^2\le4n^2\},\]
    where the right-hand side is $\overline{\beta}^{-1}(B_{2n})$ for the lifted bundle morphism $\overline{\beta}$ and closed ball $B_{2n}$ with radius $2n$; this is a compact set because $\mathcal{M}\subset\mathbb{R}^d$ is assumed to be closed.
    We have, for the decomposition of the ambient generator $\mathcal{L}_{\mathbb{R}^d}=\mathcal{L}_D+\mathcal{L}_J$ and $\mathcal{L}_Jf=\lambda (Qf-f)$,
    \begin{equation}\label{eq;decomposition-approximation}
        \mathcal{L}_{\mathbb{R}^d}f_n=\rho_n\mathcal{L}_{\mathbb{R}^d}f+f\mathcal{L}_D\rho_n+\int_Ef(z')(\rho_n(z')-\rho_n(z))\lambda(z)Q(z,dz').
    \end{equation}
    The convergence of the left-hand side in \eqref{eq;lemma;cutoff} $\int_{E\setminus\mathfrak{F}}\mathcal{L}_{\mathbb{R}^d}f_n\,d\Pi_{E\setminus\mathfrak{F}}\to\int_{E\setminus\mathfrak{F}}\mathcal{L}_{\mathbb{R}^d}f\,d\Pi_{E\setminus\mathfrak{F}}$ can be established term by term:
    \begin{description}
        \item[The first term] Since $f\in\mathcal{D}_0$, it satisfies $\mathcal{L}f\in L^1(\Pi)$. Therefore, by the Lebesgue dominated convergence theorem,
        \[\int_E\rho_n\mathcal{L}f\,d\Pi\xrightarrow{n\to\infty}\int_E\mathcal{L}f\,d\Pi.\]
        \item[The second term] Since $\mathfrak{X}$ is a second-order vector field \ref{item:assumption-tails1}, 
        \begin{align*}
            \abs{\cL_D\rho_n}&=\frac{2}{n^2}\rho'\paren{\frac{\norm{\beta(x)}^2+\norm{v}^2}{n^2}}\Paren{\brac{x,v}+\brac{v,F(x,v)}}\\
            &\le\frac{2}{n^2}C\Paren{\|v\|2n+\|F(x,v)\|2n}=\frac{C}{n}(\|v\|+\|F(x,v)\|),
        \end{align*}
        where $C:=\max_{x\in[0,4]}|\rho'(x)|$.
        Under \ref{item:assumption-tails1}, the integral can be evaluated as follows:
        \begin{align*}
            \Abs{\int_Ef\mathcal{L}_D\rho_n\,d\Pi}&\le\|f\|_\infty\int_E\abs{\cL_D\rho_n}\,d\Pi\\
            &\le\norm{f}_\infty\int_E\frac{C}{n}(\|v\|+\norm{F(z)}\,\Pi(dz)\xrightarrow{n\to\infty}0.
        \end{align*}
        \item[The third term] The $\Pi$-integral of the third term converges to $0$ by the Lebesgue dominated convergence theorem under the $\Pi$-integrability of the intensity $\lambda$ \ref{item:assumption-tails2}.
    \end{description}
    The finiteness of $J_{\mathrm{in}}$ and $J_{\mathrm{out}}$ follows from the flux balance condition combined with \ref{item:assumption-tails2}:
    \[J_{\mathrm{in}}(\Gamma_{\mathrm{in}})=J_{\mathrm{out}}(\Gamma_{\mathrm{out}})=\frac{1-\omega}{\omega}\int_{\mathfrak{F}}\Lambda\,d\Pi_{\mathfrak{F}}<\infty.\]
    We therefore conclude similarly the convergence of the right-hand side in \eqref{eq;lemma;cutoff}
    \[\int_{\Gamma_{\mathrm{in}}}(f_n)_{\mathrm{in}}\,dJ_{\mathrm{in}}-\int_{\Gamma_{\mathrm{out}}}f_n\,dJ_{\mathrm{out}}\xrightarrow{n\to\infty}\int_{\Gamma_{\mathrm{in}}}f_{\mathrm{in}}\,dJ_{\mathrm{in}}-\int_{\Gamma_{\mathrm{out}}}f\,dJ_{\mathrm{out}}\]
\end{proof}

\subsection{Proofs of corollaries}

\begin{proof}[Proof of Remark~\ref{remark;GeneralResult}]
    The assumed balance conditions imply
    \[\omega J_{\mathrm{in}}Q_{\mathrm{stick}}=(1-\omega)\Lambda\Pi_{\mathfrak F},\qquad(1-\omega)(\Lambda\Pi_{\mathfrak F})Q_{\mathrm{unstick}}=\omega J_{\mathrm{out}}.\]
    The proof of Theorem~\ref{thm;NewTheorem} uses velocity compatibility only to establish these two identities. The same Gauss--Green and approximation arguments therefore apply and establish invariance of $\Pi$.
\end{proof}

\begin{proof}[Proof of Corollary~\ref{cor;Rao-Blackwellized-rate}]
    Define the weighted target on the blown-up front face by $\Pi_{\mathfrak F}^{\mathrm{RB}}(dz)=\frac{\|v^\perp\|}{\sqrt{2/\pi}}\Pi_{\mathfrak F}(dz)$ and on the entire space by $\Pi^{\mathrm{RB}}=\omega\Pi_{E\setminus\mathfrak F}+(1-\omega)\Pi_{\mathfrak F}^{\mathrm{RB}}$. Since the lifted manifold dynamics preserves $\|v^\perp\|$, invariance of $\Pi_{\mathfrak F}$ therefore implies invariance of $\Pi_{\mathfrak F}^{\mathrm{RB}}$.
    Moreover, the velocity compatibility condition and the standard Gaussian ambient velocity law give
    \[\int \|v^\perp\|\,\nu_{F,x}(dv)=\sqrt{\frac{2}{\pi}},\]
    so this re-weighting preserves the position marginal. Finally,
    \[\Lambda_{\mathrm{RB}}\Pi_{\mathfrak F}^{\mathrm{RB}}=c\|v^\perp\|\Pi_{\mathfrak F}=\Lambda\Pi_{\mathfrak F}.\]
    Consequently, the original flux identities give
    \[\omega J_{\mathrm{in}}Q_{\mathrm{stick}}=(1-\omega)\Lambda_{\mathrm{RB}}\Pi_{\mathfrak F}^{\mathrm{RB}},\qquad(1-\omega)\Lambda_{\mathrm{RB}}\Pi_{\mathfrak F}^{\mathrm{RB}}Q_{\mathrm{unstick}}=\omega J_{\mathrm{out}}.\]
    The general flux-balance result~\ref{remark;GeneralResult} establishes invariance of $\Pi^{\mathrm{RB}}$.
\end{proof}

\begin{proof}[Proof of Corollary~\ref{cor;deterministic-sticking}]
    We augment the frozen boundary with an age variable $a\in[0,T)$:
    \[\widehat{\mathfrak F}=\mathfrak F\times[0,T),\qquad\widehat E=(E\setminus\mathfrak F)\sqcup\widehat{\mathfrak F}.\]
    We lift the manifold dynamics to $\widehat{\mathfrak F}$ by increasing the age at unit speed. Sticking initialises the age at zero, while reaching $a=T$ triggers $Q_{\mathrm{unstick}}$. Define the lifted target by
    \[\widehat\Pi=\omega\Pi_{E\setminus\mathfrak F}+(1-\omega)\Pi^{\mathrm{RB}}_{\mathfrak{F}}\otimes\frac{da}{T},\qquad 0\leq a<T,\]
    whose projection onto $E$ is $\Pi_{\mathrm{RB}}$. Let $\widehat{\mathcal L}$ be the extended generator. Writing $f_0$ for the value at age zero and $f_T$ for the limit at $a=T$, $f\in\mathcal{D}(\widehat{\mathcal{L}})$ satisfies the boundary conditions
    \[f^{\mathrm{in}}=Q_{\mathrm{stick}}f_0,\qquad f_T=Q_{\mathrm{unstick}}f.\]
    The class
    \[\widehat D_0=\bigl\{f\in D(\widehat{\mathcal L}):f\text{ is bounded},\quad\widehat{\mathcal L}f\in L^1(\widehat\Pi)\bigr\}\]
    separates probability measures by the same argument as Corollary~\ref{cor-measure-separation}. The estimates of that proof also apply in the product coordinates~\eqref{eq;IntertwiningRelation} with the additional unit age drift since the manifold dynamics are independent of the conserved normal coordinate, whose Gaussian law is now weighted by $\|v^\perp\|$.
    
    For $f\in\widehat D_0$, the ambient flow-wise Gauss--Green formula~\ref{lemma;ExtendedGaussGreen} and the associated approximation argument give
    \[\omega\int_{E\setminus\mathfrak F}\widehat{\mathcal L}f\,d\Pi_{E\setminus\mathfrak F}=\omega\int_{\mathfrak F} f_0\,d(J_{\mathrm{in}}Q_{\mathrm{stick}})-\omega\int_{\Gamma_{\mathrm{out}}} f\,dJ_{\mathrm{out}}.
    \]
    On the augmented frozen boundary, applying Dynkin's formula to the lifted manifold process and integrating it against its invariant measure $\Pi^{\mathrm{RB}}_{\mathfrak{F}}$ over $a\in[0,T]$ yield
    \[\int_0^T\int_{\mathfrak F}\widehat{\mathcal L}f(z,a)\,\Pi^{\mathrm{RB}}_{\mathfrak{F}}(dz)\,da=\int_{\mathfrak F}(f_T-f_0)\,d\Pi^{\mathrm{RB}}_{\mathfrak{F}}.\]
    Boundedness of $f$ and integrability of $\widehat{\mathcal L}f$ justify localization and passage to the terminal value $f_T$. By Corollary~\ref{cor;Rao-Blackwellized-rate}, the flux identities read
    \[\omega J_{\mathrm{in}}Q_{\mathrm{stick}}=\frac{1-\omega}{T}\Pi^{\mathrm{RB}}_{\mathfrak{F}},\qquad\frac{1-\omega}{T}\Pi^{\mathrm{RB}}_{\mathfrak{F}} Q_{\mathrm{unstick}}=\omega J_{\mathrm{out}}.\]
    Adding the two component contributions and using the boundary
    conditions therefore gives
    \[\int_{\widehat E}\widehat{\mathcal L}f\,d\widehat\Pi=0,\qquad f\in\widehat D_0.\]
    The same extended-generator criterion as in Theorem~\ref{thm;NewTheorem} proves invariance of $\widehat\Pi$.
\end{proof}

\begin{proof}[Proof of Corollary~\ref{cor-one-sided-case}]
    For every $x\in\mathcal{M}$, let $\mathcal{V}_x^+:=\{v\in\mathcal{V}_x\mid\langle v,\partial_r(x)\rangle<0\}$ denote the half-fibre of velocities directed inwards (towards $\mathcal{M}$).
    The measures $\nu_x^{\mathrm{in}}$ and $\nu_x^{\mathrm{out}}$ are now the restrictions of the ambient velocity law to the incoming and outgoing half-fibres, respectively, having total mass $1/2$ respectively.
    Since both kernels preserve the position and $\|v^\perp\|$, the velocity compatibility conditions imply
    \[(\|v^\perp\|\nu_x^{\mathrm{in}})Q_{\mathrm{stick},x}=\frac12 \|v^\perp\|\nu_{\mathfrak F,x},\qquad\left(\frac12 \|v^\perp\|\nu_{\mathfrak F,x}\right)Q_{\mathrm{unstick},x}=\|v^\perp\|\nu_x^{\mathrm{out}}.\]
    Multiplying by $\pi_{\mathbb R^d}(x)\mathcal H_{\mathcal M}^{d-1}(dx)$ therefore gives
    \[(J_{\mathrm{in}}Q_{\mathrm{stick}})(dx,dv)=\frac12 \|v^\perp\|\pi_{\mathbb R^d}(x)\mathcal H_{\mathcal M}^{d-1}(dx)\nu_{\mathfrak F,x}(dv),\qquad J_{\mathrm{in}}Q_{\mathrm{stick}}Q_{\mathrm{unstick}}=J_{\mathrm{out}}.\]
    Consequently, the rate
    \[\Lambda(x,v)=\frac{\omega}{2(1-\omega)}\|v^\perp\|\frac{\pi_{\mathbb R^d}(x)}{\pi_{\mathcal M}(x)}\]
    satisfies
    \[\omega J_{\mathrm{in}}Q_{\mathrm{stick}}=(1-\omega)\Lambda\Pi_{\mathfrak F}.\]
    Thus both flux-balance conditions in Remark~\ref{remark;GeneralResult} hold, and the claimed invariance follows.
\end{proof}

We have an alternative approach to Corollary~\ref{cor-one-sided-case}, in which we formally reduce the one-sided case to the two-sided case. We artificially consider the two faces $\mathcal{M}^+$ and $\mathcal{M}^-$, where one of the faces will never be hit directly from $U$. In this approach, we can retain the rate \eqref{eq;6Manifoldthawingrate} for unsticking, but after the first sticking event the process is interpreted to restick to the other face of the submanifold.
Equivalently, the sampler returns to $U$ at the second arrival of the inhomogeneous Poisson process with rate \eqref{eq;6Manifoldthawingrate}.

\section{Experimental details}\label{appendix-ExperimentalDetails}

The samplers were implemented in Julia; experiments used Julia~1.13.3 The CPU-budget comparisons in Sections~\ref{subsec;QuantumStateTomography} and~\ref{subsec;BayesianPCA} used an Apple M5 processor (10 CPU cores, 64-bit ARM) running macOS~26.6.2. Each timed sampling process used one Julia sampling thread and one BLAS thread. The code, experimental settings, and reproduction instructions are available at \url{https://github.com/162348/sticky-manifold-pdmp}.

\subsection{Gaussian mixture model (Section~\ref{subsec-GMM-illustration})}

In Section~\ref{subsec-GMM-illustration}, we discussed a two-component Gaussian mixture model that gives rise to the submanifold $\mathcal{M}_{\pi/4}$ using the notation in Section~\ref{subsec;LinearSubspaces}.

\subsubsection{Reversible-jump proposals}

The reversible-jump proposals are given by the following kernels:
\[P_{\text{split}}\left(\begin{pmatrix}\mu\\\mu\end{pmatrix},-\right)=N_2\left(\begin{pmatrix}\mu\\\mu\end{pmatrix},\frac{1}{2}\begin{pmatrix}1&-1\\-1&1\end{pmatrix}\right),\]
\[P_{\text{merge}}\left(\begin{pmatrix}\mu_1\\\mu_2\end{pmatrix},-\right)=\delta_{\left(\frac{\mu_1+\mu_2}{2},\frac{\mu_1+\mu_2}{2}\right)}.\]
In words, the split kernel proposes a displacement of law $N(0,1)$ in the normal direction to $\mathcal{M}_{\pi/4}$, while the merge kernel deterministically projects the sampler to the foot of the perpendicular. Through a coordinate change $(t,n)^\top=(\frac{\mu_1+\mu_2}{\sqrt2},\frac{\mu_1-\mu_2}{\sqrt2})^\top$, the detailed balance condition between the two points $(t,0)$ and $(t,n)$ requires
\[(1-\omega)\pi_{\mathcal{M}}(t)L(t,0)\,dt\,P_{\text{split}}(t,dn)\alpha_{\text{split}}(t,n)=\omega \pi_{\mathbb{R}^2}(t,n)L(t,n)\,dn\,P_{\text{merge}}((t,n),dt)\alpha_{\text{merge}}(t,n)\]
\[\therefore\qquad(1-\omega)L(t,0)\phi(n)\alpha_{\text{split}}(t,n)=\omega\phi(n)L(t,n)\alpha_{\text{merge}}(t,n)\]
to hold, where $\alpha_{\text{split}}$ and $\alpha_{\text{merge}}$ are acceptance probabilities, $L(t,n)$ the likelihood, and $\phi(x)$ denotes the density of the standard Gaussian distribution. Observe that the proposal density and the prior distribution match and cancel out. A natural choice is then
\[\alpha_{\text{split}}(n)=\frac{\omega}{1-\omega}\frac{L(t,n)}{L(t,0)}\land1,\quad\alpha_{\text{merge}}(n)=\frac{1-\omega}{\omega}\frac{L(t,0)}{L(t,n)}\land1.\]
The resulting RJ-MCMC makes proposals using these kernels after every random walk Metropolis step, which is $N_2(0,I_2/4)$ in the ambient space and $N_1(0,1/4)$ in the manifold chart $\mu$.

\subsubsection{Algorithm design}

The SM-FECMC uses the uniform velocity distribution and the identity map on sticking and unsticking just as in Example~\ref{example;SM-identity-sticking}. However, the ambient velocity is the uniform distribution on the unit hypersphere $\operatorname{Unif}(\mathbb{S}^1)$. This still satisfies the velocity compatibility condition~\eqref{eq;VelocityCompatibilityCondition} since the FECMC within $\mathcal{M}_{\pi/4}$ leaves the pushforward of $\operatorname{Unif}(\mathbb{S}^1)$ on $T(\mathcal{M}_{\pi/4})$ invariant. We do not perform any orthogonal switch within the FECMC algorithm since the problem is two dimensional.

The reversible-jump FECMC uses the same reversible-jump proposal kernels and acceptance probability as the reversible-jump random walk Metropolis--Hastings. The velocity is retained after splitting to keep the momentum persistent.

\subsubsection{Prior specification and the thawing rate}

The trajectories in Figure~\ref{fig:6PDMPGMM} were generated from the following prior specification:
\[\omega=0.8,\quad\pi_{\mathbb{R}^2}(\mu)=\frac{1}{2\pi}e^{-\frac{\|\mu\|^2}{2}},\quad\pi_{\mathcal{M}_{\pi/4}}(\mu_1)=\frac{1}{\sqrt{2\pi}}e^{-\mu^2_1}.\]
Note that for the coordinate chart $\varphi:\mu\mapsto\mu_1$ of $\mathcal{M}_{\pi/4}$, the Hausdorff measure is $\varphi_*\mathcal{H}^1_{\mathcal{M}_{\pi/4}}(d\mu)=\sqrt2\,d\mu_1$. Thus, the pushforward of the manifold measure is
\[\varphi_*\pi_{\mathcal{M}_{\pi/4}}\mathcal{H}^1_{\mathcal{M}_{\pi/4}}=\frac{1}{\sqrt{\pi}}e^{-\mu_1^2}d\mu_1,\]
namely $N_1(0,1/2)$. The rate now reads
\[\Lambda(x)=\|v^\perp_\sigma\|\frac{\omega}{1-\omega}\frac{\pi_{\mathbb{R}^2}(x)}{\pi_{\mathcal{M}_{\pi/4}}(x)}=\|v^\perp_\sigma\|\frac{\omega}{1-\omega}\frac{1}{\sqrt{2\pi}}.\]

\subsubsection{Data generation}

We generated $n=10$ data points from
\[Y_i\overset{\mathrm{i.i.d.}}{\sim}\frac{1}{2}N(0.1,1)+\frac{1}{2}N(-0.1,1).\]
The likelihood is thus
\[L=\prod_{i=1}^{10}
\left(\tfrac12\phi(y_i-\mu_1)+\tfrac12\phi(y_i-\mu_2)\right).\]

\subsection{Tilting linear subspaces}\label{subsec;LinearSubspaces}

\subsubsection{Main results}

\begin{figure}[!t]
    \centering
    \includegraphics[width=\linewidth]{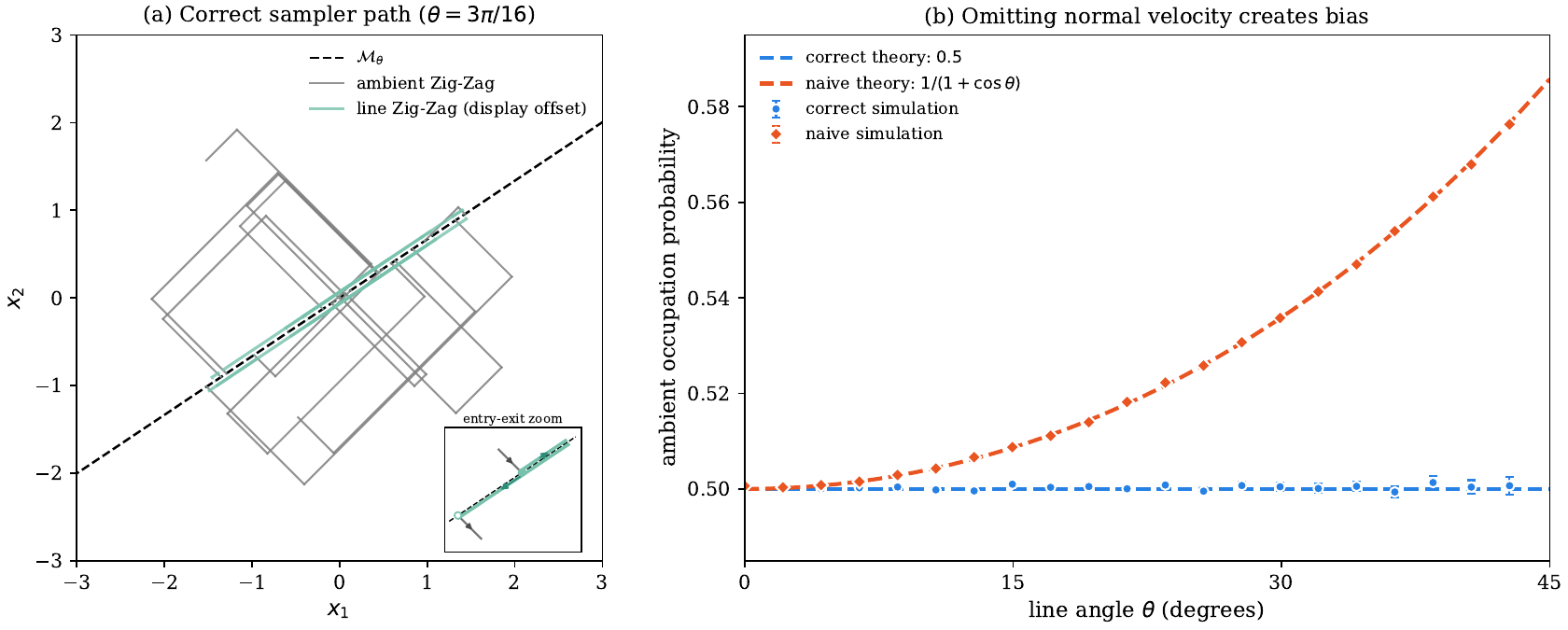}
    \caption{(a) A trajectory of the sticky zig-zag sampler with $\theta=3\pi/16$. The dynamics on $\mathcal{M}$ are displaced slightly in the normal direction for visibility. The inset represents one sticking event and one unsticking event. (b) Mean estimates of the ambient mass $\omega=1/2$ over 100 independent runs, with error bars of $\pm1.96$ Monte Carlo standard errors computed over the replicates. The 21 simulated angles are equally spaced in $[0,0.95\pi/4]$.}
    \label{fig:LinerSubspaces}
\end{figure}

Consider a one-dimensional linear subspace in $\mathbb{R}^2$ with angle $\theta$, $\mathcal{M}_\theta:=\{x_2=x_1\tan\theta\}$, and a target
\[\pi_\theta(dx)=\omega\gamma^2(x)\,dx+(1-\omega)\gamma^1(x)\mathcal{H}^{1}_{\mathcal{M}_\theta}(dx),\]
where $\gamma^d$ denotes the $d$-dimensional standard Gaussian distribution.
%In addition to the mixture model example in Section~\ref{subsec-GMM-illustration}, such geometry can also arise, for example, by applying a linear preconditioner to an axis constraint $\mathcal{M}_0=\{x_2=0\}$, possibly induced by a spike-and-slab prior.
The unsticking rate \eqref{eq;6Manifoldthawingrate} is $\Lambda_\theta(x)=\frac{\|v^\perp_\sigma\|}{\sqrt{2\pi}}\frac{\omega}{1-\omega}$ (Example~\ref{example;SM-Zigzag}).
% \begin{equation*}
%     \Lambda_\theta(x) = \|v^\perp_\sigma\|\frac{\omega}{1-\omega}\frac{(2\pi)^{-1}\exp(-\frac{1}{2}\|x\|^2)}{(2\pi)^{-1/2}\exp(-\frac{1}{2}\|x\|^2)} =\frac{\|v^\perp_\sigma\|}{\sqrt{2\pi}}\frac{\omega}{1-\omega}.
% \end{equation*}
%where the density evaluations have cancelled out, except for the additional normalising constant $1/\sqrt{2\pi}$.
When $\theta=0$, this coincides with the coordinate-aligned sticky zig-zag rate \citep[Equation~(1.3)]{Bierkens2023a}. To validate our theory, we compare our algorithm with a naïve sampler that retains the rate of \citet[Equation~(1.3)]{Bierkens2023a} at every angle $\theta\in[0,\pi/4)$, omitting the velocity factor $\|v^\perp_\sigma\|$. We set $\omega=1/2$ and employ the zig-zag dynamics in both the ambient space and the manifold. Figure~\ref{fig:LinerSubspaces} (b) shows that the two samplers coincide at $\theta=0$, while the bias in the naïve sampler increases as $\theta$ approaches $\pi/4$. The sticky manifold rate leads to unbiased estimates, illustrating the validity and necessity of the correction factor $\|v^\perp\|$ implied by our theory.

\subsubsection{Algorithm design}

The SM-Zigzag algorithm follows Example~\ref{example;SM-Zigzag}. Each run starts in the ambient space, with independent position and velocity drawn from $\gamma^2$ and $\operatorname{Unif}(\{\pm1\}^2)$, respectively. After a burn-in of 5000 time units, we estimate $\omega$ by the fraction of continuous time spent in the ambient space over a
further 50,000 time units. Event times are simulated analytically \citep[Example 1]{Bouchard2018}.

\subsubsection{Remarks}

\begin{description}
    \item[Analytic expression of the bias] \mbox{}\\
    There are only four velocity patterns $v\in\{\pm1\}^2$ in the two-dimensional zig-zag sampler. Only two patterns of them produce manifold hitting, whose norms are $\cos\theta+\sin\theta$ and $\cos\theta-\sin\theta$. As the ambient velocity distribution assigns them with equal probability, its average is $\operatorname{E}[|V^\perp|]=\cos\theta$. The correct rate \eqref{eq;6Manifoldthawingrate} is thus $\cos\theta$ times the naïve rate on average. For the naïve sampler, the detailed balance condition on the marginal model indicator process is
    \[p_\theta\int_{\partial( E\setminus\mathfrak{F})}\gamma^2(x)\|v^\perp\|\,\mathcal{H}^1_{\mathcal{M}_\theta}(x)\nu_{E\setminus\mathfrak{F}}(dv)=(1-p_\theta)\int_{\mathfrak{F}}\Lambda(x)\gamma^1(x)\,\mathcal{H}^1_{\mathcal{M}_\theta}\nu_{\mathfrak{F}}(dv)\]
    for the state $(p_\theta,1-p_\theta)$. This suggests the invariant distribution is
    \[p_\theta=\frac{\omega}{\omega+(1-\omega)\cos\theta}.\]
    \item[What happens at $\theta=\pi/4$] \mbox{}\\
    This edge case is qualitatively different in that the velocity compatibility condition \eqref{eq;VelocityCompatibilityCondition} no longer holds in its current form, because two of the velocity patterns, $(1,1)$ and $(-1,-1)$, are tangent to the line and therefore $\nu_x^{\mathrm{in}}(\Gamma_{\mathrm{in},x})=1/2$.
    Based on the general result (Remark~\ref{remark;GeneralResult}) and
    \[\nu_x^{\mathrm{in}}Q_{\mathrm{stick},x}=\frac{1}{2}\nu_{\mathfrak{F},x},\]
    the rate should be
    \[\Lambda=\frac{1}{2}\|v_\sigma^\perp\|\frac{\omega}{1-\omega}\frac{\pi_{\mathbb{R}^d}}{\pi_{\mathcal{M}}}.\]
    This is essentially because, half of the velocity patterns, $(\pm1,\pm1)$, do not produce boundary hitting in $\theta=\pi/4$, which reduces the amount of the incoming probability flow $J_{\mathrm{in}}$. The rate therefore has to be half to account for this with prolonged manifold sticking time.
\end{description}

\subsection{Concentric sphere example (Section~\ref{subsec;ConcentricSpheres})}

\subsubsection{Algorithm design}

For all four SM-PDMP algorithms, the within-manifold dynamics are identical: geodesic flow with global velocity refreshment at rate $\lambda=1$, which can be seen as the geodesic bouncy particle sampler applied to the uniform distribution \citep{Yang2024,Bell2024}. The three ambient dynamics also perform global refreshment at rate $1$. At a sphere hit, write $n=x/R$ and decompose the incoming velocity as $v=u+an$, retaining the tangent velocity $u$ and signed normal component $a$. On the sphere, only the tangential velocity is refreshed, while FECMC + normal refreshment also redraws the ambient normal velocity at each sphere refresh.
The tabulated runs select each initial stratum with probability $1/2$ and draw positions and auxiliary velocities from its conditional stationary distribution.

\subsubsection{Additional results on Rao--Blackwellisation}\label{appendix-subsubsection;LongSojourn}

Without normal refreshment or Rao--Blackwellisation, the normal speed $\|v^\perp\|$ at sphere entry has the flux-weighted density
\[f(\|v^\perp\|)=\|v^\perp\|\exp\left({-\frac{\|v^\perp\|^2}2}\right).\]
under stationarity \citep[Remark~1.9.4]{Baccelli-Bremaud2003}. The sphere sticking time satisfies $\tau\,|\,\|v^\perp\|\sim\operatorname{Exp}(\chi_d(R)\|v^\perp\|)$, thereby giving $\operatorname{Var}[\tau]=\infty$. This variance explosion is not confined to this particular setting but rather stems from the fact that the sticking time distribution is a scale mixture of the exponential distribution. Normal refreshment makes this variance finite while preserving the mean sticking time and stationary transition intensity:
\[\operatorname{E}[\tau]=\frac{1}{\chi_d(R)}\sqrt{\frac\pi2},\qquad\operatorname{E}[J_T]=T\chi_d(R)\sqrt{\frac2\pi},\]
where $J_T$ counts ambient-sphere transitions in both directions during a stationary run of duration $T$. Thus, normal refreshment changes higher order moments while preserving the mean transition count.

\begin{figure}[!ht]
    \centering
    \includegraphics[width=\linewidth]{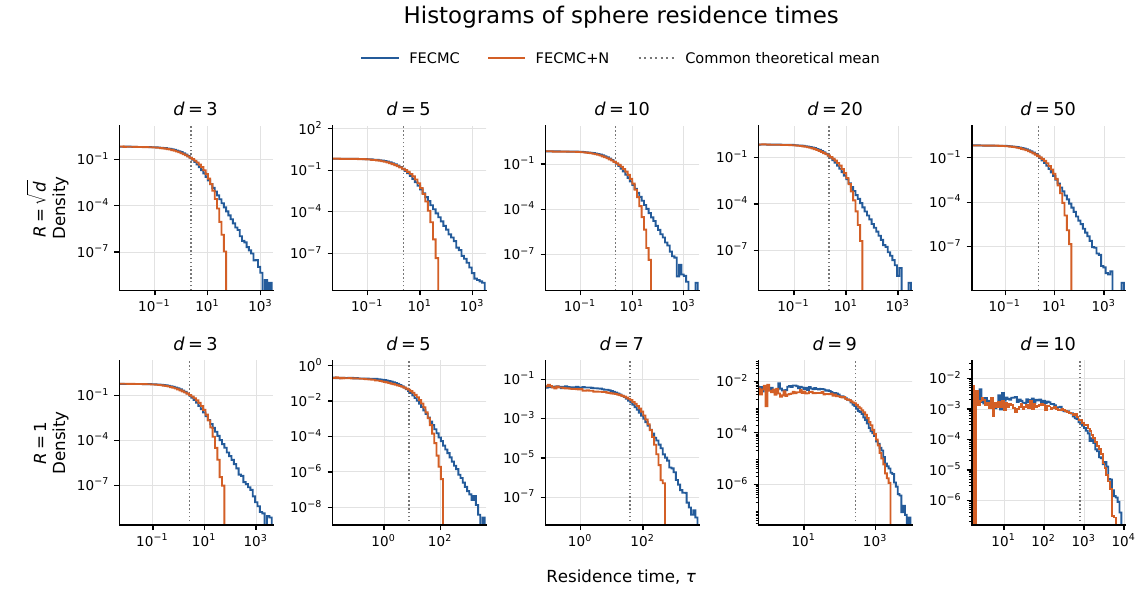}
    \caption{Histograms of sphere residence times for FECMC (blue) and FECMC+N (orange). Top: $R=\sqrt{d}$ with $d=3,5,10,20,50$; bottom: $R=1$ with $d=3,5,7,9,10$. Each condition uses 1000 independent stationary-start runs of duration $T=10^4$. Both axes are logarithmic. Dotted vertical lines indicate the theoretical mean, common to both samplers.}
    \label{fig:StickingTimeHistogram}
\end{figure}

Figure~\ref{fig:StickingTimeHistogram} shows the sticking-time distributions for standard SM-FECMC and SM-FECMC with normal refreshment. %The empirical variance is $s_\tau^2=(n-1)^{-1}\sum_{i=1}^n(\tau_i-\bar\tau)^2$, using completed sphere entry-to-exit stays pooled across runs.
We observe that the heavy tails of long manifold sojourns disappear in the normally refreshed versions. Based on our experience, we repeatedly observe situations where small $\|v^\perp\|$ causes extremely long manifold stays that exhaust the given computational budget. This phenomenon has typically been observed when we look into very biased runs, causing an inflated RMSE.
Shorter manifold stays are also shrunk towards the mean, while this effect is more subtle and less evident in the plots. Extremely short stays are retained because no normal velocity refreshment takes place during them.

\begin{figure}[!ht]
    \centering
    \includegraphics[width=\linewidth]{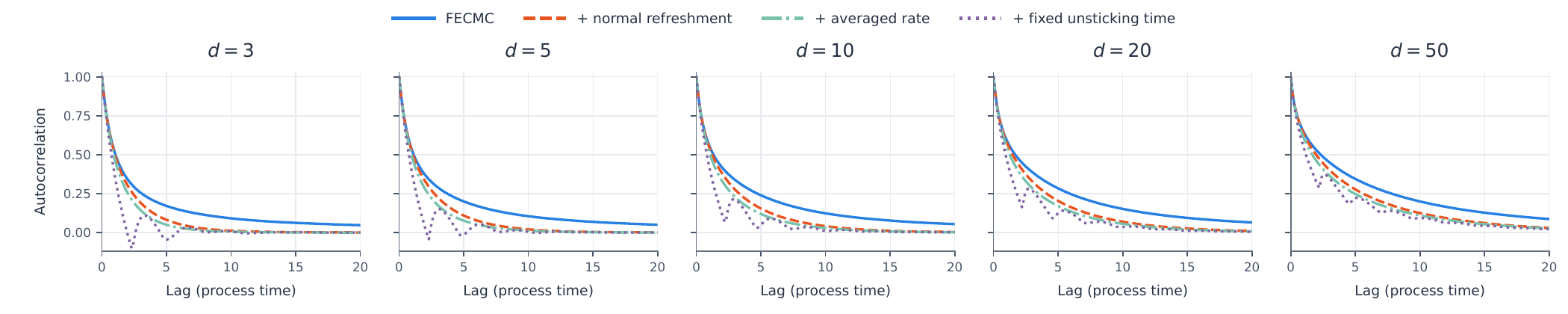}
    \caption{Autocorrelation of the model-indicator process for the four FECMC variants as in Table~\ref{tab:four-variants-fecmc}, with $R=\sqrt d$. Curves show averages over 1000 stationary-start runs of duration $T=10^4$.}
    \label{fig:S2_autocorrelation}
\end{figure}

A notable property in the autocorrelation function is observed for the Rao--Blackwellised version with fixed sticking times (Figure~\ref{fig:S2_autocorrelation}).

\subsection{Quantum state tomography example (Section~\ref{subsec;QuantumStateTomography})}\label{appendix-subsec;QuantumStateTomography}

\subsubsection{Algorithm design}

FECMC uses straight flows in the ball and geodesic flows on the sphere, without global velocity refreshment nor orthogonal switch. The radial RJ proposal uses $R\sim\operatorname{Beta}(6,1)$. Within RJ-FECMC, upon unsticking from $s\in\mathbb{S}^2$, the sampler reconstructs $v=u+\epsilon |v^\perp|s$, with a uniformly chosen sign $\epsilon\in\{-1,+1\}$. %RJ-RWMH uses Gaussian random walks in the ball and isotropic geodesic random walks on the sphere; proposals outside the ball are rejected.
RJ-HMC uses the identity mass matrix, specular boundary reflections within the ball \citep[Section~5.5.1.5]{Neal2011}, and geodesic HMC on the sphere \citep{Byrne2013}. Thinning uses locally constructed analytic bounds. The CPU time comparison includes only trajectory simulation and recording, while excluding observable integration. FECMC estimates use continuous-time averages evaluated by analytic integration along trajectories. Vector RMSEs use the usual Euclidean norm.

\subsubsection{Tuning of HMC}\label{appendx-subsec;TuningHMC_quantum}

% RJ-RWMH uses fixed scalar step sizes $h_B=0.20$ and $h_S=0.25$ across all four $N=100$ scenarios, informed by preliminary ESS-per-second tuning and a subsequent comparison of candidate ball step sizes (Tables~\ref{tab:quantum_isotropic_ess_scales}). The scenario-specific RJ attempt probabilities and RJ-FECMC rates were previously calibrated in independent pilots to approximately match the accepted model-transition frequency of RB-SM-FECMC.

% \begin{table}[!ht]
% \centering
% \caption{Isotropic scale selection by mean-ESS per elapsed second on $B^3$ and $\mathbb{S}^2$.}
% \label{tab:quantum_isotropic_ess_scales}
% \begin{tabular}{rrrrrrr}
% \toprule
% $\alpha$ & $h_B$ & $h_S$ & Accept $B^3$ & Accept $\mathbb S^2$ & ESS/s $B^3$ & ESS/s $\mathbb S^2$ \\
% \midrule
% 1.00 & 0.16250 & $0.24375$ & 22.3\% & 34.9\% & $1.37\times10^{5}$ & $2.02\times10^{5}$ \\
% 0.99 & 0.16875 & 0.26250 & 23.2\% & 32.1\% & $1.38\times10^{5}$ & $2.04\times10^{5}$ \\
% 0.98 & 0.18125 & 0.27500 & 20.9\% & 31.1\% & $1.32\times10^{5}$ & $2.06\times10^{5}$ \\
% 0.97 & 0.15625 & 0.22500 & 26.9\% & 37.9\% & $1.41\times10^{5}$ & $2.19\times10^{5}$ \\
% \bottomrule
% \end{tabular}
% \end{table}

For HMC, we conducted pilot runs with identity mass matrices. For each scenario and each integration step $T\in\{1,2,3,4,5,10\}$, we selected a common step size $\epsilon$ used across $B^3$ and $\mathbb{S}^2$ by an adaptive grid search. The objective was the minimum of the ball and sphere ESSs of the three Bloch coordinates per CPU time. The initial grid comprised 27 values from $0.001$ to $0.8$. The validation point estimates were highest at $T=3$ in all scenarios (Table~\ref{tab:quantum-hmc-pilot-settings}), with selected step sizes ranging from $0.079$ to $0.091$.

\begin{table}[!ht]
\centering
\caption{HMC tuning for each scenario and integration-step count $T$. $S$ is the smallest ESS per CPU second over three coordinates on either $\mathbb{S}^2$ or $B^3$, displayed in units of $10^5$.}
\label{tab:quantum-hmc-pilot-settings}
\small\vspace{1mm}
\setlength{\tabcolsep}{4pt}
\begin{tabular}{r r r r r}
\toprule
& \multicolumn{2}{c}{$\alpha=1.00$}
& \multicolumn{2}{c}{$\alpha=0.99$} \\
\cmidrule(lr){2-3}\cmidrule(lr){4-5}
$T$ & $\widehat\epsilon$ & $S/10^5$
    & $\widehat\epsilon$ & $S/10^5$ \\
\midrule
1  & 0.18250 & 3.97   & 0.19750 & 4.91   \\
2  & 0.12500 & $5.57$ & 0.13000 & 8.20   \\
3  & 0.07875 & 5.64   & 0.08625 & $8.89$ \\
4  & 0.05875 & 5.45   & 0.06625 & $8.38$ \\
5  & 0.05000 & $5.01$ & 0.05625 & $7.54$ \\
10 & 0.02500 & 3.71   & 0.02750 & 5.92   \\
\bottomrule
\end{tabular}
\end{table}

We here note that HMC is sensitive to this step-size tuning and it may be difficult to find the optimal step size with available quantities such as ESS.

\begin{figure}[!hbt]
    \centering
    \includegraphics[width=\linewidth]{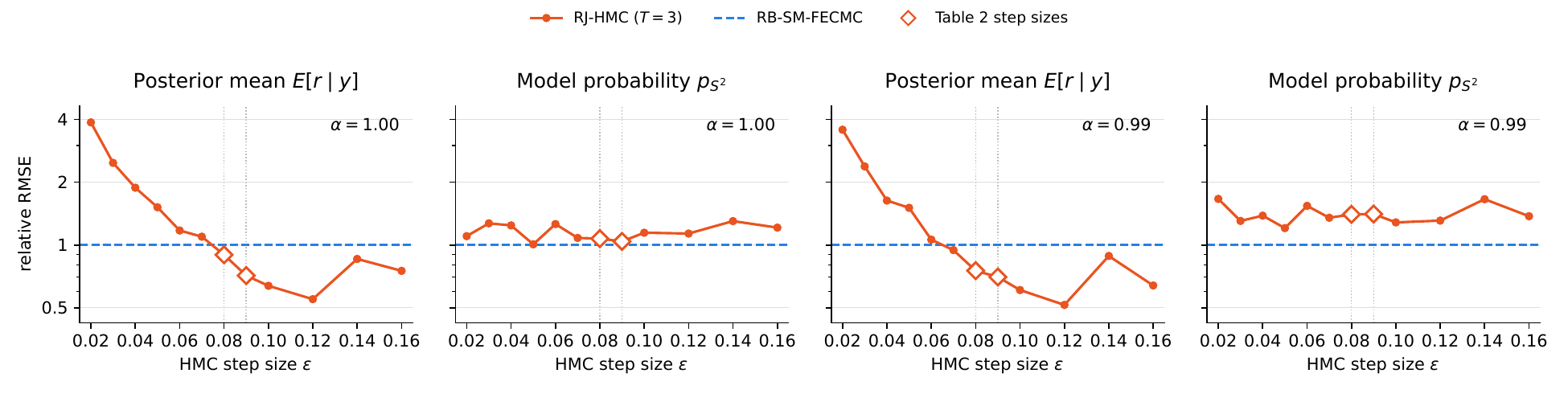}
    \caption{Step-size sensitivity of RJ-HMC with $T=3$ integration steps. $\epsilon=0.08, 0.09$ used in Table~\ref{tab:quantum_1cpu_second} were suggested from pilot runs based on ESS per CPU time, which took approximately one hour per each setting; see Appendix~\ref{appendx-subsec;TuningHMC_quantum} for details. $\epsilon=0.12$ that yields the smallest RMSE in the plot was not selected in the pilot runs.}
    \label{fig:HMC_sensitivity_quantum}
\end{figure}

\subsubsection{Prior specification and the thawing rate}

Since the uniform ball and sphere densities are $3/(4\pi)$ and $1/(4\pi)$, the one-sided unsticking rate is
\[\frac{1}{2}\|v^\perp\|\frac{3/4\pi}{1/4\pi}=\|v^\perp\|\frac{3}{2},\]
where the factor $1/2$ arises from the one-sidedness of the boundary (Corollary~\ref{cor-one-sided-case}). The Rao--Blackwellised sticking time is therefore $\sqrt{2\pi}/3$.

\subsubsection{Initialisation and RMSE computation}

Initial states are generated by conditional posterior rejection sampling with model probabilities computed by numerical quadrature, which also supplies the reference values for RMSE. This uses rejection sampling with global quadratic likelihood envelopes. Reference values are computed by numerical quadrature in spherical coordinates. Numerical stability is assessed by increasing the quadrature orders; the resulting changes are negligible relative to the reported Monte Carlo RMSEs.

\subsection{Bayesian PCA example (Section~\ref{subsec;BayesianPCA})}

\subsubsection{Prior specification and length-scale differences}\label{appendix-subsubsec;PriorBCA}

We use the spectral lift of Example~\ref{exp;HigherDimensionalRankCones} with $d=20$. Relative to $dr\,\mathcal{H}(dQ)$, where $\mathcal{H}$ is the Hausdorff measure derived from the Frobenius metric (or equivalently normalized Haar measure) on $O(20)$, the prior on the rank-$k$ stratum has density
\begin{equation}\label{eq;PCA-prior-on-eigenvalues}
    f_k(r)=Z_k^{-1}\exp\!\left(-\frac{k+1}{2}\sum_{i=1}^k r_i\right)\prod_{1\le i<j\le k}(r_i-r_j),\qquad Z_k=\frac{2^k\prod_{j=0}^{k-1}j!}{(k+1)^{k(k+1)/2}},
\end{equation}
with $f_0=Z_0=1$ \citep[Corollary~3.2.19 p.107]{Muirhead1982}. The Vandermonde determinant factor introduces divergence in the gradient and the Hessian of the potential $U_k:=-\log f_k$ along the order constraints of $C_k$. This makes the boundaries from the order constraints inaccessible for the PDMP and the exact Hamiltonian dynamics.

\begin{remark}\label{remark;Pourzanjani}
    \cite[Section~5.3]{Pourzanjani+2021} consider a related parametrisation to the spectral lift, while defining a uniform (improper) prior on the restricted (unbounded) region $C_k$. This yields a factor of inverse Vandermonde determinant in the original matrix space, producing divergences of the density near regions with overlapping eigenvalues.
\end{remark}

\begin{figure}[!ht]
    \centering
    \includegraphics[width=\linewidth]{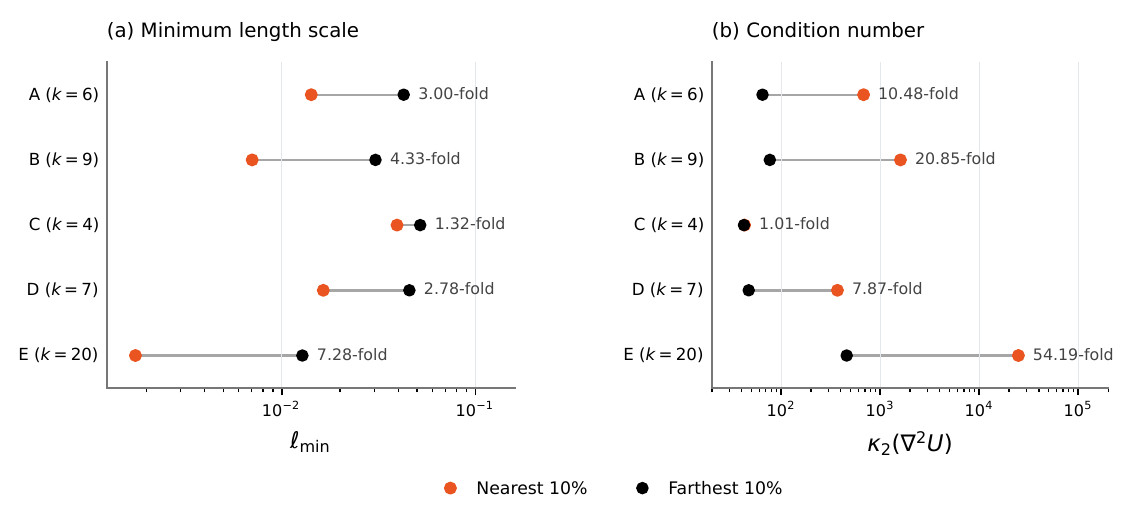}
    \caption{Variation of the shortest length scale (left) and the condition number (right)  within the main rank stratum for each scenario. The two end-points are the medians computed from the nearest and farthest 10\% of the output of the reference runs.}
    \label{fig:LengthScaleDifferences}
\end{figure}

The left panel in Figure~\ref{fig:LengthScaleDifferences} shows how the typical shortest length scale $\ell_{\min}:=1/\sqrt{\lambda_{\max}(\nabla^2U)}$ of the posterior potential $U$ differs within the main stratum of each scenario, while the right panel similarly shows that of the condition number $\kappa_2:=\lambda_{\max}(\nabla^2U)/\lambda_{\min}(\nabla^2 U)$. From the eight reference RB-SM-FECMC chains (Section~\ref{appendix-subsubsec;Reference}), the medians of the nearest and farthest 10\% of the discretised samples to the order constraint boundaries are plotted. As a result, the average acceptance probability of HMC proposals on typical points varies within a stratum (Table~\ref{tab:ppca-position-rejection}). Each entry in Table~\ref{tab:ppca-position-rejection} averages estimates at eight fixed positions, randomly selected from each chain in the nearest or farthest area. At each position, rejection probability is estimated by averaging it over 128 Gaussian momentum draws.

\begin{table}[t]
\centering
\caption{Estimated within-stratum HMC rejection probabilities at the posterior modal rank. All rows use identity mass, three leapfrog steps and $\epsilon\simeq0.01234$.}
\label{tab:ppca-position-rejection}
\begin{tabular}{crrr}
\toprule
Scenario & Rank & Near boundary & Far from boundary \\
\midrule
A & 6  & 29.5\% & 3.5\% \\
B & 9  & 36.0\% & 10.6\% \\
C & 4  & 2.9\%  & 1.2\% \\
D & 7  & 17.0\% & 5.0\% \\
E & 20 & 99.0\% & 77.3\% \\
\bottomrule
\end{tabular}
\end{table}

Such multiscale structure limits the potential of HMC. Stable step sizes are governed by the most constrained direction, which may be suboptimal for other parts of the posterior and for dynamics on $O(20)$, especially when the condition number is large. This is a mechanism commonly reported for HMC \citep[Section~5.4.2.2]{Neal2011}.

\subsubsection{Thawing rates}

We use Gaussian velocity law $(v_r,(\sqrt{2}\Omega_{ij})_{i<j})\sim N(0,I_{k+190})$ for $\Omega\in T_I(O(20))$, i.e., $\Omega^\top=-\Omega$, such that the geodesic flow is described as $\dot r=v_r$ and $\dot Q=Q\Omega$ and the velocity compatibility condition is met at every interface.

At $r_{k+1}=0$, Theorem~\ref{thm;NewTheorem}, Corollaries~\ref{cor;Rao-Blackwellized-rate} and~\ref{cor-one-sided-case} together give the Rao--Blackwellised rate
\[\Lambda_{k\to k+1}(r)=\frac{1}{2}\frac{f_{k+1}(r,0)}{f_k(r)}\sqrt{\frac2\pi}=\frac{1}{\sqrt{2\pi}}\frac{Z_k}{\,Z_{k+1}}\exp\!\left(-\frac12\sum_{i=1}^k r_i\right)\prod_{i=1}^k r_i,\qquad k=0,1,\cdots19.\]

\subsubsection{Invariance proof}

Theorem~\ref{thm;NewTheorem} assumes an Euclidean ambient space, which precludes the spectral lift parametrisation. However, the invariance argument using the extended Gauss--Green formula (Lemma~\ref{lemma;ExtendedGaussGreen}) in the proof also extends to product ambient spaces $N=U\times K$, where $U\subset\mathbb R^d$ is open and $K$ is a compact Riemannian manifold without boundary, with sticking interfaces of the form $\mathcal{M}\times K$. Imposing the intrinsic counterparts of the previous assumptions and require any additional boundary terms to vanish, the flowwise Gauss--Green argument applies in local coordinates with respect to the Riemannian volume density, and a partition of unity yields the same identity. Under the product metric, the unit normal is of the form $(\boldsymbol{n}_\mathcal{M},0)$, so the normal flux factor is again $|\langle v_r,n_\mathcal{M}\rangle|$.

\subsubsection{Algorithm design and reversible-jump proposals}

Both PDMPs use FECMC dynamics on the spectral lift. No global refreshment is used, while orthogonal switch is performed on every reflection event. Thinning uses locally constructed analytic bounds. RB-SM-FECMC discards the incoming radial velocity upon sticking and re-samples the outgoing velocity from the Rayleigh density $v e^{-v^2/2}\;(v>0)$ upon thawing. RJ-FECMC and RJ-HMC reflect at $r_k=0$. RJ-HMC alternates a fixed-rank HMC update with the identity mass matrix on the spectral lift with the same birth--death kernel as RJ-FECMC. The orthogonal frame evolves along geodesics on $O(20)$ within FECMC and HMC \citep{Byrne2013}.

Both reversible-jump algorithms propose births of new eigenvalues from densities
\[q_k(s\mid r)=\frac{a_k e^{-a_k s}}{1-e^{-a_k r_k}},\qquad 0<s<r_k,\qquad a_k=\frac{k+2}{2},\]
with $r_0=\infty$. Death proposals delete the smallest eigenvalue. RJ-FECMC uses Poisson clocks with rates $\lambda_k$, same for each direction.
Writing the likelihood by $L(G)=p(Y\,|\,G)$, the birth acceptance probabilities are $\min(1,A_k)$, where
\[A_k=\frac{f_{k+1}(r,s)\,L(\beta_{k+1}((r,s),Q))}{f_k(r)\,L(\beta_k(r,Q))\,q_k(s\mid r)}.\]
The reverse death uses $\min\{1,A_k^{-1}\}$.

\subsubsection{Tuning of HMC}\label{appendx-subsec;TuningHMC}

For each scenario, pilot experiments compared $T\in\{2,3,5,10,20\}$ and searched for $\epsilon$ using a logarithmic grid followed by local refinement based on effective sample size per CPU second. Each candidate was evaluated using four chains of the complete RJ-HMC sampler (Table~\ref{tab:HMC_tune}).
Guided by these comparisons, we chose four representative configurations $(T,\epsilon)\in\{(3,0.03),(3,0.04),(3,0.05),(10,0.007)\}$ for the main experiments.

\begin{table}[!htb]
\centering
\small
\caption{Summary of the RJ-HMC pilot experiments, reporting the harmonic mean of ESSs over all components of the three quantities per process CPU second. Bold values indicate the largest validation point estimate within each scenario.}
\label{tab:HMC_tune}
\begin{tabular}{@{}cc*{5}{c}@{}}
\toprule
Scenario & Quantity & $T=2$ & $T=3$ & $T=5$ & $T=10$ & $T=20$ \\
\midrule
A & \shortstack[c]{$\epsilon$\\ESS/CPU sec} & \shortstack{$0.0538$\\719.7} & \shortstack{$0.0494$\\\textbf{922.6}} & \shortstack{$0.0381$\\705.4} & \shortstack{$0.0247$\\469.5} & \shortstack{$0.0123$\\292.2} \\
\addlinespace
B & \shortstack[c]{$\epsilon$\\ESS/CPU sec} & \shortstack{$0.0349$\\173.2} & \shortstack{$0.0320$\\\textbf{345.6}} & \shortstack{$0.0247$\\342.3} & \shortstack{$0.0160$\\323.2} & \shortstack{$0.00951$\\217.5} \\
\addlinespace
C & \shortstack[c]{$\epsilon$\\ESS/CPU sec} & \shortstack{$0.0538$\\883.4} & \shortstack{$0.0587$\\\textbf{994.0}} & \shortstack{$0.0381$\\799.1} & \shortstack{$0.0349$\\503.0} & \shortstack{$0.0269$\\275.2} \\
\addlinespace
D & \shortstack[c]{$\epsilon$\\ESS/CPU sec} & \shortstack{$0.0453$\\536.3} & \shortstack{$0.0415$\\\textbf{596.3}} & \shortstack{$0.0293$\\503.2} & \shortstack{$0.0147$\\360.9} & \shortstack{$0.00951$\\232.8} \\
\addlinespace
E & \shortstack[c]{$\epsilon$\\ESS/CPU sec} & \shortstack{$0.0135$\\36.1} & \shortstack{$0.0123$\\43.9} & \shortstack{$0.00951$\\46.2} & \shortstack{$0.00673$\\\textbf{50.7}} & \shortstack{$0.00476$\\49.3} \\
\addlinespace
O & \shortstack[c]{$\epsilon$\\ESS/CPU sec} & \shortstack{$0.0987$\\\textbf{1318.0}} & \shortstack{$0.0698$\\1148.5} & \shortstack{$0.0587$\\916.8} & \shortstack{$0.0494$\\545.3} & \shortstack{$0.0320$\\292.3} \\
\bottomrule
\end{tabular}
\end{table}

\begin{table}[t]
\centering
\caption{Total measured sampling CPU time for searching and validating all five integration step counts $T\in\{2,3,5,10,20\}$ in each scenario.}
\label{tab:ppca-hmc-pilot-cost}
\begin{tabular}{lcr}
\toprule
Scenario & Selected $(T,\epsilon)$ & Sampling CPU time (min) \\
\midrule
A & $(3,\,0.04935)$   & 84  \\
B & $(3,\,0.032)$     & 74  \\
C & $(3,\,0.05869)$   & 88  \\
D & $(3,\,0.04150)$   & 80  \\
E & $(10,\,0.006727)$ & 238 \\
O & $(2,\,0.09870)$   & 102 \\
\bottomrule
\end{tabular}
\end{table}

\subsubsection{Data generation, initialisation and RMSE computation}\label{appendix-subsubsec;Reference}

\begin{figure}[!ht]
    \centering
    \includegraphics[width=\linewidth]{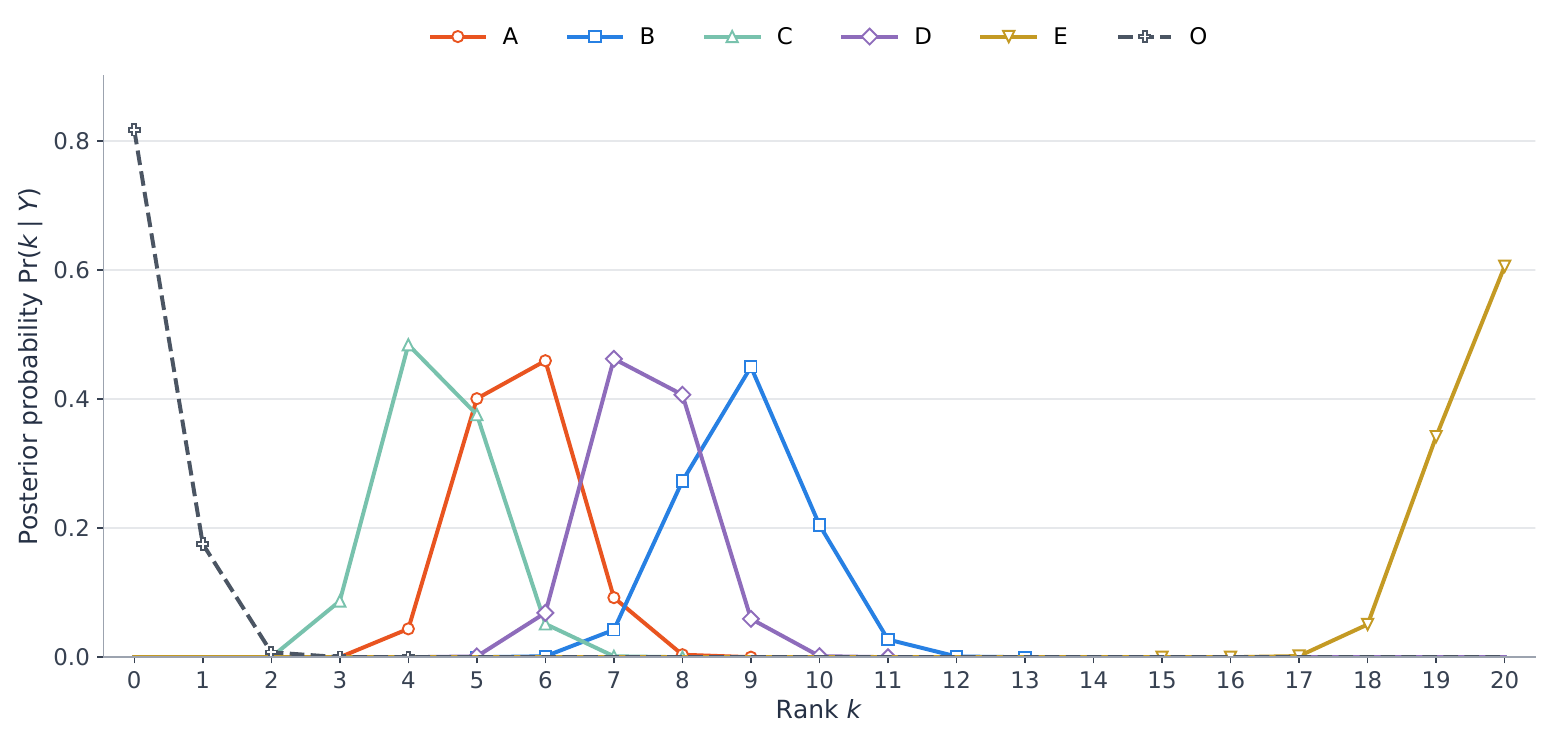}
    \caption{Estimated posterior rank probabilities $\widehat{p_k}\;(k=0,\ldots,20)$ from the long reference runs for the six scenarios A–E and O.}
    \label{fig:PosteriorRankProbability}
\end{figure}

We generate each dataset with $n=80$ and $\sigma=0.5$:
\[X=Q_\star D_\star^{1/2}Z,\qquad Y=X+\sigma E,\]
where $D_\star$ the diagonal matrix with the specified signal eigenvalues, $Q_\star$ is uniform (Haar-distributed) eigendirections on $O(20)$, and $Z,E$ have independent standard Gaussian entries. %, so that \[y_i|G_\star\overset{\textrm{i.i.d.}}{\sim} N_{20}(0,G_\star+\sigma^2I_{20}),\qquad i\in[n],\] with $G_\star:=Q_\star D_\star Q_\star^\top$.
Each dataset is shared by all methods and replicates. The eigenvalue scenarios are selected to cover different patterns of posterior rank uncertainty (Figure~\ref{fig:PosteriorRankProbability}).

For scenarios A--E, both initial positions and reference values were
obtained using independent long RB-SM-FECMC runs from dispersed starts.
Pilot runs are used to determine burn-in time $T_{\textrm{burn-in}}$ that satisfies (1) rank-normalised split $\widehat R<1.01$ \citep{Vehtari+2021}, (2) pooled effective sample sizes were at least 1,000, (3) time averages between first and second halves of $[T_{\textrm{burn-in}},2T_{\textrm{burn-in}}]$ match with a prescribed tolerance, and (4) each rank with estimated probability between 0.001 and 0.999 has at least 20 visits across chains.
We then ran 100 independent preparation chains per scenario and,
after confirming the same diagnostics, selected one position uniformly
from $[T_{\textrm{burn-in}},2T_{\textrm{burn-in}}]$.
These positions were shared across all samplers.

Reference values were computed from eight separate RB-SM-FECMC chains, up to both (i) $T_{\mathrm{burn-in}}$ has passed and (ii) the criteria (1)-(4) have been met. Batch-means estimators are then used to estimate Monte Carlo standard errors. These reference simulations are prolonged until the MCSEs are less than 10\% of the smallest RMSE across methods for every scenario and estimand (Table~\ref{tab:ReferenceCheck}). RMSEs are the square roots of the averages of the following squared errors over 100 replicates:
\[\|\widehat{\boldsymbol p}-\boldsymbol p_{\mathrm{ref}}\|_2^2,\qquad\|\widehat G-G_{\mathrm{ref}}\|_F^2,\qquad\frac1n\|(\widehat B-B_{\mathrm{ref}})Y\|_F^2,
\]
where $\boldsymbol p=(p_0,\ldots,p_{20})$ and $B=\operatorname{E}[G(G+\sigma^2I)^{-1}\,|\, Y]$.

\begin{table}[!ht]
\centering
\caption{Precision of the independent RB-SM-FECMC reference computation. Each MCSE is followed by its percentage of the smallest RMSE among the six methods for that scenario and observable.}
\label{tab:ReferenceCheck}
\small
\setlength{\tabcolsep}{4pt}
\renewcommand{\arraystretch}{1.18}
\begin{tabular}{@{}crrrrrr@{}}
\toprule
Scenario & \shortstack{Horizon\\$H$} & \shortstack{Burn-in\\$b$} & \shortstack{Rank-vector\\MCSE (\%)} & \shortstack{Mean-$G$\\MCSE (\%)} & \shortstack{Denoising\\MCSE (\%)} & $\widehat R_{\max}$ \\
\midrule
A & 65536 & 4096 & $7.51\times 10^{-4}\;(8.39\%)$ & $1.09\times 10^{-3}\;(7.47\%)$ & $3.26\times 10^{-4}\;(6.60\%)$ & 1.00090 \\
B & 43008 & 4096 & $1.08\times 10^{-3}\;(8.44\%)$ & $1.39\times 10^{-3}\;(8.00\%)$ & $7.55\times 10^{-4}\;(7.84\%)$ & 1.00046 \\
C & 135168 & 8192 & $7.40\times 10^{-4}\;(8.64\%)$ & $5.94\times 10^{-4}\;(5.47\%)$ & $2.81\times 10^{-4}\;(5.86\%)$ & 1.00020 \\
D & 51200 & 4096 & $7.14\times 10^{-4}\;(7.64\%)$ & $1.24\times 10^{-3}\;(6.94\%)$ & $5.29\times 10^{-4}\;(6.92\%)$ & 1.00077 \\
E & 24576 & 2048 & $2.10\times 10^{-3}\;(9.87\%)$ & $3.24\times 10^{-3}\;(6.29\%)$ & $8.93\times 10^{-4}\;(6.31\%)$ & 1.00086 \\
O & 2150400 & 4096 & $4.15\times 10^{-4}\;(7.14\%)$ & $2.67\times 10^{-5}\;(5.79\%)$ & $4.13\times 10^{-5}\;(5.80\%)$ & 1.00000 \\
\bottomrule
\end{tabular}
\end{table}

\clearpage
\subsubsection{Additional results}\label{appendix-subsec;AdditionalResultPCA}

\begin{figure}[!ht]
    \centering
    \includegraphics[width=\linewidth]{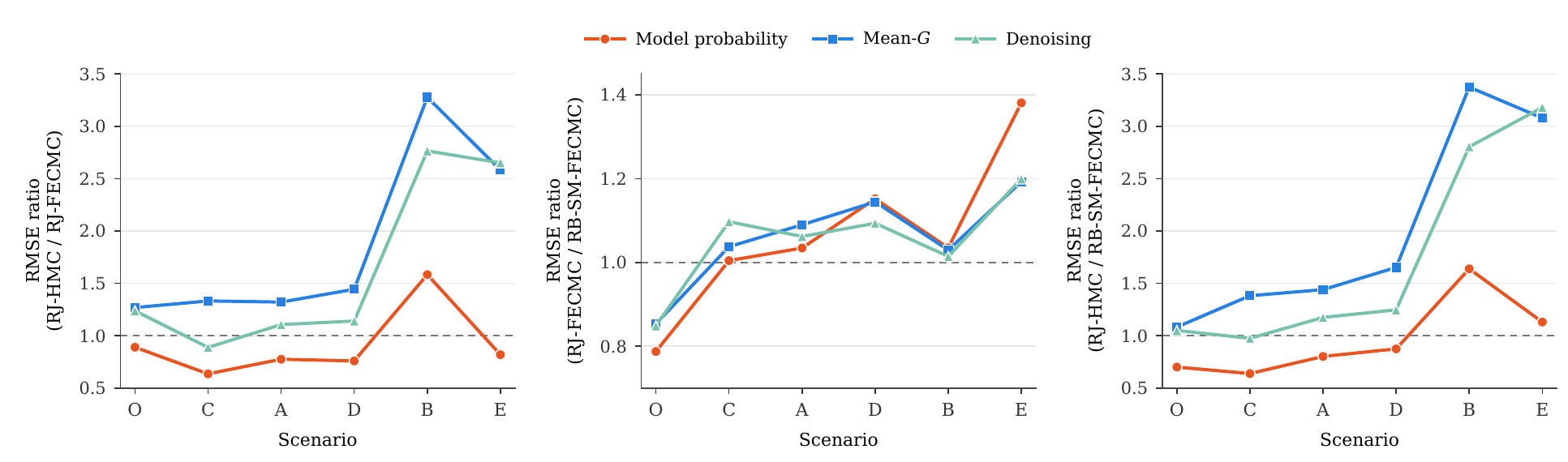}
    \caption{Relative RMSEs across the six PPCA scenarios. Left: RJ-HMC relative to RJ-FECMC; middle: RJ-FECMC relative to RB-SM-FECMC; right: RJ-HMC relative to RB-SM-FECMC. For each scenario and quantity, the RJ-HMC ratio uses the smallest observed RMSE among all configurations.}
    \label{fig:RelativeRMSEs}
\end{figure}

Here we restate the full results (Table~\ref{tab:ppca-d20-six-methods-three-cpu-rematched-full}), with Table~\ref{tab:ppca-hmc-acceptance-rmse-X} and Figure~\ref{fig:RelativeRMSEs} showing RMSE ratios for the posterior mean of the true signal $x$. The modal posterior rank $k^*$ increases in the order of scenario O, C, A, D, B, and E.

\begin{table}[!ht]
\centering
\caption{Bayesian PCA comparison using 100 paired replicates per scenario and a three-second process-CPU stopping threshold. RJ-FECMC and RJ-HMC are tuned to match the model-transition count of RB-SM-FECMC using independent pilot and validation runs.}
\label{tab:ppca-d20-six-methods-three-cpu-rematched-full}
\small
\setlength{\tabcolsep}{3pt}
\renewcommand{\arraystretch}{1.08}\vspace{1mm}
\begin{tabular}{@{}clrrrr@{}}
\toprule
Scenario & Sampler & \shortstack{Rank-vector\\RMSE} & \shortstack{Mean-$G$\\RMSE} & \shortstack{Denoising\\RMSE} & \shortstack{Mean rank\\transitions} \\
\midrule
\multirow{6}{*}{\shortstack{A\\{\footnotesize $(2,1.5,1,0.5,0,\ldots,0)$}\\{\footnotesize $k^\star=6,\ \widehat p_{k^\star}=0.459$}}} & RB-SM-FECMC & $1.12\times 10^{-2}$ & $\mathbf{1.46\times 10^{-2}}$ & $\mathbf{4.95\times 10^{-3}}$ & $3.96\times 10^{4}$ \\
 & RJ-FECMC & $1.15\times 10^{-2}$ & $1.60\times 10^{-2}$ & $5.25\times 10^{-3}$ & $3.62\times 10^{4}$ \\
 & RJ-HMC ($T=3,\ \epsilon=0.03$) & $1.18\times 10^{-2}$ & $2.76\times 10^{-2}$ & $7.16\times 10^{-3}$ & $3.85\times 10^{4}$ \\
 & RJ-HMC ($T=3,\ \epsilon=0.04$) & $1.05\times 10^{-2}$ & $2.31\times 10^{-2}$ & $6.26\times 10^{-3}$ & $3.96\times 10^{4}$ \\
 & RJ-HMC ($T=3,\ \epsilon=0.05$) & $\mathbf{8.95\times 10^{-3}}$ & $2.11\times 10^{-2}$ & $5.81\times 10^{-3}$ & $4.02\times 10^{4}$ \\
 & RJ-HMC ($T=10,\ \epsilon=0.007$) & $1.98\times 10^{-2}$ & $5.15\times 10^{-2}$ & $1.31\times 10^{-2}$ & $3.85\times 10^{4}$ \\
\midrule
\multirow{6}{*}{\shortstack{B\\{\footnotesize $(1,1,1,0.1,\ldots,0.1)$}\\{\footnotesize $k^\star=9,\ \widehat p_{k^\star}=0.450$}}} & RB-SM-FECMC & $\mathbf{1.27\times 10^{-2}}$ & $\mathbf{1.74\times 10^{-2}}$ & $\mathbf{9.63\times 10^{-3}}$ & $2.85\times 10^{4}$ \\
 & RJ-FECMC & $1.32\times 10^{-2}$ & $1.79\times 10^{-2}$ & $9.77\times 10^{-3}$ & $2.97\times 10^{4}$ \\
 & RJ-HMC ($T=3,\ \epsilon=0.03$) & $3.03\times 10^{-2}$ & $1.05\times 10^{-1}$ & $6.44\times 10^{-2}$ & $3.01\times 10^{4}$ \\
 & RJ-HMC ($T=3,\ \epsilon=0.04$) & $4.13\times 10^{-2}$ & $1.23\times 10^{-1}$ & $7.44\times 10^{-2}$ & $3.22\times 10^{4}$ \\
 & RJ-HMC ($T=3,\ \epsilon=0.05$) & $3.90\times 10^{-2}$ & $1.24\times 10^{-1}$ & $7.52\times 10^{-2}$ & $3.23\times 10^{4}$ \\
 & RJ-HMC ($T=10,\ \epsilon=0.007$) & $2.09\times 10^{-2}$ & $5.86\times 10^{-2}$ & $2.70\times 10^{-2}$ & $2.86\times 10^{4}$ \\
\midrule
\multirow{6}{*}{\shortstack{C\\{\footnotesize $(1,1,1,0.1,0.1,0.1,0,\ldots,0)$}\\{\footnotesize $k^\star=4,\ \widehat p_{k^\star}=0.484$}}} & RB-SM-FECMC & $1.34\times 10^{-2}$ & $\mathbf{1.09\times 10^{-2}}$ & $4.92\times 10^{-3}$ & $3.52\times 10^{4}$ \\
 & RJ-FECMC & $1.35\times 10^{-2}$ & $1.13\times 10^{-2}$ & $5.40\times 10^{-3}$ & $3.82\times 10^{4}$ \\
 & RJ-HMC ($T=3,\ \epsilon=0.03$) & $1.17\times 10^{-2}$ & $2.27\times 10^{-2}$ & $6.82\times 10^{-3}$ & $3.59\times 10^{4}$ \\
 & RJ-HMC ($T=3,\ \epsilon=0.04$) & $9.03\times 10^{-3}$ & $1.77\times 10^{-2}$ & $5.33\times 10^{-3}$ & $3.73\times 10^{4}$ \\
 & RJ-HMC ($T=3,\ \epsilon=0.05$) & $\mathbf{8.56\times 10^{-3}}$ & $1.50\times 10^{-2}$ & $\mathbf{4.80\times 10^{-3}}$ & $3.71\times 10^{4}$ \\
 & RJ-HMC ($T=10,\ \epsilon=0.007$) & $2.12\times 10^{-2}$ & $4.30\times 10^{-2}$ & $1.28\times 10^{-2}$ & $3.61\times 10^{4}$ \\
\midrule
\multirow{6}{*}{\shortstack{D\\{\footnotesize $(1,1,1,1,1,0,\ldots,0)$}\\{\footnotesize $k^\star=7,\ \widehat p_{k^\star}=0.462$}}} & RB-SM-FECMC & $1.07\times 10^{-2}$ & $\mathbf{1.79\times 10^{-2}}$ & $\mathbf{7.64\times 10^{-3}}$ & $5.85\times 10^{4}$ \\
 & RJ-FECMC & $1.23\times 10^{-2}$ & $2.05\times 10^{-2}$ & $8.36\times 10^{-3}$ & $5.64\times 10^{4}$ \\
 & RJ-HMC ($T=3,\ \epsilon=0.03$) & $1.04\times 10^{-2}$ & $3.60\times 10^{-2}$ & $1.05\times 10^{-2}$ & $5.81\times 10^{4}$ \\
 & RJ-HMC ($T=3,\ \epsilon=0.04$) & $\mathbf{9.34\times 10^{-3}}$ & $3.10\times 10^{-2}$ & $9.53\times 10^{-3}$ & $5.91\times 10^{4}$ \\
 & RJ-HMC ($T=3,\ \epsilon=0.05$) & $1.07\times 10^{-2}$ & $2.96\times 10^{-2}$ & $9.80\times 10^{-3}$ & $5.91\times 10^{4}$ \\
 & RJ-HMC ($T=10,\ \epsilon=0.007$) & $1.70\times 10^{-2}$ & $6.12\times 10^{-2}$ & $1.80\times 10^{-2}$ & $5.54\times 10^{4}$ \\
\midrule
\multirow{6}{*}{\shortstack{E\\{\footnotesize $(1,0.95,0.9,\ldots,0.05)$}\\{\footnotesize $k^\star=20,\ \widehat p_{k^\star}=0.606$}}} & RB-SM-FECMC & $\mathbf{2.13\times 10^{-2}}$ & $\mathbf{5.15\times 10^{-2}}$ & $\mathbf{1.42\times 10^{-2}}$ & $6.45\times 10^{4}$ \\
 & RJ-FECMC & $2.94\times 10^{-2}$ & $6.15\times 10^{-2}$ & $1.70\times 10^{-2}$ & $6.91\times 10^{4}$ \\
 & RJ-HMC ($T=3,\ \epsilon=0.03$) & $1.65\times 10^{-1}$ & $1.05\times 10^{0}$ & $3.10\times 10^{-1}$ & $7.28\times 10^{4}$ \\
 & RJ-HMC ($T=3,\ \epsilon=0.04$) & $3.71\times 10^{-1}$ & $1.85\times 10^{0}$ & $5.82\times 10^{-1}$ & $7.52\times 10^{4}$ \\
 & RJ-HMC ($T=3,\ \epsilon=0.05$) & $4.09\times 10^{-1}$ & $1.99\times 10^{0}$ & $6.35\times 10^{-1}$ & $7.32\times 10^{4}$ \\
 & RJ-HMC ($T=10,\ \epsilon=0.007$) & $2.41\times 10^{-2}$ & $1.59\times 10^{-1}$ & $4.50\times 10^{-2}$ & $6.71\times 10^{4}$ \\
\midrule
\multirow{4}{*}{\shortstack{O\\{\footnotesize $(0,\ldots,0)$}\\{\footnotesize $k^\star=0,\ \widehat p_{k^\star}=0.816$}}} & RB-SM-FECMC & $8.30\times 10^{-3}$ & $5.40\times 10^{-4}$ & $8.39\times 10^{-4}$ & $3.76\times 10^{4}$ \\
 & RJ-FECMC & $6.53\times 10^{-3}$ & $\mathbf{4.61\times 10^{-4}}$ & $\mathbf{7.13\times 10^{-4}}$ & $3.58\times 10^{4}$ \\
 & RJ-HMC ($T=2,\ \epsilon=0.1$) & $\mathbf{5.82\times 10^{-3}}$ & $5.84\times 10^{-4}$ & $8.82\times 10^{-4}$ & $3.92\times 10^{4}$ \\
 & RJ-HMC ($T=3,\ \epsilon=0.05$) & $9.75\times 10^{-3}$ & $8.28\times 10^{-4}$ & $1.26\times 10^{-3}$ & $3.77\times 10^{4}$ \\
\bottomrule
\end{tabular}
\end{table}

\begin{table}[!ht]
\centering
\caption{HMC acceptance rates and relative RMSEs.}
\label{tab:ppca-hmc-acceptance-rmse-X}
\small
\setlength{\tabcolsep}{12pt}\vspace{1mm}
  \resizebox{\linewidth}{!}{%
\begin{tabular}{@{}ccrrrrrr@{}}
\toprule
Scenario
& \shortstack{Posterior\\modal rank $k^\star$}
& $T$ & $\epsilon$
& \shortstack{acceptance\\rate of HMC}
& \shortstack{RMSE ratio\\for $G$}
& \shortstack{RMSE ratio\\for $X$}
& \shortstack{acceptance\\rate of RJ} \\
\midrule
\multirow{2}{*}{O} & \multirow{2}{*}{0}
& 2 & $0.1$ & $0.907$ & $\mathbf{1.083}$ & $\mathbf{1.050}$ & $0.069$ \\
& & 3 & $0.05$ & $0.974$ & $1.534$ & $1.506$ & $0.069$ \\
\midrule
\multirow{4}{*}{C} & \multirow{4}{*}{4}
& 3 & $0.03$ & $0.761$ & $2.092$ & $1.388$ & $0.242$ \\
& & 3 & $0.04$ & $0.625$ & $1.625$ & $1.084$ & $0.242$ \\
& & 3 & $0.05$ & $0.499$ & $\mathbf{1.382}$ & $\mathbf{0.975}$ & $0.242$ \\
& & 10 & $0.007$ & $0.958$ & $3.956$ & $2.611$ & $0.242$ \\
\midrule
\multirow{4}{*}{A} & \multirow{4}{*}{6}
& 3 & $0.03$ & $0.654$ & $1.887$ & $1.447$ & $0.256$ \\
& & 3 & $0.04$ & $0.504$ & $1.579$ & $1.264$ & $0.256$ \\
& & 3 & $0.05$ & $0.407$ & $\mathbf{1.440}$ & $\mathbf{1.175}$ & $0.256$ \\
& & 10 & $0.007$ & $0.931$ & $3.516$ & $2.657$ & $0.256$ \\
\midrule
\multirow{4}{*}{D} & \multirow{4}{*}{7}
& 3 & $0.03$ & $0.477$ & $2.012$ & $1.374$ & $0.255$ \\
& & 3 & $0.04$ & $0.317$ & $1.730$ & $\mathbf{1.247}$ & $0.255$ \\
& & 3 & $0.05$ & $0.206$ & $\mathbf{1.652}$ & $1.282$ & $0.255$ \\
& & 10 & $0.007$ & $0.853$ & $3.418$ & $2.353$ & $0.255$ \\
\midrule
\multirow{4}{*}{B} & \multirow{4}{*}{9}
& 3 & $0.03$ & $0.298$ & $6.073$ & $6.692$ & $0.289$ \\
& & 3 & $0.04$ & $0.138$ & $7.095$ & $7.724$ & $0.288$ \\
& & 3 & $0.05$ & $0.054$ & $7.161$ & $7.804$ & $0.290$ \\
& & 10 & $0.007$ & $0.829$ & $\mathbf{3.374}$ & $\mathbf{2.804}$ & $0.290$ \\
\midrule
\multirow{4}{*}{E} & \multirow{4}{*}{20}
& 3 & $0.03$ & $3.40\times 10^{-4}$ & $20.420$ & $21.868$ & $0.157$ \\
& & 3 & $0.04$ & $6.40\times 10^{-6}$ & $35.793$ & $41.091$ & $0.170$ \\
& & 3 & $0.05$ & $3.81\times 10^{-8}$ & $38.542$ & $44.884$ & $0.163$ \\
& & 10 & $0.007$ & $0.183$ & $\mathbf{3.079}$ & $\mathbf{3.179}$ & $0.151$ \\
\bottomrule
\end{tabular}}
\end{table}

RJ-HMC shows varying acceptance probabilities across different rank strata (Table~\ref{tab;AcceptanceProbability}). This is due to the differences in length scale, which is reflected in the increased number of rejections from order constraints violation. The Vandermonde determinant in the prior \eqref{eq;PCA-prior-on-eigenvalues} serves as the potential barrier to the order constraints, so exact Hamiltonian flows would not reach these boundaries.

\begin{table}[!ht]
\centering
\caption{Acceptance probabilities of the RJ-HMC with $(T,\epsilon)=(3,0.05)$ in scenarios C and D.}
\label{tab;AcceptanceProbability}
\small
\setlength{\tabcolsep}{5pt}
\renewcommand{\arraystretch}{1.08}
\begin{tabular}{@{}ccrrr@{}}
\toprule
Scenario & Rank $k$
& \shortstack{Posterior rank\\probability (\%)}
& \shortstack{HMC acceptance\\rate (\%)}
& \shortstack{Rejection due to\\ordering violations (\%)} \\
\midrule
C & 3 &  8.7 & 69.9 &  3.8 \\
  & 4 & 48.4 & 62.1 &  3.8 \\
  & 5 & 37.6 & 35.6 & 28.1 \\
  & 6 &  5.2 &  7.4 & 80.8 \\
\midrule
D & 7 & 46.2 & 33.0 & 28.3 \\
  & 8 & 40.7 &  4.1 & 86.6 \\
\bottomrule
\end{tabular}
\end{table}

Comparison between SM-FECMC and RJ-FECMC reveals an advantage of the sticky manifold transition. Additional computational efforts spent on RJ proposals are reflected in reduced trajectory lengths in process time; see Table~\ref{tab:ppca-relative-trajectory-length-rmse}. The rejection-free property works particularly favourably in scenarios D and E, while the two between-model transition mechanisms are comparable in scenarios A, B, and C, where RMSEs for model probability estimation are comparable between SM-FECMC and RJ-FECMC. Things are more nuanced between SM-FECMC and RJ-HMC because within-stratum dynamics are also different.

\begin{table}[tbp]
  \centering
  \caption{Relative trajectory lengths based on process time and model-probability RMSEs of RJ-FECMC compared with RB-SM-FECMC. The RJ-HMC column reports the smallest RMSE ratio among the RJ-HMC configurations}
  \label{tab:ppca-relative-trajectory-length-rmse}\vspace{1mm}
  \resizebox{\linewidth}{!}{%
\begin{tabular}{@{}ccccccccc@{}}
    \toprule
    & & &
      \multicolumn{2}{c}{\shortstack{Model probability\\RMSE ratio}}
      & \multicolumn{2}{c}{\shortstack{Mean-$G$\\RMSE ratio}}
      & \multicolumn{2}{c}{\shortstack{Denoising\\RMSE ratio}} \\
    \cmidrule(lr){4-5}
    \cmidrule(lr){6-7}
    \cmidrule(lr){8-9}
    Scenario
    & \shortstack{Mean trajectory\\length (\%)}
    & \shortstack{RJ-FECMC\\RJ acceptance\\probability}
    & RJ-FECMC & RJ-HMC
    & RJ-FECMC & RJ-HMC
    & RJ-FECMC & RJ-HMC \\
    \midrule
    A & 85.5 & 0.2560 & 1.034 & 0.802 & 1.090 & 1.440 & 1.062 & 1.175 \\
    B & 97.5 & 0.2892 & 1.035 & 1.639 & 1.030 & 3.374 & 1.014 & 2.804 \\
    C & 89.9 & 0.2415 & 1.004 & 0.638 & 1.037 & 1.382 & 1.097 & 0.975 \\
    D & 78.2 & 0.2550 & 1.152 & 0.874 & 1.144 & 1.652 & 1.094 & 1.247 \\
    E & 73.4 & 0.2161 & 1.381 & 1.131 & 1.192 & 3.079 & 1.199 & 3.179 \\
    O & 96.6 & 0.1174 & 0.787 & 0.701 & 0.853 & 1.083 & 0.849 & 1.050 \\
    \bottomrule
  \end{tabular}}
\end{table}

\clearpage
\subsection{Survival extrapolation example (Section~\ref{subsec;survival-extrapolation})}

\subsubsection{Algorithm design}

The ambient dynamics are FECMC \citep{Michel2020} with Gaussian velocity. No orthogonal switch is performed, while global refreshment is performed with rate 0.5. Within each submanifold, the arc length parametrisation is used within manifold PDMP (Algorithm~\ref{alg;ManifoldBPS}). As this parametrisation is isometric, we continue to use standard Gaussian velocity on it, as in Example~\ref{example;SM-identity-sticking}.
Event times are computed by thinning with analytic bounds.

\subsubsection{The model}

The accelerated failure time model is specified with a treatment-dependent location parameter as
\[\mu_i=\alpha+\beta z_i\]
with respect to the treatment indicator $z_i$, where $z_i=0$ denotes the control group and $z_i=1$ the treatment group for patient $i$. The separate model fits $\theta_g=(\mu_g,\eta_g,\lambda_g)$ to both groups $g\in\{0,1\}$ respectively. Restricted mean survival time at time $\tau$ is defined by
\[\operatorname{RMST}(\tau):=\int^\tau_0S(t)\,dt,\]
$S(t)$ is the survival function.

\subsubsection{Prior specification and the thawing rates}

In the ambient generalised gamma component, we assign mutually independent priors
\[\alpha\sim N(\log2.5,1.5^2),\quad\beta\sim N(0,1),\quad\eta:=\log\sigma\sim N(0,0.75^2),\quad\lambda\sim N(0,1).\]
For each submodel $i\in\{\mathrm{LN,W},\Gamma,\mathrm{A}\}$, we define the priors as pull-backs through the associated inclusion map $i:\mathcal{M}_i\to\mathbb{R}^4$. In other words, they share the same density up to a normalising factor, with respect to the Hausdorff measure $\mathcal{H}^3_{\mathcal{M}_i}$ such that
\[d\mathcal{H}^3_{\mathcal{M}_i}=\sqrt{1+(r'_i(\eta))^2}\,d\alpha d\beta d\eta,\]
with $\lambda=r_i(\eta)$ being the constraints for the submodel $i$.
The Rao--Blackwellised unsticking time (Corollary~\ref{cor;deterministic-sticking}) is therefore
\[\tau_i=\sqrt{\frac{\pi}{2}}\frac{\omega_i}{\omega_{\mathrm{GG}}}\frac{1}{Z_i},\]
where
\[Z_i:=\int_{\mathbb{R}^3}\phi(\alpha;\log2.5,1.5^2)\phi(\beta;0,1)\phi(\eta;0,0.75^2)\phi(r_i(\eta);0,1)\,\mathcal{H}^3_{\mathcal{M}_i}(d\alpha d\beta d\eta).\]

\subsubsection{Data, initialisation, and output analysis}

We analysed all 810 records in the \texttt{ta174} dataset as distributed with the R package \texttt{survHE} \citep{Baio2020}. We use the observed overall survival times $t_i$ in years, death indicators $\delta_i$, and treatment indicators $z_i$ for all 810 patients. The data comprise 93 deaths and 717 right-censored observations, with a maximum follow-up of four years. The likelihood is
\[
L(\theta)=\prod_{i=1}^{810}
f(t_i\mid z_i,\theta)^{\delta_i}
S(t_i\mid z_i,\theta)^{1-\delta_i}.
\]
The sampler is initialised in the GG component at the posterior mode obtained by Nelder--Mead optimisation. The initial velocity is randomly drawn from $N(0,I_4)$, and the first 200 units of process time are discarded as burn-in. The Monte Carlo standard errors are reported in Tables~\ref{tab:survival-model-probability-mcse} and \ref{tab:survival-rmst-mcse}.

\begin{table}[!ht]
\centering
\caption{Posterior model probabilities with Monte Carlo standard errors for each survival model.}
\label{tab:survival-model-probability-mcse}\vspace{1mm}
\small
\setlength{\tabcolsep}{5pt}
\renewcommand{\arraystretch}{1.1}
\begin{tabular}{@{}lrrrrr@{}}
\toprule
Model prior & Gen-Gamma & Log-normal & Weibull & Gamma & Ammag \\
\midrule
\addlinespace[2pt]
\multicolumn{6}{@{}l}{\textbf{Treatment arms modelled together}} \\
Uniform prior & $8.91\;(0.136)$ & $10.54\;(0.242)$ & $15.07\;(0.304)$ & $10.39\;(0.200)$ & $55.09\;(0.464)$ \\
Informative prior & $6.66\;(0.088)$ & $30.93\;(0.719)$ & $11.75\;(0.284)$ & $8.16\;(0.165)$ & $42.49\;(0.651)$ \\
\midrule
\addlinespace[2pt]
\multicolumn{6}{@{}l}{\textbf{Treatment arms modelled separately}} \\
\multicolumn{6}{@{}l}{\textit{Control}} \\
Uniform prior & $10.50\;(0.125)$ & $11.68\;(0.233)$ & $18.64\;(0.277)$ & $13.17\;(0.185)$ & $46.00\;(0.418)$ \\
Informative prior & $7.89\;(0.094)$ & $34.29\;(0.784)$ & $13.71\;(0.309)$ & $10.12\;(0.179)$ & $34.00\;(0.637)$ \\
\addlinespace[2pt]
\multicolumn{6}{@{}l}{\textit{Treatment}} \\
Uniform prior & $9.21\;(0.111)$ & $9.53\;(0.213)$ & $16.17\;(0.327)$ & $10.97\;(0.212)$ & $54.12\;(0.485)$ \\
Informative prior & $7.07\;(0.070)$ & $28.62\;(0.611)$ & $12.87\;(0.320)$ & $8.68\;(0.214)$ & $42.76\;(0.520)$ \\
\bottomrule
\end{tabular}
\end{table}

\begin{table}[!ht]
\centering
\caption{Monte Carlo standard errors of estimates of the 30-year restricted mean survival time difference.}
\label{tab:survival-rmst-mcse}
\small
\setlength{\tabcolsep}{5pt}
\renewcommand{\arraystretch}{1.1}
\begin{tabular}{@{}lrrrrr@{}}
\toprule
& \multicolumn{5}{c}{Model averaged} \\
\cmidrule(lr){2-6}
Model prior & \shortstack{Posterior\\mean}
  & MCSE & \shortstack{Relative\\MCSE (\%)}
  & \shortstack{MCSE range\\across batch counts}
  & \shortstack{Half-run\\difference} \\
\midrule
\addlinespace[2pt]
\multicolumn{6}{@{}l}{\textbf{Treatment arms modelled together}} \\
Uniform prior & $2.7422$ & $0.00468$ & $0.17$ & $0.00329$--$0.00478$ & $-0.00679$ \\
Informative prior & $2.5094$ & $0.00956$ & $0.38$ & $0.00894$--$0.00999$ & $-0.01537$ \\
\midrule
\addlinespace[2pt]
\multicolumn{6}{@{}l}{\textbf{Treatment arms modelled separately}} \\
Uniform prior & $3.6319$ & $0.02126$ & $0.59$ & $0.01917$--$0.02141$ & $-0.01335$ \\
Informative prior & $2.6499$ & $0.04274$ & $1.61$ & $0.03750$--$0.04625$ & $+0.08710$ \\
\bottomrule
\end{tabular}
\end{table}

\subsubsection{Additional results}

Prior and posterior model probabilities are shown in Table \ref{tab:model-probability-prior-sensitivity}. The results in Section \ref{subsec;survival-extrapolation} for the independent models for treatment arms are based on the informative prior for the control arm and uniform prior for the treatment arm. We also include the posterior probabilities for when the treatment arm also uses the informative prior.

\begin{table}[!ht]
\centering
\caption{Prior and posterior model probabilities for the joint AFT model and the treatment-only and control-only models.}
\label{tab:model-probability-prior-sensitivity}
\setlength{\tabcolsep}{9pt}\vspace{1mm}
\begin{tabular}{lrrrrrr}
\toprule
& \multicolumn{2}{c}{Joint model} & \multicolumn{2}{c}{Treatment only} & \multicolumn{2}{c}{Control only} \\
\cmidrule(lr){2-3}\cmidrule(lr){4-5}\cmidrule(lr){6-7}
Model & Prior & Posterior & Prior & Posterior & Prior & Posterior \\
\midrule
\addlinespace[2pt]
\multicolumn{7}{l}{\textbf{Uniform prior}} \\
GG & $0.200$ & $0.0891$ & $0.200$ & $0.0921$ & $0.200$ & $0.1050$ \\
Log-normal & $0.200$ & $0.1054$ & $0.200$ & $0.0953$ & $0.200$ & $0.1168$ \\
Weibull & $0.200$ & $0.1507$ & $0.200$ & $0.1617$ & $0.200$ & $0.1864$ \\
Gamma & $0.200$ & $0.1039$ & $0.200$ & $0.1097$ & $0.200$ & $0.1317$ \\
Ammag & $0.200$ & $0.5509$ & $0.200$ & $0.5412$ & $0.200$ & $0.4600$ \\
\midrule
\addlinespace[2pt]
\multicolumn{7}{l}{\textbf{Informative prior}} \\
GG & $0.125$ & $0.0666$ & $0.125$ & $0.0707$ & $0.125$ & $0.0789$ \\
Log-normal & $0.500$ & $0.3093$ & $0.500$ & $0.2862$ & $0.500$ & $0.3429$ \\
Weibull & $0.125$ & $0.1175$ & $0.125$ & $0.1287$ & $0.125$ & $0.1371$ \\
Gamma & $0.125$ & $0.0816$ & $0.125$ & $0.0868$ & $0.125$ & $0.1012$ \\
Ammag & $0.125$ & $0.4249$ & $0.125$ & $0.4276$ & $0.125$ & $0.3400$ \\
\bottomrule
\end{tabular}
\end{table}

Tha full table of Table~\ref{tab:rmst-prior-sensitivity_full} is as follows:

\begin{table}[!ht]
\centering
\caption{Estimates of 30-year restricted mean survival time (years). Entries are posterior means (95\% credible intervals).}
\label{tab:rmst-prior-sensitivity_full}
\setlength{\tabcolsep}{7pt}\vspace{1mm}
\resizebox{\linewidth}{!}{%
\begin{tabular}{lrrrrrrr}
\toprule
 & & & & & & \multicolumn{2}{c}{Model averaged} \\
\cmidrule(lr){7-8}
Model & Gen-Gamma & Log-normal & Weibull & Gamma & Ammag & Uniform prior & Informative prior \\
\midrule
\addlinespace[2pt]
\multicolumn{8}{l}{\textbf{Treatment arms modelled together}} \\
Control    & $15.63\;(8.82,\,20.70)$ & $18.16\;(15.74,\,20.48)$ & $13.10\;(9.53,\,16.95)$ & $13.14\;(10.08,\,16.55)$ & $13.08\;(8.49,\,18.02)$ & $13.85\;(8.81,\,19.48)$ & $14.82\;(9.02,\,19.96)$ \\
Treatment  & $17.89\;(11.56,\,22.05)$ & $19.85\;(17.47,\,22.13)$ & $15.96\;(12.04,\,19.82)$ & $15.99\;(12.49,\,19.61)$ & $16.04\;(11.33,\,20.41)$ & $16.59\;(11.65,\,21.22)$ & $17.33\;(11.93,\,21.63)$ \\
Difference & $2.26\;(-0.56,\,5.73)$ & $1.70\;(-0.55,\,3.89)$ & $2.86\;(-0.58,\,6.32)$ & $2.85\;(-0.58,\,6.28)$ & $2.97\;(-0.64,\,6.72)$ & $2.74\;(-0.61,\,6.44)$ & $2.51\;(-0.58,\,6.24)$ \\
\midrule
\addlinespace[2pt]
\multicolumn{8}{l}{\textbf{Treatment arms modelled separately}} \\
Control    & $14.81\;(7.86,\,20.53)$ & $17.19\;(14.17,\,20.08)$ & $12.25\;(8.30,\,16.78)$ & $12.44\;(9.02,\,16.41)$ & $12.74\;(8.35,\,17.95)$ & $13.35\;(8.45,\,19.14)$ & $14.33\;(8.70,\,19.55)$ \\
Treatment  & $18.25\;(11.31,\,22.98)$ & $20.10\;(17.17,\,22.73)$ & $16.33\;(11.37,\,20.75)$ & $16.28\;(12.08,\,20.42)$ & $16.55\;(11.37,\,21.34)$ & $16.98\;(11.53,\,21.90)$ & $16.98\;(11.53,\,21.90)$ \\
Difference & $3.45\;(-5.71,\,12.34)$ & $2.92\;(-1.20,\,6.93)$ & $4.08\;(-2.56,\,10.21)$ & $3.84\;(-1.89,\,9.31)$ & $3.80\;(-3.59,\,10.67)$ & $3.63\;(-4.42,\,11.01)$ & $2.65\;(-5.28,\,10.58)$ \\
\bottomrule
\end{tabular}}
\end{table}

\end{document}

%% file: preamble.tex
\definecolor{MyGreen}{cmyk}{1,0.07,0.10,0.10}\PassOptionsToPackage{colorlinks,pagebackref,linkcolor=MyGreen,urlcolor=MyGreen,citecolor=MyGreen}{hyperref}\usepackage{url}\usepackage[colorlinks,pagebackref,linkcolor=MyGreen,urlcolor=MyGreen,citecolor=MyGreen]{hyperref}\usepackage[all]{xy}\usepackage{booktabs}\usepackage{makecell}\usepackage{multirow}
\usepackage{lmodern}\usepackage{algorithm}
\usepackage{algpseudocode}\newcommand{\tpar}{{\!\!\scalebox{0.6}{\rotatebox{45}{$\parallel$}}}}
\newcommand{\abs}[1]{\lvert#1\rvert}\newcommand{\Abs}[1]{\left|#1\right|}\newcommand{\norm}[1]{\|#1\|}\newcommand{\Brace}[1]{\left\{#1\right\}}\newcommand{\paren}[1]{\left(#1\right)}\newcommand{\Paren}[1]{\biggr(#1\biggl)}
\newcommand{\ov}[1]{\overline{#1}}\newcommand{\brac}[1]{\langle#1\rangle}

\usepackage{etoolbox}
\renewcommand{\do}[1]{\csdef{b#1}{\mathbb{#1}}}\docsvlist{A,B,C,D,E,F,G,H,I,J,K,L,M,N,O,P,Q,R,S,T,U,V,W,X,Y,Z}
\renewcommand{\do}[1]{\csdef{c#1}{\mathcal{#1}}}\docsvlist{A,B,C,D,E,F,G,H,I,J,K,L,M,N,O,P,Q,R,S,T,U,V,W,X,Y,Z}
\renewcommand{\do}[1]{\csdef{f#1}{\mathfrak{#1}}}\docsvlist{A,B,C,D,E,F,G,H,I,J,K,L,M,N,O,P,Q,R,S,T,U,V,W,X,Y,Z}
\newcommand{\R}{\mathbb{R}}

\newcommand{\mono}{\hookrightarrow}

\newcommand{\RelaxCommands}[1]{\renewcommand*{\do}[1]{\expandafter\let\csname ##1\endcsname\relax}\docsvlist{#1}}
\RelaxCommands{P}
\newcommand{\DeclareMathOperators}[1]{\renewcommand*{\do}[1]{\expandafter\DeclareMathOperator\csname ##1\endcsname{##1}}\docsvlist{#1}}
\DeclareMathOperators{P,E,V,C,id,grad,dist,diam}

\usepackage{xcolor}\definecolor{amethyst}{rgb}{0.6, 0.4, 0.8}

\numberwithin{equation}{section}
\newtheorem{theorem}{Theorem}[section]
\newtheorem{proposition}[theorem]{Proposition}
\newtheorem{assumption}[theorem]{Assumption}
\newtheorem{lemma}[theorem]{Lemma}
\newtheorem{corollary}[theorem]{Corollary}
\theoremstyle{definition}
\newtheorem{example}[theorem]{Example}
\newtheorem{remark}[theorem]{Remark}
\newtheorem{definition}[theorem]{Definition}

\usepackage{enumitem}\setlist[enumerate,1]{label=(\arabic*), itemsep=0pt}
\usepackage{float}

\usepackage{authblk}
\renewcommand{\d}{\mathrm{d}}

\newcommand{\keywords}[1]{%
  \par\smallskip
  \noindent\textit{Keywords:} #1
}